\documentclass[aps,prb,twocolumn,superscriptaddress,floatfix]{revtex4}

\usepackage{amscd,rotating}
\usepackage{amsmath} 
\usepackage{amssymb}
\usepackage{bm}
\usepackage{color}
\usepackage{graphicx}     
\usepackage{longtable}

\begin{document}

\title{Antiferromagnetism and singlet formation \\ in underdoped high-Tc cuprates: \\ Implications for superconducting pairing}
\author{Damien Bensimon}
\altaffiliation{Present adress: Max-Planck-Institut f\"ur Festk\"orperfor-schung,  
Heisenbergstrasse 1, 
D-70569 Stuttgart, Germany.}
\affiliation{Department of Applied Physics, University of Tokyo, 
Bunkyo-ku, 
Tokyo 113-8656, Japan}
\author{Naoto Nagaosa}
\affiliation{CREST, Department of Applied Physics, University of Tokyo, 
Bunkyo-ku, Tokyo 113-8656, Japan}
\affiliation{Correlated Electron Research Center (CERC), 
National Institute of Advanced Industrial Science and Technology 
(AIST),
Tsukuba Central 4, Tsukuba 305-8562, Japan}
\date{\today}
%\date{June 28, 2004}

\begin{abstract}
The extended $t-J$ model is theoretically studied, in the context of hole underdoped cuprates.
Based on results obtained by recent numerical studies, we identify the mean field 
state having both the antiferromagnetic and staggered flux resonating valence bond orders.  
The random-phase approximation is employed
to analyze all the possible collective modes in this mean field state.
In the static (Bardeen Cooper Schrieffer) limit justified in the weak coupling regime, we obtain the effective 
superconducting interaction between the doped holes at 
the small pockets located around $\bm{k}= (\pm \pi/2, \pm \pi/2)$.
In contrast to the spin-bag theory, which takes into acccount only the
antiferromagnetic order, this effective force is pair breaking for 
the pairing without the nodes in each of the small hole pocket, 
and is canceled out to be very small for 
the $d_{x^2-y^2}$ pairing with nodes which is realized in the real cuprates.
Therefore we conclude that no superconducting instability can occur when only the magnetic mechanism is considered.
The relations of our work with other approaches are also discussed.
\end{abstract} 

\maketitle

%%%%%%%%%%%%%%%%%%%%%%%%%%%%%%%%%%%%%%%%%%%%%%%%%%%%%%%%%%%%%%%%%%%%%%%%%%%%%%%%%%%%%%%%%%%%%%%%%%%%%%%%%%%%%%%%%%%%%%%%%%%%%%%%%%%%%%%
%%%%%%%%%%%%%%%%%%%%%%%%%%%%%%%%%%%%%%%%%%%%%%%%%%%%%%%%%%%%%%%%%%%%%%%%%%%%%%%%%%%%%%%%%%%%%%%%%%%%%%%%%%%%%%%%%%%%%%%%%%%%%%%%%%%%%%%

\section{Introduction}
\label{Intro}

Since the discovery of high temperature superconduc\-ting 
cuprates, \cite{Bednorz1986} it has been established that 
the strong Coulomb repulsion 
between electrons plays an essential role in the physics there
\cite{RMP}
and the magnetic mechanism of the 
superconductivity has been studied intensively. There are two streams 
of 
thoughts; one is the antiferromagnetic (AF) spin fluctuation exchange 
based on
the (nearly) AF ordered state, while the other is the reso\-nating 
valence bond 
(RVB) mechanism with the focus being put on the spin singlet formation 
by
the kinetic exchange interaction $J$.
The representative of the former is the spin-bag theory 
\cite{Schrieffer1989,Schrieffer1989long}
where the doped carriers into the spin density wave (SDW) state 
form small hole pockets, and exchange the AF spin fluctuation to result 
in the $d_{xy}$ superconductivity. The $z$-component of the spin 
fluctuation,
i.e., $\chi^{zz}$, gives the dominant contribution.
The RVB picture, on the other hand, puts more weight on the 
spin singlet formation and takes into account the order parame\-ter 
defined 
on the bond but usual\-ly does not consider the antiferromagnetic 
long range ordering (AFLRO). \cite{Anderson1987}
These two scenarios have been stu\-died rather separately thus far, 
and the 
relation between them remains unclear. Hsu was the first to take into
account both the AFLRO and the RVB correlation at half-filling in terms of the 
Gutzwiller approximation. \cite{Hsu1990} His picture is that the 
AFLRO occurs on top of the $d$-wave RVB or equivalently the flux state.
It is also supported by the variational Monte Carlo study on Heisenberg
model, which shows that the Gutzwiller-projected wavefunction 
$\Psi_{\rm RVB+SDW}$ starting from the coexisting $d$-wave RVB and SDW 
mean field state gives an excellent agreement with the exact 
diagonalization
concerning the ground state ener\-gy and staggered moment. \cite{tjhole} 
The varia\-tional wavefunction projecting the simple SDW state, on the 
other hand, gives higher energy. 
These results mean that both aspects, i.e., SDW and RVB, coexist in the 
Mott insulator.
At finite hole doping concentration $\mathrm{x}$, the AFLRO is rapidly 
suppressed
and the superconductivity emerges. Here an important question still 
remains, namely short range AF fluctuation dominates or the RVB 
correlation is more important.
Because there is no AFLRO and no distinction between 
$\chi^{zz}$ and $\chi^{\pm}$, this question might appear an academic
one with the reality being somewhere inbetween.
Well-known results from the variational Monte Carlo studies are that 
the variational wavefunction $\Psi_{\rm RVB+SDW}$ has the lowest energy 
therefore
RVB (superconductivity) and SDW coexist also at small do\-ping level. \cite{ogatahimeda}
From the viewpoint of the 
particle-hole SU(2) symmetry, this means that the degeneracy, 
i.e., the SU(2) gauge symmetry, between $d$-wave pairing and $\pi$-flux 
state is lifted at finite $\mathrm{x}$, and the former has lower energy. \\
\indent However this accepted view has been challenged by recent studies 
based on the exact diagonalization \cite{tjexact} and 
variational Monte Carlo method. \cite{tjhole}
These works found that the energetically most favorable state
with small pockets 
at small hole doping levels is not superconduc\-ting. This nonsuperconducting state is 
consistent with the doped SDW+$\pi$-flux state with the hole pockets 
near $\bm{k}=(\pi/2,\pm \pi/2)$.
This means that the mean field picture is not so reliable at small $\mathrm{x}$ 
because the phase fluctuation of the superconductivity is huge at 
small $\mathrm{x}$ where the charge density, which is the canonical conjugate 
operator to the phase, is suppressed.  Once the superconductivity is 
destroyed by this quantum fluctuation, the only order surviving is the 
AFLRO, and the system remains nonsuperconducting. Considering that the 
AFLRO state at $\mathrm{x}=0$ is well described by the SDW+($d$-wave  RVB or 
$\pi$-flux)
state, the relevant mean field state at small but finite $\mathrm{x}$ is 
SDW+$\pi$-flux state with small hole pockets. \\
\indent  From the experimental side, recent ARPES datas on Na-CCOC found the 
small "Fermi arc" near  $\bm{k}=(\pi/2,\pm \pi/2)$, and the $\bm{k}$-dependence of the "pseudo-gap" is quite different from that of $(\cos k_x - \cos k_y)$ 
expected for the $d_{x^2-y^2}$ pairing. \cite{Shen-PRB2003,Shen-JPSJ2003} 
This result strongly suggests that the 
pseudo-gap is  distinct from the superconduc\-ting gap, 
and the superconductivity 
comes from some other interaction(s) different from $J$. 
Although only the "arc" is observed experimentally, the Fermi 
surface can not terminate at some $\bm{k}$-points inside of the 
Brillouin zone (BZ). Considering that the 
Fermi surface disappears with a large pseudogap at the anti-nodal 
direction, i.e., near $\bm{k}=(\pi, 0)$ and $(0,\pi)$, it is natural 
to assume that the small hole pocket is formed, along half of 
which the intensity is small and/or too  broad to be observed 
experimentally. 
\cite{note}\
Then the question of the pairing symmetry arises because 
$d_{x^2-y^2}$ requires the nodes 
at the small hole pockets, which usually reduces the condensation 
energy
and is energetically unfavorable. The natural symmetry appears to be 
$d_{xy}$ without the nodes at the hole pockets, as has been 
claimed by the original spin-bag scenario. \cite{Schrieffer1989,Schrieffer1989long} \\
\indent In this paper, we study the superconductivity of the 
doped SDW+$\pi$-flux state for small $\mathrm{x}$. This state includes both 
the RVB correlation and the AFLRO. There are two important effects of 
this RVB correlation; one is to introduce the parity anomaly to the 
nodal 
Dirac fermions at $\bm{k}=(\pi/2,\pm \pi/2)$ and the other is to 
enhance the transverse spin-spin correlation $\chi^{\pm}$ compared
with the longitudinal $\chi^{zz}$. These two aspects might have 
crutial influence on the superconductivity in the underdoped region.
We have employed the $1/N$-expansion, or random-phase 
approximation (RPA), to derive the 
effective interaction between the quasi-particles along the small
hole pockets and the pairing force derived from it.  \\
\indent The plan of this paper follows. In Section \ref{Mod_MFT}, we discuss the 
model, formulation, and its mean field treatment. The Gaussian
(second order) fluctuation around the mean field saddle point 
is treated and the effective interactions between the 
quasi-particles are studied in Section 
\ref{GaussInt-Pair}. Section \ref{Disc-Conc} is
devoted to discussion and conclusions.

%%%%%%%%%%%%%%%%%%%%%%%%%%%%%%%%%%%%%%%%%%%%%%%%%%%%%%%%%%%%%%%%%%%%%%%%%%%%%%%%%%%%%%%%%%%%%%%%%%%%%%%%%%%%%%%%%%%%%%%%%%%%%%%%%%%%%%
%%%%%%%%%%%%%%%%%%%%%%%%%%%%%%%%%%%%%%%%%%%%%%%%%%%%%%%%%%%%%%%%%%%%%%%%%%%%%%%%%%%%%%%%%%%%%%%%%%%%%%%%%%%%%%%%%%%%%%%%%%%%%%%%%%%%%%

\section{Model and Mean Field Theory}
\label{Mod_MFT}

%%%%%%%%%%%%%%%%%%%%%%%%%%%%%%%%%%%%%%%%%%%%%%%%%%%%%%%%%%%%%%%%%%%%%%%%%%%%%%%%%%%%%%%%%%%%%%%%%%%%%%%%%%%%%%%%%%%%%%%%%%%%%%%%%%%%%%

\subsection{Formulation of the microscopic model}
\label{Formul-MicroMod}

\begin{figure}[t]
\includegraphics[width=62mm]{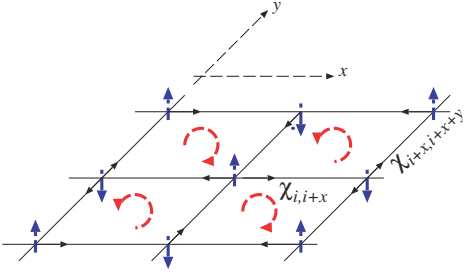}
\caption {(Color online) Square lattice exhibiting simultaneously a staggered magnetization and a flux phase.} 
\label{AFFlux-Fig}
\end{figure}

The microscopic model of cuprates we consider for the study is the well-known 
$t-J$ model. \cite{Anderson1987,Rice1988} 
The second and third nearest neighbor hopping terms are taken into account, \cite{Fukuyama1993} as they are necessary to describe the ARPES experiments measurements. \cite{ARPES-1998}
 We use the slave bosons 
representation \cite{Barnes1976,Coleman1984} of the electronic 
operators
\begin{eqnarray}
 c_{i,\sigma}^{\dagger} = f_{i,\sigma}^{\dagger}b_{i}, 
\end{eqnarray}
with the constraint
\begin{eqnarray}
 b_{i}^{\dagger}b_{i} + \sum_{\sigma=\downarrow,\uparrow} 
f_{i,\sigma}^{\dagger}f_{i,\sigma}  = 1.
\label{SB-const}
 \end{eqnarray} 
In terms of this formalism, the double occupancy of each site is 
excluded. 
\cite{AndersontSL,Fukuyama1988,Lavagna1994}
The operators $f^\dagger$, $f$ are the fermionic ones while 
$b^\dagger$, $b$ are bosonic ones (slave bosons).
\cite{KivRokSet,AndersonPRL1990} 
In all what follows we will exclusively work at 
zero temperature, and all the slave bosons will be assumed to be condensed. 
(In actual calculation, we take an extremely low temperature 
by technical reason.
The results, however, are saturated and can be regarded 
as those at zero temperature.) \\
\indent We consider the following Hamiltonian on a square lattice with the 
lattice constant put to be unity and containing $\mathrm{N_{s}}$ sites
\begin{eqnarray}
\mathcal{H} =  &-& \Big( \sum_{\langle i,j \rangle} t_{ij} +  
\sum_{\langle i,j \rangle^{'}}t_{ij}^{'}  + \sum_{\langle i,j \rangle^{''}} 
t_{ij}^{''} \Big) \nonumber \\
& &\hspace{1mm} \times \sum_{\sigma=\downarrow,\uparrow}\left[ b_{i} 
b_{j}^{\dagger} f_{i,\sigma}^{\dagger}f_{j,\sigma} + b_{j} b_{i}^{\dagger} 
f_{j,\sigma}^{\dagger}f_{i,\sigma} \right] \nonumber \\ 
          & + & J \sum_{\langle i,j \rangle } {\mathbf{S}}_{i} \cdot 
{\mathbf{S}}_{j},
\label{Hamil-model}
 \end{eqnarray} 
where $\langle \rangle$, $\langle \rangle^{'}$, $\langle \rangle^{''}$ 
denote 
nearest neighbor, second nea\-rest neighbor and third nearest 
neighbor pair,
respectively. We assume that the hopping 
contribution is uniform: $  t_{ij} = t $ , $ t_{ij}^{'} = t^{'} $ , $ t_{ij}^{''} = 
t^{''} $. For $ t' < 0 \ , \ t'' > 0$ the previous Hamiltonian deals with 
hole doped (p-type) cuprates, while the case $ t' > 0 \ , \ t'' < 0$  
concerns electron doped (n-type) cuprates. The unity of energy is: $ t 
\cong 0.4 \ \mbox{eV} =1 $, and we will 
only study the hole doped case. Following Ref.~\onlinecite{ARPES-1998} we assume $t^{'} = -0.3 $, $t^{''} = 0.2 $ 
for the extended hopping terms, and $J = 0.3 $ for the Heisenberg interaction.
 We have checked that slight variations in the numerical values of these parameters 
 do not induce any significant modification of our results, as discussed in 
 Section \ref{Calc-BCSpai}.  \\
\indent We construct the mean field theory to study this Hamiltonian. 
It has been recognized that the validity of the mean field theory is 
much less 
trivial in the theo\-ry with constraint. The simple comparison 
of energy does not offer the criterion because the mean field theory 
vio\-lates the constraint, and hence can lower the energy in the 
physically forbidden Hilbert space. Therefore the choice of the mean 
field 
theory requires a physical intuition, and we employ the 
nonsuperconducting 
saddle point according to the reasons explained in the Introduction.
In order to study simultaneously the influences of antiferromagnetism 
and flux in the resonating valence bond state, we introduce the two mean field 
parameters. The first one is a staggered magnetization
\cite{Schrieffer1989,Schrieffer1989long}
\begin{eqnarray}
  m_{i}^{s_{m}} = \langle \mathbf{S}_{i}^{s_{m}} \rangle \ , \ 
% \nonumber \\
 \mathbf{S}_{i}^{s_{m}} =  \frac{\displaystyle 1}{\displaystyle 2}  
\sum_{\sigma,\sigma^{'}} f_{i,\sigma}^{\dagger} 
\widetilde{\sigma}_{\sigma,\sigma^{'}}^{s_{m}} f_{i,\sigma^{'}} \ , \ 
\end{eqnarray} 
with $s_{m} = x$, $y$, $z$, and $\widetilde{\sigma}^{s_{m}}$ being the Pauli matrices. 
The second one is the flux phase parameter \cite{Affleck1989,Marston1989}
\begin{eqnarray}
& &  \chi_{ij} =  \langle \sum_{\sigma=\downarrow,\uparrow} 
f_{i,\sigma}^{\dagger}f_{j,\sigma}\rangle  \ , \ \chi = |\chi_{ij}|, \nonumber \\
& &\chi_{i,i+x} =  \chi \cdot \mathrm{exp}\left[ +\mathrm{i} 
\frac{\displaystyle \phi}{\displaystyle 4 } (-1)^{i} \right] \ , \nonumber \\  
& & \chi_{i,i+y} = \chi \cdot \mathrm{exp}\left[ -\mathrm{i} \frac{\displaystyle 
\phi}{\displaystyle 4 } (-1)^{i} \right]. 
\end{eqnarray} 
Fig.\! \ref{AFFlux-Fig} shows the pattern of the flux and the staggered 
magnetization on the square lattice.

We rewrite the model Hamiltonian (\ref{Hamil-model}) as
\begin{equation}
  \mathcal{H} = \mathcal{H}^{t} + \mathcal{H}^{m} + \mathcal{H}^{\chi},
\end{equation}
with
\begin{eqnarray}
  & \mathcal{H}^{t} & = \Big(  - \sum_{\langle i,j \rangle} t -  
\sum_{\langle i,j \rangle^{'}}t^{'}  - \sum_{\langle i,j \rangle^{''}} t^{''} 
\Big) \nonumber \\
& & \hspace{5mm}\times \sum_{\sigma}\left[ b_{i} b_{j}^{\dagger} 
f_{i,\sigma}^{\dagger}f_{j,\sigma} + b_{j} b_{i}^{\dagger} f_{j,\sigma}^{\dagger}f_{i,\sigma} 
\right],  \nonumber 
\end{eqnarray} 
\vfill\eject
\begin{eqnarray}
  & \mathcal{H}^{m} & = \alpha J \sum_{\langle i,j \rangle } 
\mathbf{S}_{i} \cdot \mathbf{S}_{j} \ ,  
\nonumber \\
& \mathcal{H}^{\chi} &  =  (1 - \alpha) J \sum_{\langle i,j \rangle } 
\mathbf{S}_{i} \cdot \mathbf{S}_{j}~.
\end{eqnarray} 
The mean field Hamiltonian for each parameter is given by
\begin{eqnarray}
 \mathcal{H}_{MF}^{m} =&\alpha J& \sum_{\langle i,j \rangle } 
\sum_{s_{m} = x,y,z} \big[ \ m_{j}^{s_{m}}\mathbf{S}_{i}^{s_{m}}+ m_{i}^{s_{m}}\mathbf{S}_{j}^{s_{m}} \nonumber \\
& &\hspace{21mm} - m_{i}^{s_{m}} m_{j}^{s_{m}} \big] \ , \nonumber \\
%\end{eqnarray}
%\begin{eqnarray}
 \mathcal{H}_{MF}^{\chi} =  &-& \frac{\displaystyle (1 - \alpha) 
J}{\displaystyle 2} \sum_{\langle i,j \rangle } \sum_{\sigma} \big[ 
\chi_{ij} f_{j,\sigma}^{\dagger}f_{i,\sigma} +  
\chi_{ij}^{*}f_{i,\sigma}^{\dagger}f_{j,\sigma} \big] \nonumber \\
&+& \frac{\displaystyle (1 - \alpha) 
J}{\displaystyle 2} \sum_{\langle i,j \rangle } \chi_{ij} \chi_{ij}^{*}. \nonumber
\end{eqnarray} 
In the previous expressions a parameter $\alpha$ has been introduced to divide the Heisenberg 
exchange interaction into AF part and flux part.
Here, the value of $\alpha$ is determined so as to reproduce the optimized result 
by the Gutzwiller variational method at half-filling. \cite{Hsu1990}
Hence $\alpha$ should be regarded as a variational parameter, but we
will keep it constant ($\alpha= 0.301127$, as explained in Section \ref{Sub-MFE}) independently of the do\-ping.
We work in a path integral formalism by using the 
Stratonovitch - Hubbard transformation. \cite{Strato,Hub}
The partition function is written
\begin{eqnarray}
   \mathcal{Z} = \int \mathcal{D}\bar{\Psi} \mathcal{D}\Psi 
\mathcal{D}\chi \mathcal{D}m \mathcal{D}\lambda \mathcal{D}b \ 
\mathrm{exp} \left( - \int_{0}^{\beta}d\tau \ \mathcal{L}(\tau)\right),
\end{eqnarray} 
with $\beta = 1/k_{B}T$ the thermal factor and the Lagrangian
\begin{widetext}
\begin{eqnarray}
\mathcal{L}(\tau) =
 &-& \big( \sum_{\langle i,j \rangle} t +  \sum_{\langle i,j 
\rangle^{'}}t^{'}  + \sum_{\langle i,j \rangle^{''}} t^{''} \big) \sum_{\sigma}
\left[ b_{i}(\tau) b_{j}^{\dagger}(\tau) \bar{\Psi}_{i,\sigma}(\tau) 
\Psi_{j,\sigma}(\tau)
+ b_{j}(\tau) b_{i}^{\dagger}(\tau) \bar{\Psi}_{j,\sigma}(\tau) 
\Psi_{i,\sigma}(\tau) \right] \nonumber \\ 
&+& \sum_{i,\sigma} \bar{\Psi}_{i,\sigma}(\tau) \big[ \partial_{\tau} - 
\mathrm{i}\lambda_{i}(\tau) - \mu \big] \Psi_{i,\sigma}(\tau) 
+ \sum_{i} b_{i}^{\dagger}(\tau) \big[ \partial_{\tau} - 
\mathrm{i}\lambda_{i}(\tau)] b_{i}(\tau)  \nonumber \\ 
&+& \alpha J \sum_{\langle i,j \rangle } \sum_{\sigma,\sigma^{'}} 
\sum_{s = x,y,z} 
\big[ m_{j}^{s}(\tau) \bar{\Psi}_{i,\sigma}(\tau) 
\widetilde{\sigma}_{\sigma,\sigma^{'}}^{s} \Psi_{i,\sigma^{'}}(\tau)
+ m_{i}^{s}(\tau) \bar{\Psi}_{j,\sigma}(\tau) 
\widetilde{\sigma}_{\sigma,\sigma^{'}}^{s} \Psi_{j,\sigma^{'}}(\tau) \big] \nonumber \\ 
&-& \frac{\displaystyle (1 - \alpha) J}{\displaystyle 2} \sum_{\langle 
i,j \rangle } \sum_{\sigma} 
\big[ \chi_{ij}(\tau) \bar{\Psi}_{j,\sigma}(\tau) \Psi_{i,\sigma}(\tau) 
+  \chi_{ij}^{*}(\tau) \bar{\Psi}_{i,\sigma}(\tau) 
\Psi_{j,\sigma}(\tau) \big] \nonumber \\
&-& \alpha J \sum_{\langle i,j \rangle }
 \big[m_{i}^{x}(\tau) m_{j}^{x}(\tau) + m_{i}^{y}(\tau) m_{j}^{y}(\tau) 
+  m_{i}^{z}(\tau) m_{j}^{z}(\tau)\big] 
+  \frac{\displaystyle (1 - \alpha) J}{\displaystyle 2} \sum_{\langle 
i,j \rangle } \chi_{ij}(\tau) \chi_{ij}^{*}(\tau).
\label{Lagran_MF}
\end{eqnarray} 
%\end{widetext}
In Eq.\! (\ref{Lagran_MF}) $\bar{\Psi}_{i,\sigma}$, $\Psi_{i,\sigma}$ are 
the Grassmann variables associated with the $f_{i,\sigma}^{\dagger}$, 
$f_{i,\sigma}$ operators, respectively. The Lagrange multipliers $(\mathrm{i} 
\lambda_{i})$ assure that the constraint of no double occupancy (\ref{SB-const}) is satisfied. 
The chemical potential $\mu$ associated with the fermions controls the 
electron density $n_{f}$, and hence the hole dopping level $\mathrm{x}$.
\vspace{25mm}
\end{widetext}

%%%%%%%%%%%%%%%%%%%%%%%%%%%%%%%%%%%%%%%%%%%%%%%%%%%%%%%%%%%%%%%%%%%%%%%%%%%%%%%%%%%%%%%%%%%%%%%%%%%%%%%%%%%%%%%%%%%%%%%%%%%%%%%%%%%%%%

\newpage
% \begin{figure}[t]
% \includegraphics[width=82mm]{Fig2-Brillouin.pdf}
% \caption {(Color online) First Brillouin zone and reduced (ma\-gnetic) Brillouin zone for a square lattice. } 
% \label{Brillouin-Fig}
% \end{figure}

\subsection{Saddle point hypothesis and derivation of the mean field Hamiltonian}
  To identify the saddle point solution we consider the following 
assumptions. As we work at zero temperature, all the 
holons are supposed to be condensed, i.e., 
$(b_{i})_{0} = b_{0}$ with $(b_{0})^{2} = \mathrm{x}$ where $\mathrm{x}$ is the 
hole doping concentration. 
Therefore the $f$ operators can be simply viewed as the 
renormalized electron operators.
The Lagrange multipli\-ers are considered independent of the sites, which 
signifies 
that the constraint is imposed on average at the mean field level
\begin{eqnarray}
(\mathrm{i} \lambda_{i})_{0} = \lambda_{0}.
\label{MF_lambda}
\end{eqnarray}
For the antiferromagnetic part we impose an AF order polarized in $z$-direction
\begin{eqnarray}
(m_{i}^{x})_{0} = 0 \ , \ (m_{i}^{y})_{0} = 0 \ , \ (m_{i}^{z})_{0} = 
(-1)^{i} m.
\label{MF_magne}
\end{eqnarray}
Concerning the flux contribution we impose at half-filling ($\mathrm{x}=0$) a 
$\pi$-flux scheme ($\phi=\pi$), which is energetically 
the most favorable case for a square lattice \cite{RiWieg1989}
\begin{eqnarray}
& &(\chi_{i,i+x})_{0} = (\chi_{i}^{x})_{0}= \chi \cdot \mathrm{exp}\left[ 
+\mathrm{i} \frac{\displaystyle \pi}{\displaystyle 4 } (-1)^{i} \right] \ , \nonumber \\
& &(\chi_{i,i+y})_{0} = (\chi_{i}^{y})_{0}= \chi \cdot \mathrm{exp}\left[ 
-\mathrm{i} \frac{\displaystyle \pi}{\displaystyle 4 } (-1)^{i} \right].
\label{MF_chi}
\end{eqnarray}

To write the mean field Hamiltonian $\mathcal{H}_{MF}$, we note that the 
wavevector of the AF and $\pi$-flux is $\bm{Q}=(\pi,\pi)$. 
Therefore the summation over the momentum is done in the reduced magnetic first BZ.
The borders of the ma\-gnetic (or reduced) 
BZ are given by the Fermi surface of the Hubbard model at half 
filling. The reduced BZ is characterized by the nesting property 
under the vector $\bm{Q}=(\pi,\pi)$.
% , as we can see in Fig.\! \ref{Brillouin-Fig}.
$\mathcal{H}_{MF}$ can be expressed in an appropriate spinor basis of 
the momentum space
\begin{eqnarray}
\left[ 
\begin{array}{c}
f_{\bm{k},\sigma} \\
f_{\bm{k}+\bm{Q},\sigma}
\end{array}
\right].
\label{Def_spinor}
\end{eqnarray}
\\
We obtain the mean field $t - J$ Hamiltonian written in a matrix form
\begin{widetext}
\begin{eqnarray}
  {\mathcal H}_{MF} = \sum_{\bm{k}}~{'} \sum_{\sigma} \left[ 
f_{\bm{k},\sigma}^{\dagger} \ f_{\bm{k}+\bm{Q},\sigma}^{\dagger} \right] 
                     \left[ \begin{array}{cc} \xi_{\bm{k}} & - 
\Delta_{\bm{k},\sigma}^{m} - \Delta_{\bm{k},\sigma}^{\chi} \\
                                     - (\Delta_{\bm{k},\sigma}^{m})^{*} - 
(\Delta_{\bm{k},\sigma}^{\chi})^{*} & \xi_{\bm{k}+\bm{Q}} \end{array} \right] 
                      \left[ \begin{array}{c} f_{\bm{k},\sigma} \\ 
                                              f_{\bm{k}+\bm{Q},\sigma} 
\end{array}  \right].
\label{Hamil0}
\end{eqnarray}          
\end{widetext}       
The summation $\sum_{\bm{k}}~{'}$ in $\bm{k}$-space is over the reduced 
ma\-gnetic BZ, and we have defined the following 
energies
\begin{eqnarray}
\epsilon_{\bm{k}} &=& -2t \big[\cos({k}_{x})+\cos({k}_{y})\big] - 4 
t^{'}\big[\cos({k}_{x}).\cos({k}_{y})\big] \nonumber \\
& &- 2 t^{''}\big[\cos(2{k}_{x})+\cos(2{k}_{y})\big]~, \nonumber \\
\label{Def-xik}
\xi_{\bm{k}} &=& \epsilon_{\bm{k}}\mathrm{x} - (\lambda_{0}+\mu)
 \nonumber \\
& &- (1-\alpha) J \chi \mathrm{cos}(\phi/4) 
\big[\mathrm{cos}({k}_{x}) + \mathrm{cos}({k}_{y})\big]~, \\ 
\label{Def-xikQ}
\xi_{\bm{k}+\bm{Q}} &=& \epsilon_{\bm{k}+\bm{Q}}\mathrm{x} - (\lambda_{0}+\mu)
 \nonumber \\
& &+ (1-\alpha) J \chi \mathrm{cos}(\phi/4) 
\big[\mathrm{cos}({k}_{x}) + \mathrm{cos}({k}_{y})\big]~.
\end{eqnarray}
The order parameters associated with the staggered ma\-gnetization and $\phi$-flux are respectively
\begin{eqnarray}
\label{Def-delta}
  \Delta_{\bm{k},\sigma}^{m} &=& 2 \alpha J m \sigma \ , \\ 
\Delta_{\bm{k},\sigma}^{\chi} &=& \textrm{i}(1-\alpha) J \chi \mathrm{sin}(\phi/4) 
\big[\mathrm{cos}({k}_{x}) - \mathrm{cos}({k}_{y})\big] \ , \nonumber
\end{eqnarray}
for  $\sigma = \pm 1$.
${\mathcal H}_{MF}$ can be diagonalized as
\begin{eqnarray}
{\mathcal H}_{MF} =  \sum_{\bm{k}}~{'} \sum_{\sigma} \left[ E_{\bm{k}}^{up} 
\gamma_{1\bm{k},\sigma}^{\dagger} \gamma_{1\bm{k},\sigma} + E_{\bm{k}}^{low}
\gamma_{2\bm{k},\sigma}^{\dagger} \gamma_{2\bm{k},\sigma} \right], \nonumber
\end{eqnarray} 
by a Bogoliubov-Valatin unitary transformation \cite{Bogo-trans,Val-trans}
\begin{eqnarray}
\left[  \begin{array}{c} f_{\bm{k},\sigma} \\ f_{\bm{k}+\bm{Q},\sigma} \end{array} 
\right]
= \left[ \begin{array}{cc} u_{\bm{k},\sigma} &  v_{\bm{k}} \\
                            - v_{\bm{k}}  & u_{\bm{k},\sigma}^{*} \end{array} 
\right] 
   \left[ \begin{array}{c}  \gamma_{1\bm{k},\sigma} \\  \gamma_{2\bm{k},\sigma} 
\end{array} \right].
\label{Gammaop}
\end{eqnarray}

\begin{figure}[t]
\includegraphics[width=82mm]{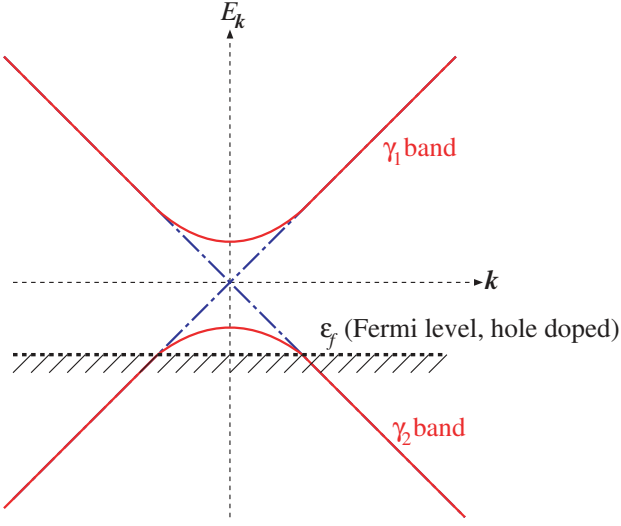}
\caption {(Color online) Upper and lower Hubbard bands associated with the eigenvalues $E_{\bm{k}}^{up}$ and $E_{\bm{k}}^{low}$
of ${\mathcal H}_{MF}$.} 
\label{Hubbard-Fig}
\end{figure}

In the previous expression $\gamma^\dagger_{1\bm{k},\sigma}$,  
$\gamma_{1\bm{k},\sigma}$
($\gamma^\dagger_{2\bm{k},\sigma}$,  $\gamma_{2\bm{k},\sigma}$ ) are the 
creation and annihilation operators of the upper (lower)
Hubbard band, respectively, with the correspon\-ding energy eigenvalues (see Fig.\! \ref{Hubbard-Fig})
\begin{eqnarray}
\label{Def-Eup}
E_{\bm{k}}^{up} & = & + \frac{1}{2} 
\sqrt{\big[\xi_{\bm{k}}-\xi_{\bm{k}+\bm{Q}} \big]^{2} + 4 
(|\Delta_{\bm{k},\sigma}^{m}|^{2}+|\Delta_{\bm{k},\sigma}^{\chi}|^{2})} \nonumber \\
& & +\frac{\xi_{\bm{k}}+\xi_{\bm{k}+\bm{Q}}}{2} \ , \\
\label{Def-Elow}
E_{\bm{k}}^{low} & = & - \frac{1}{2} 
\sqrt{\big[\xi_{\bm{k}}-\xi_{\bm{k}+\bm{Q}} \big]^{2} + 4 
(|\Delta_{\bm{k},\sigma}^{m}|^{2}+|\Delta_{\bm{k},\sigma}^{\chi}|^{2})} \nonumber \\
& & +\frac{\xi_{\bm{k}}+\xi_{\bm{k}+\bm{Q}}}{2} \ .
\end{eqnarray}
The elements of the unitary matrix are given by 
\begin{eqnarray}
& &u_{\bm{k},\sigma} = \mathrm{cos}(\theta_{\bm{k}}) \ . \ \mathrm{e}^{\mathrm{i} 
\phi_{\bm{k},\sigma}} \ , \ v_{\bm{k}} = \mathrm{sin}(\theta_{\bm{k}})~, 
\label{BogoCoeff}
\end{eqnarray}
with the trigonometric factors
\begin{eqnarray}
& &\mathrm{cos}(\theta_{\bm{k}}) \nonumber \\
&=& 
\sqrt{\frac{1}{2}\left(1 + 
\frac{\xi_{\bm{k}}-\xi_{\bm{k}+\bm{Q}}}{\sqrt{\left(\xi_{\bm{k}}-\xi_{\bm{k}+\bm{Q}}\right)^{2} +  4 
(|\Delta_{\bm{k},\sigma}^{m}|^{2}+|\Delta_{\bm{k},\sigma}^{\chi}|^{2})}}\right)}, \nonumber
\end{eqnarray}
\begin{eqnarray}
& &\mathrm{sin}(\theta_{\bm{k}}) \nonumber \\
&=& 
\sqrt{\frac{1}{2}\left(1 - 
\frac{\xi_{\bm{k}}-\xi_{\bm{k}+\bm{Q}}}{\sqrt{\left(\xi_{\bm{k}}-\xi_{\bm{k}+\bm{Q}}\right)^{2} +  4 
(|\Delta_{\bm{k},\sigma}^{m}|^{2}+|\Delta_{\bm{k},\sigma}^{\chi}|^{2})}}\right)}, \nonumber 
\end{eqnarray}
\begin{eqnarray}
\mathrm{cos}(\phi_{\bm{k},\sigma}) =
\frac{\Delta_{\bm{k},\sigma}^{m}}{\sqrt{|\Delta_{\bm{k},\sigma}^{m}|^{2}+|\Delta_{\bm{k},\sigma}^{\chi}|^{2})}} \ , \nonumber
\end{eqnarray}  
\begin{eqnarray}
\mathrm{sin}(\phi_{\bm{k},\sigma}) = 
\frac{-\textrm{i}\Delta_{\bm{k},\sigma}^{\chi}}{\sqrt{|\Delta_{\bm{k},\sigma}^{m}|^{2}+|\Delta_{\bm{k},\sigma}^{\chi}|^{2})}}.
\nonumber 
\end{eqnarray} 

It is noted here that the unitary transformation contains the complex phase
factor $\mathrm{e}^{\mathrm{i} \phi_{\bm{k},\sigma}}$, which becomes important when we solve the BCS
equation in Section \ref{GaussInt-Pair}. In the field-theory terminology, this offers an example of 
''parity anomaly'' in (2+1)D.  \cite{QFTPAnomaly}

%%%%%%%%%%%%%%%%%%%%%%%%%%%%%%%%%%%%%%%%%%%%%%%%%%%%%%%%%%%%%%%%%%%%%%%%%%%%%%%%%%%%%%%%%%%%%%%%%%%%%%%%%%%%%%%%%%%%%%%%%%%%%%%%%%%%%%%

\subsection{Saddle point solution and mean field equations}
\label{Sub-MFE}

  In the path integral language, the saddle point action is
\begin{eqnarray}
{\mathcal S}_{0} = -  \sum_{\bm{k}}~{'} \sum_{\sigma} 
\sum_{\textrm{i}\omega_{n}} & &
                      \left[ 
\bar{\Psi}_{\bm{k},\sigma}(\mathrm{i}\omega_{n})  \ \bar{\Psi}_{\bm{k}+\bm{Q},\sigma}(\mathrm{i}\omega_{n}) \right]
\nonumber \\
&\times& \tilde{\mathcal G}_{0}^{-1}(\bm{k}, \sigma, 
\mathrm{i}\omega_{n}) \nonumber \\ 
&\times& \left[ \begin{array}{c} 
\Psi_{\bm{k},\sigma}(\mathrm{i}\omega_{n}) \\ 
\Psi_{\bm{k}+\bm{Q},\sigma}(\mathrm{i}\omega_{n}) \end{array}  \right],
\end{eqnarray} 
with $\textrm{i}\omega_{n}$ the fermionic Matsubara frequencies and 
$ \tilde{\mathcal G}_{0}$ the free Green's 
functions' matrix
\begin{eqnarray}
& &\tilde{\mathcal G}_{0}(\bm{k}, \sigma, \mathrm{i}\omega_{n}) \nonumber \\
&=& \frac{1}{\omega_{n}^{2} + \mathrm{i}\omega_{n}(\xi_{\bm{k}}+\xi_{\bm{k}+\bm{Q}})-  
\xi_{\bm{k}} \xi_{\bm{k}+\bm{Q}} + 
|\Delta_{\bm{k},\sigma}^{m}+\Delta_{\bm{k},\sigma}^{\chi}|^{2}} \nonumber \\ 
& &\times \left[ \begin{array}{cc} -(\mathrm{i}\omega_{n} - \xi_{\bm{k}+\bm{Q}}) &  
\Delta_{\bm{k},\sigma}^{m} + \Delta_{\bm{k},\sigma}^{\chi} \\ \Delta_{\bm{k},\sigma}^{m} 
- \Delta_{\bm{k},\sigma}^{\chi} & -( \mathrm{i}\omega_{n} - \xi_{\bm{k}}) 
\end{array} \right].
\label{Green_0}
\end{eqnarray}

\begin{figure}[t]
\includegraphics[width=86.6mm]{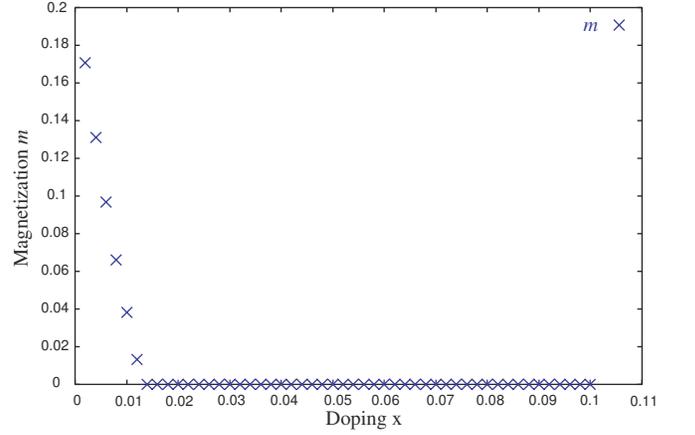}
\caption {(Color online) Dependence of the modulus of the staggered magnetization $m$ obtained at the saddle-point, as a function of the doping $\mathrm{x}=1-n_{f}$.}
\label{MAGNE-Fig}
\end{figure}

% \begin{figure}[t]
% \includegraphics[width=86.6mm]{Fig6-LAG-CHEM_x.pdf}
% \caption {(Color online) Dependence of the Lagrange multiplier $\lambda_{0}$ and chemical potential $\mu$ at the saddle point, 
% as a function of the doping $\mathrm{x}=1-n_{f}$.}
% \label{LAMBDAMU-Fig}
% \end{figure}

Now we expand the action up to the second order with respect to the
deviation from the saddle point solution. 
At the first order we can derive the self-consistent mean field equations.
The details are given in Appendix \ref{Appen-MFEq}, 
and we give here the final results

\begin{eqnarray}
\label{MagneAF-MFEq}
1 - \frac{4}{\mathrm{N_{s}}} \alpha J \sum_{\bm{k}}~{'}&\bigg\{ &  
\frac{1}{E_{\bm{k}}^{up}-E_{\bm{k}}^{low}} \bigg\} = 0~, 
\end{eqnarray}
\begin{eqnarray}
\frac{1}{\mathrm{N_{s}}} \sum_{\bm{k}}~{'}\bigg\{\frac{\big[ 
\mathrm{cos}({k}_{x}) + \mathrm{cos}({k}_{y})\big]( 
\xi_{\bm{k}+\bm{Q}}-\xi_{\bm{k}})\mathrm{cos}(\phi/4)}{E_{\bm{k}}^{up}-E_{\bm{k}}^{low}} \hspace{4mm} \\
 -\frac{ 2 \textrm{i}\big[ \mathrm{cos}({k}_{x}) - 
\mathrm{cos}({k}_{y})\big] \Delta_{\bm{k},\sigma}^{\chi} \mathrm{sin}(\phi/4)}
{E_{\bm{k}}^{up}-E_{\bm{k}}^{low}} \bigg\} = \chi~,  \nonumber
\end{eqnarray}
\begin{eqnarray}
\frac{1}{\mathrm{N_{s}}} \sum_{\bm{k}}~{'}\bigg\{
\frac{ \big[ \mathrm{cos}({k}_{x}) + \mathrm{cos}({k}_{y})\big]( 
\xi_{\bm{k}+\bm{Q}}-\xi_{\bm{k}})}{E_{\bm{k}}^{up}-E_{\bm{k}}^{low}} \bigg\} \hspace{4mm} \\
= \chi \mathrm{cos}(\phi/4)~, \nonumber \hspace{50mm}
\end{eqnarray}
\begin{eqnarray}  
\label{Lambda-MFEq}
\frac{4}{\mathrm{N_{s}}} \sum_{\bm{k}}~{'}\bigg\{ 
\tilde{t}_{\bm{k}}
\frac{( \xi_{\bm{k}+\bm{Q}}-\xi_{\bm{k}})}{E_{\bm{k}}^{up}-E_{\bm{k}}^{low}}
\hspace{41.5mm}  \\
\hspace{10mm} + (\tilde{t}_{\bm{k}}^{\hspace{0.05cm}'}+\tilde{t}_{\bm{k}}^{\hspace{0.05cm}''}) \frac{( \xi_{\bm{k}+\bm{Q}}+\xi_{\bm{k}}- 2 
E_{\bm{k}}^{low})}{E_{\bm{k}}^{up}-E_{\bm{k}}^{low}} \bigg\} = -\lambda_{0}~, \nonumber
\end{eqnarray}
where the quantities $\xi_{\bm{k}}$, $\xi_{\bm{k+Q}}$, $\Delta_{\bm{k},\sigma}^{\chi}$, $E_{\bm{k}}^{up}$, $E_{\bm{k}}^{low}$,
 $\tilde{t}_{\bm{k}}$, $\tilde{t}_{\bm{k}}^{\hspace{0.05cm}'}$ and $\tilde{t}_{\bm{k}}^{\hspace{0.05cm}''}$
 are given in Eqs.\! (\ref{Def-xik}), (\ref{Def-xikQ}), (\ref{Def-delta}), (\ref{Def-Eup}), (\ref{Def-Elow}),
(\ref{Def-tk}), (\ref{Def-tkp}) and (\ref{Def-tkpp}), respectively.

We adjust the electron number $n_{f}$ by the chemical potential $\mu$. At zero
temperature the upper Hubbard band is empty, while the lower band is partially filled 
as
\begin{eqnarray}
\frac{1}{\mathrm{N_{s}}}\sum_{\bm{k}}\frac{1}{\mathrm{exp}\big[ \beta E_{\bm{k}}^{low} \big] + 
1} = 1 - \mathrm{x},
\end{eqnarray}
where ($\sum_{\bm{k}}$) is extended over the first BZ.

\begin{figure*}[t]
\includegraphics[width=174mm]{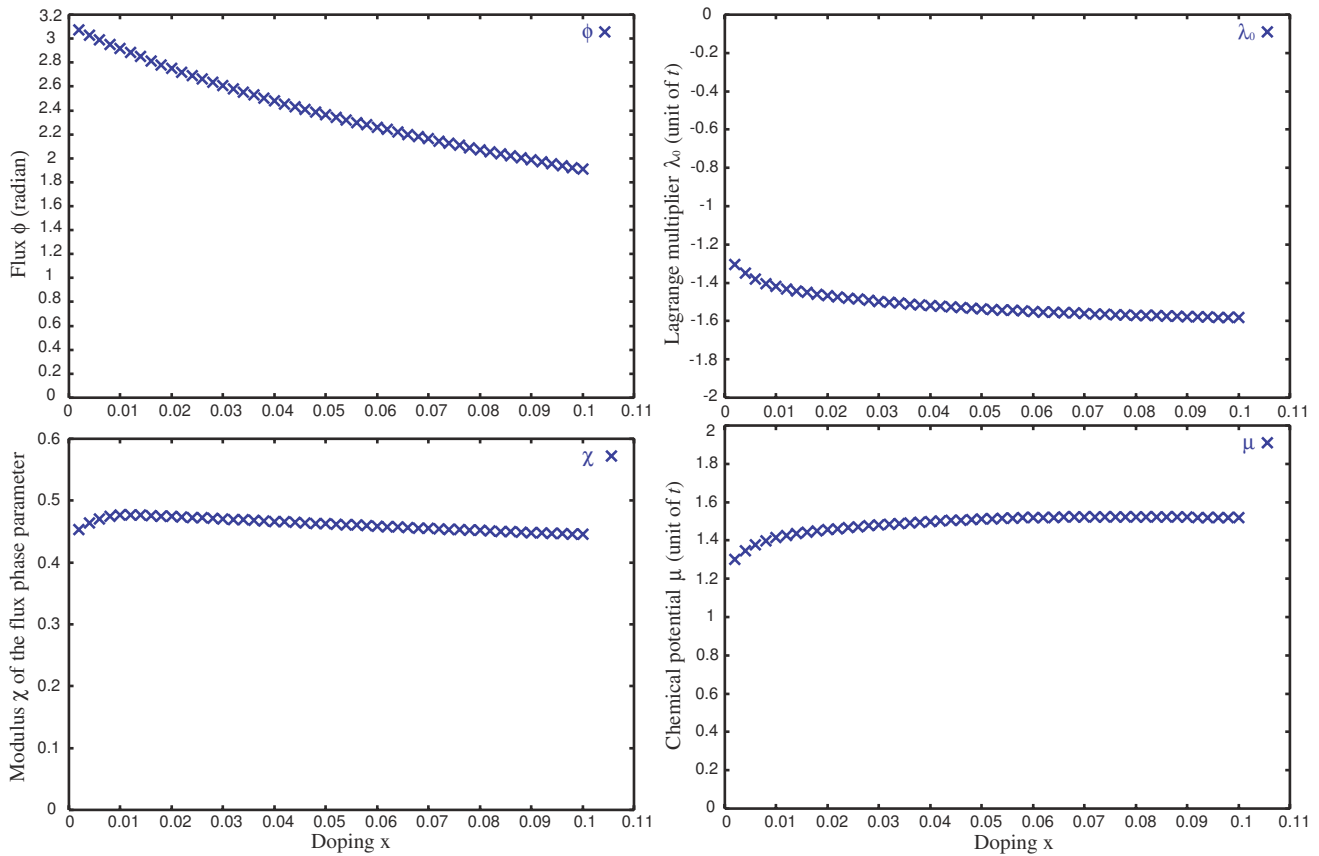}
\caption {(Color online) Dependence of the  flux parameters $\phi$ and $\chi$, 
Lagrange multiplier $\lambda_{0}$ and chemical potential $\mu$
obtained at the saddle point, 
as a function of the doping $\mathrm{x}=1-n_{f}$.} 
\label{FLUX-LAG-CHEM-Fig}
\end{figure*}

  We have solved numerically these equations by discretizing the reduced 
BZ in 2 millions of points, providing us a precision on 
the obtained values better than $10^{-5}$. In particular at 
half-filling ($\mathrm{x} = 0$) we have found
\begin{eqnarray}
\alpha = 0.301127,
\label{Val_Alpha}
\end{eqnarray}
by imposing the relation: $ m = 0.5 \chi$ previously obtained by Hsu. \cite{Hsu1990} 
Away from half-filling this relation between $m$ and $\chi$ will be changed
 but the value of $\alpha$ (\ref{Val_Alpha}) is assumed to be 
the same on all the range of doping. This assumption means that the weight assigned to each of the decoupling terms of the Heisenberg exchange interaction is kept constant as a function of the doping. This assumption does not change the essential
features of our results presented below.

\begin{figure*}[t]
\includegraphics[width=178mm]{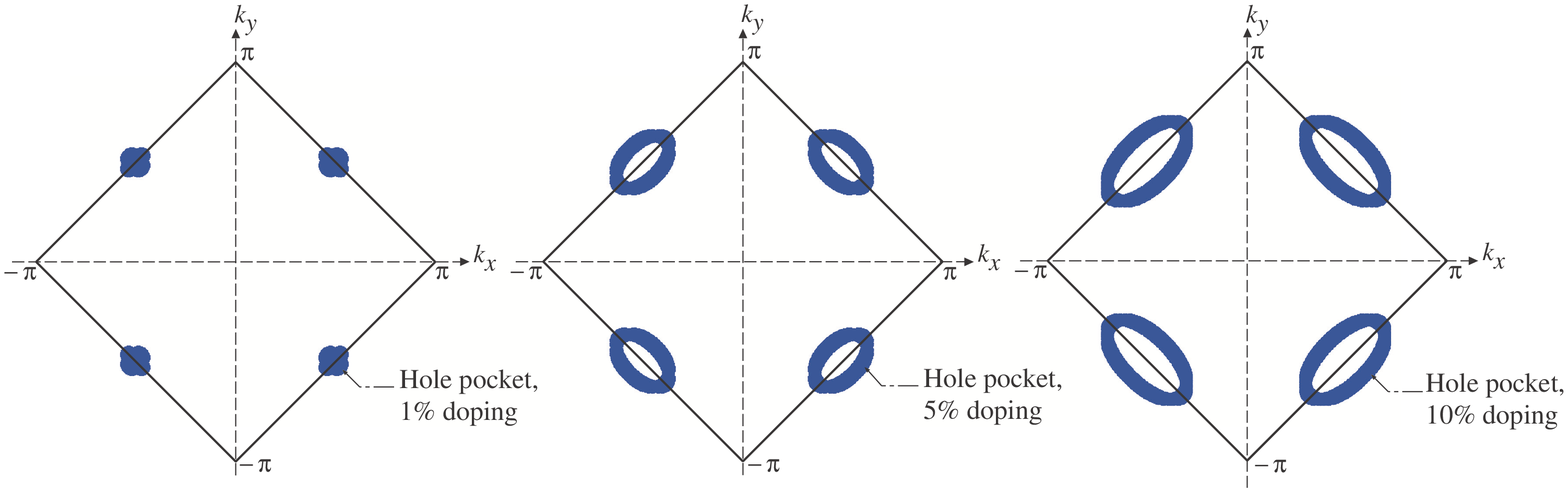}
\caption {(Color online) Representation of the Fermi surface for three different values of the doping: $\mathrm{x} = 0.01, 0.05, 0.1$. }
\label{FERMISurf-Fig}
\end{figure*}

We present the numerically obtained values of the mean field parameters 
as a function of the doping $\mathrm{x}$ in
Figs.\!  \ref{MAGNE-Fig} and \ref{FLUX-LAG-CHEM-Fig}. We see that the
N\'eel state disappears at a very small value of the doping, typically 1.5 \%, 
which is in agreement with the experimental phase diagram of hole doped cuprates.
This is in sharp contrast to the previous 
studies of the $t-J$ model, which found that the AF state could remain until doping values of around 15 \%. \cite{Fukuyama1992} The flux 
$\phi$ decreases rela\-tively linearly from $\pi$ at half-filling, and reaches 
$\phi \simeq 2$ for a doping of 10 \%; this result is similar to the data
obtained by Hsu \textit{et al.}. \cite{HsuAffl-1991}

%\vspace{2mm}

 We also show the shape of the Fermi surface for three doping cases: $\mathrm{x} 
= 0.01, 0.05, 0.1$ in Fig.\! \ref{FERMISurf-Fig}, where the
Fermi surface is composed of arcs which delimit the
hole pockets located around the four nodes $\bm{k}=(\pm\pi/2,\pm\pi/2)$.

%\vspace{2mm}

In the next section we will consider the quantum fluctua\-tion around this mean field solution. The second order Gaussian fluctuation corresponds to the RPA. 
We will calculate the ``$\grave{a}\ la\ BCS$'' \cite{BCS,Schrief-book} pairing 
potential in terms of the exchange of these fluctuations,
to see the effects of the coexistence of antiferromagnetism and the staggered flux 
on the pairing force in underdoped region.

%%%%%%%%%%%%%%%%%%%%%%%%%%%%%%%%%%%%%%%%%%%%%%%%%%%%%%%%%%%%%%%%%%%%%%%%%%%%%%%%%%%%%%%%%%%%%%%%%%%%%%%%%%%%%%%%%%%%%%%%%%%%%%%%%%%%%%%
%%%%%%%%%%%%%%%%%%%%%%%%%%%%%%%%%%%%%%%%%%%%%%%%%%%%%%%%%%%%%%%%%%%%%%%%%%%%%%%%%%%%%%%%%%%%%%%%%%%%%%%%%%%%%%%%%%%%%%%%%%%%%%%%%%%%%%%

\section{Gaussian Fluctuation and BCS Pairing Interaction  between Quasi-Particles}
\label{GaussInt-Pair}

%%%%%%%%%%%%%%%%%%%%%%%%%%%%%%%%%%%%%%%%%%%%%%%%%%%%%%%%%%%%%%%%%%%%%%%%%%%%%%%%%%%%%%%%%%%%%%%%%%%%%%%%%%%%%%%%%%%%%%%%%%%%%%%%%%%%%%

\subsection{Second order action and correlation functions}

In this Section we expand the action with respect to the fluctuations
of the mean field parameters up to second order starting from
the saddle point solution for the $t - J$ model 
($\ref{Hamil0}$). 
By analogy with a diagrammatic language such a treatment is equivalent 
to 
taking into account all the bubbles associated with the fluctuating 
bosonic 
fields. \cite{Nagaosa_books} This approach allows us to 
calculate 
the diffe\-rent correlation 
functions (or susceptibilities, both being 
linked 
via the fluctuation-dissipation theorem \cite{Kubo_book}) between
those fields. 

Integrating over the fermionic Grassman variables and expanding the action with
respect to the fluctuations of the bosonic modes up to the second order, we obtain
\begin{widetext}
\begin{eqnarray}
& &{\mathcal S}_{2}
= \int_{0}^{\beta} d\tau \bigg\{ - \sum_{i} 
(b_{0})^{2} \big[2 \cdot \mathrm{i} \delta \lambda_{i} \cdot \delta b_{i}+ 
\lambda_{0}\cdot (\delta b_{i})^{2}\big] 
-  \alpha J \sum_{\langle i,j \rangle } \sum_{s_{m} = x,y,z} 
\big[\delta m_{i}^{s_{m}} \cdot \delta m_{j}^{s_{m}}\big] 
+ \frac{(1-\alpha)}{2} J \sum_{\langle i,j \rangle }
\big[   \delta \chi_{ij} \cdot \delta \chi_{ij}^{*}   \big] \hspace{3mm} \nonumber  \\ 
& &\hspace{22mm}- \big(\sum_{\langle i,j \rangle} t + \sum_{\langle i,j 
\rangle^{'}}t^{'}  + \sum_{\langle i,j \rangle^{''}} t^{''} \big) (b_{0})^{2} 
 \times \sum_{\sigma} 
\delta b_{i} \cdot \delta b_{j} \big[ \langle\bar{\Psi}_{i,\sigma} 
\Psi_{j,\sigma}\rangle 
+\langle \bar{\Psi}_{j,\sigma} \Psi_{i,\sigma}\rangle \big] \bigg\} 
+ \frac{1}{2} \mathrm{Tr}\big[ \tilde{\mathcal G}_{0} \tilde{\mathcal V}_{1} 
\tilde{\mathcal G}_{0} \tilde{\mathcal V}_{1} \big]~.
\label{Def_S2}
\end{eqnarray} 
%\end{widetext}
Evaluating the bubbles $\frac{1}{2}  \mathrm{Tr}\big[ \tilde{\mathcal G}_{0} 
\tilde{\mathcal V}_{1} \tilde{\mathcal G}_{0} \tilde{\mathcal V}_{1} \big]$ 
as explained in 
Appendix \ref{Appen-GaussAct}, the quadratic action is given by
\begin{eqnarray}
{\mathcal S}_{2} &=&
\sum_{\bm{q}}~{'} \sum_{\bm{{q}_{1}},\bm{{q}_{2}}=\bm{q},\bm{q}+\bm{Q}} \sum_{\mathrm{i}\omega_{\ell}} 
\sum_{i,j=1}^{9} 
\delta X_{i}(\bm{{q}_{1}},\mathrm{i}\omega_{\ell}) 
%\nonumber \\
%& &\times  
\left({\mathcal 
M}_{i,j}(\bm{q},\bm{{q}_{1}},\bm{{q}_{2}},\mathrm{i}\omega_{\ell})+
\frac{1}{2} {\Pi}_{i,j}(\bm{q},\bm{{q}_{1}},\bm{{q}_{2}},\mathrm{i}\omega_{\ell})
\right)
% \nonumber \\
%& &\times 
\delta X_{j}(-\bm{{q}_{2}},-\mathrm{i}\omega_{\ell}), \hspace{2mm}
\label{Result_S2}
\end{eqnarray}
where the first order fluctuations of the bosonic fields $\delta X_i$ are defined in Eq.\! (\ref{Def_VecXq}), 
and the matrix elements ${\mathcal M}_{i,j}$ and ${\Pi}_{i,j}$ are detailed
in Eqs.\! (\ref{DefMatrixM}) and (\ref{Def_PiMatrix}), respectively. In the previous formula, $\mathrm{i}\omega_{\ell}$ denotes the bosonic 
Matsubara frequencies. 

In a path-integral formalism, the second order term of the effective 
action gives easy access to the different correlation functions 
between bosonic fields renormalized at a RPA level \cite{Nagaosa_books}
% \vspace{4mm}
% \end{widetext}
% \newpage

\begin{eqnarray}
& &\tilde{\mathcal C}(\bm{q},\bm{{q}_{1}},-\bm{{q}_{2}},\mathrm{i}\omega_{\ell}) 
=
\bigg[ {\mathcal C}_{i,j}(\bm{q},\bm{{q}_{1}},-\bm{{q}_{2}},\mathrm{i}\omega_{\ell}) \bigg]_{1\leq i,j \leq 9} 
% \nonumber \\ 
% &=&
 = \bigg[ \langle
X_{i}(\bm{{q}_{1}},\mathrm{i}\omega_{\ell}) X_{j}(-\bm{{q}_{2}},-\mathrm{i}\omega_{\ell})
\rangle_{RPA} \bigg]_{1\leq i,j \leq 9}, \nonumber \\
% \end{eqnarray}
% \begin{eqnarray}
& &\bigg[ \langle
X_{i}(\bm{{q}_{1}},\mathrm{i}\omega_{\ell}) X_{j}(-\bm{{q}_{2}},-\mathrm{i}\omega_{\ell})
\rangle_{RPA} \bigg]_{1\leq i,j \leq 9} 
% \nonumber \\ 
% &=&
= \left(
\tilde {\mathcal M}(\bm{q},\bm{{q}_{1}},\bm{{q}_{2}},\mathrm{i}\omega_{\ell}) + 
\frac{1}{2} \tilde{\Pi}(\bm{q},\bm{{q}_{1}},\bm{{q}_{2}},\mathrm{i}\omega_{\ell})
\right)^{-1}~.
\label{RPA_Corre_f}
\end{eqnarray}
The behaviour of the obtained transverse spin-spin correlation function 
$\chi^{\pm}$ is in agreement
with well-known results concerning the Heisenberg antiferroma\-gnets, as 
discussed in Appendix \ref{Appen-Haldane}.
\vfill\eject
\newpage
\end{widetext}
% \vfill\eject
% \newpage

%%%%%%%%%%%%%%%%%%%%%%%%%%%%%%%%%%%%%%%%%%%%%%%%%%%%%%%%%%%%%%%%%%%%%%%%%%%%%%%%%%%%%%%%%%%%%%%%%%%%%%%%%%%%%%%%%%%%%%%%%%%%%%%%%%%%%%%

\subsection{Derivation of the effective action}

We can now explicitly calculate the ''$\grave{a}\ la\ BCS$'' 
\cite{BCS,Schrief-book} pairing potential exchanging all the collective modes
des\-cribed in the previous section. 
Following Schrieffer 
$\it{et \ al.}$, \cite{Schrieffer1989,Schrieffer1989long} we adapt their approach to the 
framework of the $t-J$ model. We neglect the retardation effects 
specific 
to the Eliashberg theory \cite{Schrief-book,Eliashberg} and build a pairing 
potential 
by assuming the static limit ($\mathrm{i}\omega_{\ell}=0$).
Therefore the 
frequency dependence will not be mentioned from now on. This is justified
in the weak coupling region, where the adiabatic approximation
$k_B T_c, \Delta_{SC} << \hbar \omega_D$ is satisfied ( $T_c$: transition 
temperature, $\Delta_{SC}$: superconducting order parameter, $\omega_D$:
frequency of the exchan\-ging bosons).
This approach is not sufficient to describe the superconducting state in underdoped cuprates where the features of the strong coupling effect are observed experimentally. \cite{Nernst} However, as will be shown below, the magnetic mechanism based on the 
generalized spin-bag theory gives only a very small pairing force or is pair breaking. Therefore this weak coupling approximation is justified {\it a posteriori}, even though it does not describe the real cuprates. \\
\indent We start with the linear interaction between the fermions and
the bosonic fields (\ref{Moment_S1_fe})
\begin{eqnarray}
{\mathcal S}_{1}^{fe} =  \sum_{\bm{k},\bm{q}}~{'} \sum_{\sigma^{'},\sigma} & &
\left[ \bar{\Psi}_{\bm{k}+\bm{q},\sigma^{'}}  \ \ \bar{\Psi}_{\bm{k}+\bm{q}+\bm{Q},\sigma^{'}} 
\right] \nonumber \\
&\times& \tilde{{\mathcal V}}_{1}(\bm{k}+\bm{q},\sigma^{'};\bm{k},\sigma) \nonumber \\
&\times& \left[ \begin{array}{c} \Psi_{\bm{k},\sigma} \\ 
                             \Psi_{\bm{k}+\bm{Q},\sigma} \end{array}  \right].
\label{Pa_S1Psi_fe}
\end{eqnarray} 
Using the notations defined in Appendix \ref{Appen-GaussAct}, the interaction matrix $\tilde{\mathcal V}_{1}$ 
(\ref{Def_V1}) can be rewritten as
\begin{eqnarray}
& &\tilde{\mathcal V}_{1}(\bm{k}+\bm{q},\sigma^{'};\bm{k},\sigma) \\
&=& \sum_{\bm{{q}_{1}}=\bm{q},\bm{q}+\bm{Q}} \sum_{i,=1}^{9} 
\Big(\sum_{a^{'}=1,2} c_{i,a^{'}}^{\sigma^{'},\sigma}(\bm{k},\bm{{q}_{1}}) 
\tilde{s}_{i,a^{'}}(\bm{{q}_{1}})\Big) \nonumber \\
& & \hspace{32mm} \times \delta X_{i}(\bm{{q}_{1}}), \nonumber 
\label{V1_vect}
\end{eqnarray}
where $\delta X_{i}$, $c_{i,a^{'}}$, $\tilde{s}_{i,a^{'}}$ are given by Eqs.\!
(\ref{Def_VecXq}), (\ref{Def_VecCoeffq}), (\ref{Def_VecCoeffqQ}), and
(\ref{Def_VecPauliq}),
respectively. \\
\indent We remember that $\bar{\Psi}$, $\Psi$ are the Grassmann variables 
associated with the $f^{\dagger}$, $f$ spinon operators. As we are 
interested 
in the interactions between two $\gamma_{2}$ fermions of the lower 
Hubbard 
band, we introduce $\bar{\Phi}$, $\Phi$  the Grassmann variables 
associated with the $\gamma^{\dagger}_{2}$, $\gamma_{2}$ operators. By neglecting the contribution 
of 
the upper Hubbard band we obtain from the diagonalization of the 
mean-field 
Hamiltonian (\ref{Gammaop})
\begin{eqnarray}
  \Psi_{\bm{k},\sigma} &=& \mathrm{sin}(\theta_{\bm{k}}) \Phi_{\bm{k},\sigma} \ , \nonumber \\  
\Psi_{\bm{k}+\bm{Q},\sigma} &=& \mathrm{e}^{-\mathrm{i} \phi_{\bm{k},\sigma}} 
\mathrm{cos}(\theta_{\bm{k}}) \Phi_{\bm{k},\sigma}. 
\label{Simp_Gammaop} 
\\ \nonumber
\end{eqnarray}
The first order action related to the $\gamma_{2}$ 
operators is given by
\begin{eqnarray}
{\mathcal S}_{1}^{fe} 
&=&  \sum_{\bm{k},\bm{q}}~{'} \sum_{\sigma^{'},\sigma} 
\bar{\Phi}_{\bm{k}+\bm{q},\sigma^{'}} \nonumber \\
& &\times \Bigg[ \mathrm{sin}(\theta_{\bm{k}+\bm{q}})  \ \ \ \mathrm{e}^{ 
\mathrm{i}\phi_{\bm{k}+\bm{q},\sigma^{'}}}\mathrm{cos}(\theta_{\bm{k}+\bm{q}})  \Bigg] \nonumber \\
& &\times \Bigg\{ \sum_{\bm{{q}_{1}},=\bm{q},\bm{q}+\bm{Q}} \ \sum_{i,=1}^{9} 
\Big(\sum_{a^{'}=1,2} c_{i,a^{'}}^{\sigma^{'},\sigma}(\bm{k},\bm{{q}_{1}}) 
\tilde{s}_{i,a^{'}}(\bm{{q}_{1}})\Big)  \nonumber \\
& &  \hspace{39mm}  \times \delta X_{i}(\bm{{q}_{1}}) \Bigg\} \nonumber \\
& &\times \left[ \begin{array}{c}  \mathrm{sin}(\theta_{\bm{k}}) \\ 
                             \mathrm{e}^{-\mathrm{i} \phi_{\bm{k},\sigma}} 
\mathrm{cos}(\theta_{\bm{k}}) \end{array} \right] \nonumber \\
& &\times \Phi_{\bm{k},\sigma}.
\label{Pa_S1Phi_fe}
\end{eqnarray} 
To incorporate the interaction effects at a RPA level, we build an 
effective action by adding to ${\mathcal S}_{1}^{fe}$ the second order in $\delta X_{i}$ contribution
${\mathcal S}_{2}$ , i.e. using Eqs.\! (\ref{Result_S2}) and (\ref{RPA_Corre_f}),
\begin{eqnarray}
\label{Pa_S2_RPA}
{\mathcal S}_{2} &=&
\sum_{\bm{q}}~{'} \sum_{\bm{{q}_{1}},\bm{{q}_{2}}=\bm{q},\bm{q}+\bm{Q}} \ \sum_{i,j=1}^{9} 
\delta X_{i}(\bm{{q}_{1}}) \hspace{25mm} \\
& &\hspace{28mm} \times {\mathcal C}_{i,j}^{-1} (\bm{q},\bm{{q}_{1}},-\bm{{q}_{2}}) 
%\nonumber \\ &\times& 
\cdot \delta  X_{j}(-\bm{{q}_{2}}), \nonumber
\end{eqnarray} 
then the effective action ${\mathcal S}^{eff}$ is given by
\begin{eqnarray} 
{\mathcal S}^{eff}= {\mathcal S}_{1}^{fe}+{\mathcal S}_{2}.
 \label{Pa_SEFF}
\end{eqnarray} 
\indent In order to obtain  the pairing potential, we integrate out the 
bosonic fields $\delta X_{i}$ in ${\mathcal S}^{eff}$. 
For convenience we define $\vec{\phi}(\bm{{q}_{1}})$ for 
$\bm{{q}_{1}}=\bm{q},\bm{q}+\bm{Q}$ as 
\begin{eqnarray}
\vec{\phi}(\bm{{q}_{1}}) = \big[ {\phi}_{i} (\bm{{q}_{1}}) \big]_{1\leq i \leq 9} \ , \nonumber
\end{eqnarray}

\begin{widetext}
\begin{eqnarray}
{\phi}_{i} (\bm{{q}_{1}}) 
%\hspace{8mm} & & \nonumber \\
= \frac{1}{2} \sum_{\bm{{k}_{1}}}~{'} 
\sum_{\sigma_{1}^{'},\sigma_{1}} & &\bar{\Phi}_{\bm{{k}_{1}}+\bm{q},\sigma_{1}^{'}} 
%\nonumber \\
%&\times& 
\left[ \mathrm{sin}(\theta_{\bm{{k}_{1}}+\bm{q}})  \ \ \ \mathrm{e}^{\mathrm{i} 
\phi_{\bm{{k}_{1}}+\bm{q},\sigma_{1}^{'}}} \mathrm{cos}(\theta_{\bm{{k}_{1}}+\bm{q}})  \right] 
%\nonumber \\
%&\times& 
\Big(\sum_{a_{1}^{'}=1,2} 
c_{i,a_{1}^{'}}^{\sigma_{1}^{'},\sigma_{1}}(\bm{{k}_{1}},\bm{{q}_{1}}) \tilde{s}_{i,a_{1}^{'}}(\bm{{q}_{1}})\Big)\nonumber \\
&\times&\bigg[ \begin{array}{c}  \mathrm{sin}(\theta_{\bm{{k}_{1}}}) \\ 
                             \mathrm{e}^{- 
\mathrm{i}\phi_{\bm{{k}_{1}},\sigma_{1}}} \mathrm{cos}(\theta_{\bm{{k}_{1}}}) \end{array} \bigg] 
\Phi_{\bm{{k}_{1}},\sigma_{1}}.
\label{Phi_vect}
\end{eqnarray} 
% We obtain
% \begin{eqnarray} 
% {\mathcal S}^{eff} 
% %\nonumber \\
% &=& \sum_{\bm{q}}~{'} \sum_{\bm{{q}_{1}},\bm{{q}_{2}}=\bm{q},\bm{q}+\bm{Q}} \delta 
% \vec{X}(\bm{{q}_{1}})\tilde{\mathcal C}^{-1} (\bm{q},\bm{{q}_{1}},-\bm{{q}_{2}}) \delta 
% \vec{X}(-\bm{{q}_{2}})
% % \nonumber \\
% %& &
% + \sum_{\bm{q}}~{'} \sum_{\bm{{q}_{1}}=\bm{q},\bm{q}+\bm{Q}} \vec{\phi}(\bm{{q}_{1}})\delta 
% \vec{X}(\bm{{q}_{1}}) \nonumber \\
% & &+ \sum_{\bm{q}}~{'} \sum_{\bm{{q}_{2}}=\bm{q},\bm{q}+\bm{Q}} \vec{\phi}(-\bm{{q}_{2}})\delta 
% \vec{X}(-\bm{{q}_{2}}) \ ,
% \label{Pa_SEFF_Phi}
% \end{eqnarray} 
After integrating out the bosonic fields we have
\begin{eqnarray} 
{\mathcal S}^{eff} = -\sum_{\bm{q}}~{'} \sum_{\bm{{q}_{1}},\bm{{q}_{2}}=\bm{q},\bm{q}+\bm{Q}} 
\vec{\phi}(-\bm{{q}_{2}}) \tilde{\mathcal C}(\bm{q},\bm{{q}_{1}},-\bm{{q}_{2}}) \vec{\phi}(\bm{{q}_{1}}), \nonumber
\end{eqnarray} 
or equivalently from Eq.\! (\ref{Phi_vect})
%\begin{widetext}
\begin{eqnarray} 
{\mathcal S}^{eff} &=& - \frac{1}{4} \sum_{\bm{q}}~{'} 
\sum_{\bm{{q}_{1}},\bm{{q}_{2}}=\bm{q},\bm{q}+\bm{Q}} \ \sum_{i,j=1}^{9} \ \sum_{\bm{{k}_{1}},\bm{k_{2}}}~{'} 
\sum_{\sigma_{1}^{'},\sigma_{1},\sigma_{2}^{'},\sigma_{2}}
\bar{\Phi}_{\bm{{k}_{2}}-\bm{q},\sigma_{2}^{'}} {\Phi}_{\bm{{k}_{2}},\sigma_{2}} 
\bar{\Phi}_{\bm{{k}_{1}}+\bm{q},\sigma_{1}^{'}} \Phi_{\bm{{k}_{1}},\sigma_{1}}  \nonumber \\
& &\hspace{7mm} \times 
\Bigg\{ \bigg[ \mathrm{sin}(\theta_{\bm{{k}_{2}}-\bm{q}})  \ \ \ 
\mathrm{e}^{\mathrm{i} \phi_{\bm{{k}_{2}}-\bm{q},\sigma_{2}^{'}}} \mathrm{cos}(\theta_{\bm{{k}_{2}}-\bm{q}}) \bigg]
\Big(\sum_{a_{2}^{'}=1,2} 
c_{i,a_{2}^{'}}^{\sigma_{2}^{'},\sigma_{2}}(\bm{{k}_{2}},-\bm{{q}_{2}}) \tilde{s}_{i,a_{2}^{'}}(-\bm{{q}_{2}})\Big) 
\left[ \begin{array}{c}  \mathrm{sin}(\theta_{\bm{{k}_{2}}}) \\ 
                             \mathrm{e}^{-\mathrm{i} 
\phi_{\bm{{k}_{2}},\sigma_{2}}} \mathrm{cos}(\theta_{\bm{{k}_{2}}}) \end{array} \right] \Bigg\} \nonumber \\
& &\hspace{7mm} \times \ {\mathcal C}_{i,j}(\bm{q},\bm{{q}_{1}},-\bm{{q}_{2}}) \nonumber \\
& &\hspace{7mm} \times \Bigg\{ \bigg[ \mathrm{sin}(\theta_{\bm{{k}_{1}}+\bm{q}})  \ \ \ 
\mathrm{e}^{\mathrm{i} \phi_{\bm{{k}_{1}}+\bm{q},\sigma_{1}^{'}}} \mathrm{cos}(\theta_{\bm{{k}_{1}}+\bm{q}}) \bigg]
\Big(\sum_{a_{1}^{'}=1,2} 
c_{j,a_{1}^{'}}^{\sigma_{1}^{'},\sigma_{1}}(\bm{{k}_{1}},\bm{{q}_{1}}) \tilde{s}_{j,a_{1}^{'}}(\bm{{q}_{1}})\Big)
\left[ \begin{array}{c}  \mathrm{sin}(\theta_{\bm{{k}_{1}}}) \\ 
                             \mathrm{e}^{-\mathrm{i} 
\phi_{\bm{{k}_{1}},\sigma_{1}}} \mathrm{cos}(\theta_{\bm{{k}_{1}}}) \end{array} \right] \Bigg\}.
\label{Pa_SEFF_det}
\end{eqnarray} 
\end{widetext}

%%%%%%%%%%%%%%%%%%%%%%%%%%%%%%%%%%%%%%%%%%%%%%%%%%%%%%%%%%%%%%%%%%%%%%%%%%%%%%%%%%%%%%%%%%%%%%%%%%%%%%%%%%%%%%%%%%%%%%%%%%%%%%%%%%%%%%%

\subsection{Calculation of the  BCS pairing interaction}
\label{Calc-BCSpai}

  To have a BCS form in the previous effective action 
(\ref{Pa_SEFF_det}) we impose the following relations
\begin{eqnarray} 
*& &\bm{k_{1}} = \bm{k} \ , \bm{k_{2}} = -\bm{k} \ , \ \bm{q} = \bm{{k}^{'}}-\bm{k}, \nonumber \\
& &\sigma_{1}^{'} = \uparrow \ , \ \sigma_{2}^{'} =\downarrow  \ , \ 
\sigma_{2} = \downarrow \ , \ \sigma_{1} =  \uparrow.
\label{BCS_criteriaZZ1} \\
*& &\bm{k_{1}} = -\bm{k} \ , \bm{k_{2}} = \bm{k} \ , \ \bm{q} = \bm{k}-\bm{{k}^{'}}, \nonumber \\
& &\sigma_{1}^{'} = \downarrow \ , \ \sigma_{2}^{'} =\uparrow  \ , \ 
\sigma_{2} = \uparrow \ , \ \sigma_{1} =  \downarrow.
\label{BCS_criteriaZZ2} \\
*& &\bm{k_{1}} = -\bm{k} \ , \bm{k_{2}} = \bm{k} \ , \ \bm{q} = \bm{{k}^{'}}+\bm{k}, \nonumber \\
& &\sigma_{1}^{'} = \uparrow \ , \ \sigma_{2}^{'} =\downarrow  \ , \ 
\sigma_{2} = \uparrow \ , \ \sigma_{1} =  \downarrow.
\label{BCS_criteriaPM1} \\
*& &\bm{k_{1}} = \bm{k} \ , \bm{k_{2}} = -\bm{k} \ , \ \bm{q} = -\bm{{k}^{'}}-\bm{k}, \nonumber \\
& &\sigma_{1}^{'} = \downarrow \ , \ \sigma_{2}^{'} =\uparrow  \ , \ 
\sigma_{2} = \downarrow \ , \ \sigma_{1} =  \uparrow.
\label{BCS_criteriaPM2} 
\end{eqnarray} 
The configurations (\ref{BCS_criteriaZZ1}) and (\ref{BCS_criteriaZZ2}) are related to the $zz$ contribution, while the forms (\ref{BCS_criteriaPM1}) and (\ref{BCS_criteriaPM2}) give the $\pm$ contribution.
Therefore the BCS effective action is
\begin{eqnarray} 
{\mathcal S}^{eff}_{BCS} = \sum_{\bm{k},\bm{k^{'}}}~{'} & &
  {V}_{BCS}(\bm{k},\bm{{k}^{'}}) \nonumber \\
&\times& \bar{\Phi}_{\bm{{k}^{'}},\uparrow} \bar{\Phi}_{-\bm{{k}^{'}},\downarrow} 
{\Phi}_{-\bm{k},\downarrow} {\Phi}_{\bm{k},\uparrow}.
\label{BCS_SEFF}
\end{eqnarray} 
where
\begin{equation}
 {V}_{BCS}(\bm{k},\bm{{k}^{'}}) =
 {V}_{BCS}^{zz}(\bm{k},\bm{{k}^{'}}) + {V}_{BCS}^{\pm}(\bm{k},\bm{{k}^{'}}) 
\label{Pairing_tot}
\end{equation}
The $zz$ BCS - type pairing potential ${V}_{BCS}^{zz}$ is obtained by putting the configurations (\ref{BCS_criteriaZZ1}) and (\ref{BCS_criteriaZZ2}) in the effective action (\ref{Pa_SEFF_det})
 \begin{widetext}
\begin{eqnarray} 
& &\hspace{3mm}{V}_{BCS}^{zz}(\bm{k},\bm{{k}^{'}}) =\sum_{\bm{{q}_{1}},\bm{{q}_{2}}=\bm{k^{'}}-\bm{k},\bm{k^{'}}-\bm{k}+\bm{Q}} 
\ \sum_{i,j=1}^{9} 
\left( - \frac{1}{4} \right) \nonumber \\
& &\hspace{5mm} \times \Bigg\{ \ \bigg[ \mathrm{sin}(\theta_{-\bm{k}^{'}})  \ \ \ \mathrm{e}^{\mathrm{i} 
\phi_{-\bm{k}^{'},\downarrow}} \mathrm{cos}(\theta_{-\bm{k}^{'}}) \bigg]
\Big(\sum_{a_{2}^{'}=1,2} 
c_{i,a_{2}^{'}}^{\downarrow,\downarrow}(-\bm{k},-\bm{{q}_{2}}) \tilde{s}_{i,a_{2}^{'}}(-\bm{{q}_{2}})\Big)
\left[ \begin{array}{c}  \mathrm{sin}(\theta_{-\bm{k}}) \\ 
                            \mathrm{e}^{-\mathrm{i} 
\phi_{-\bm{k},\downarrow}} \mathrm{cos}(\theta_{-\bm{k}}) \end{array} \right] 
{\mathcal C}_{i,j}(\bm{{k}^{'}}-\bm{k},\bm{{q}_{1}},-\bm{{q}_{2}}) \nonumber \\
& &\hspace{14mm} \times \bigg[ \mathrm{sin}(\theta_{\bm{{k}^{'}}})  \ \ \ \mathrm{e}^{\mathrm{i} 
\phi_{\bm{{k}^{'}},\uparrow}} \mathrm{cos}(\theta_{\bm{{k}^{'}}}) \bigg]
\Big(\sum_{a_{1}^{'}=1,2} c_{j,a_{1}^{'}}^{\uparrow,\uparrow}(\bm{k},\bm{{q}_{1}}) 
\tilde{s}_{j,a_{1}^{'}}(\bm{{q}_{1}})\Big)
\left[ \begin{array}{c}  \mathrm{sin}(\theta_{\bm{k}}) \\ 
                             \mathrm{e}^{-\mathrm{i} \phi_{\bm{k},\uparrow}} 
\mathrm{cos}(\theta_{\bm{k}}) \end{array} \right]  \nonumber \\%
& &\hspace{7mm} + \bigg[ \mathrm{sin}(\theta_{\bm{{k}^{'}}})  \ \ \ \mathrm{e}^{\mathrm{i} 
\phi_{\bm{{k}^{'}},\uparrow}} \mathrm{cos}(\theta_{\bm{{k}^{'}}}) \bigg]
\Big(\sum_{a_{2}^{'}=1,2} 
c_{i,a_{2}^{'}}^{\uparrow,\uparrow}(\bm{k},\bm{{q}_{2}}) \tilde{s}_{i,a_{2}^{'}}(\bm{{q}_{2}})\Big)
\left[ \begin{array}{c}  \mathrm{sin}(\theta_{\bm{k}}) \\ 
                             \mathrm{e}^{-\mathrm{i} 
\phi_{\bm{k},\uparrow}} \mathrm{cos}(\theta_{\bm{k}}) \end{array} \right] 
{\mathcal C}_{i,j}(\bm{k}-\bm{k}^{'},-\bm{{q}_{1}},\bm{{q}_{2}}) \nonumber \\
& &\hspace{14mm} \times \bigg[ \mathrm{sin}(\theta_{-\bm{k}^{'}})  \ \ \ \mathrm{e}^{\mathrm{i} 
\phi_{-\bm{k}^{'},\downarrow}} \mathrm{cos}(\theta_{-\bm{k}^{'}}) \bigg]
\Big(\sum_{a_{1}^{'}=1,2} c_{j,a_{1}^{'}}^{\downarrow,\downarrow}(-\bm{k},-\bm{{q}_{1}}) 
\tilde{s}_{j,a_{1}^{'}}(-\bm{{q}_{1}})\Big) 
\left[ \begin{array}{c}  \mathrm{sin}(\theta_{-\bm{k}}) \\ 
                             \mathrm{e}^{-\mathrm{i} \phi_{-\bm{k},\downarrow}} 
\mathrm{cos}(\theta_{-\bm{k}}) \end{array} \right]  \Bigg\}. 
\label{BCSZZ_PAIRING} 
\end{eqnarray} 
\end{widetext}

 \begin{figure*}[t]
\includegraphics[width=179mm]{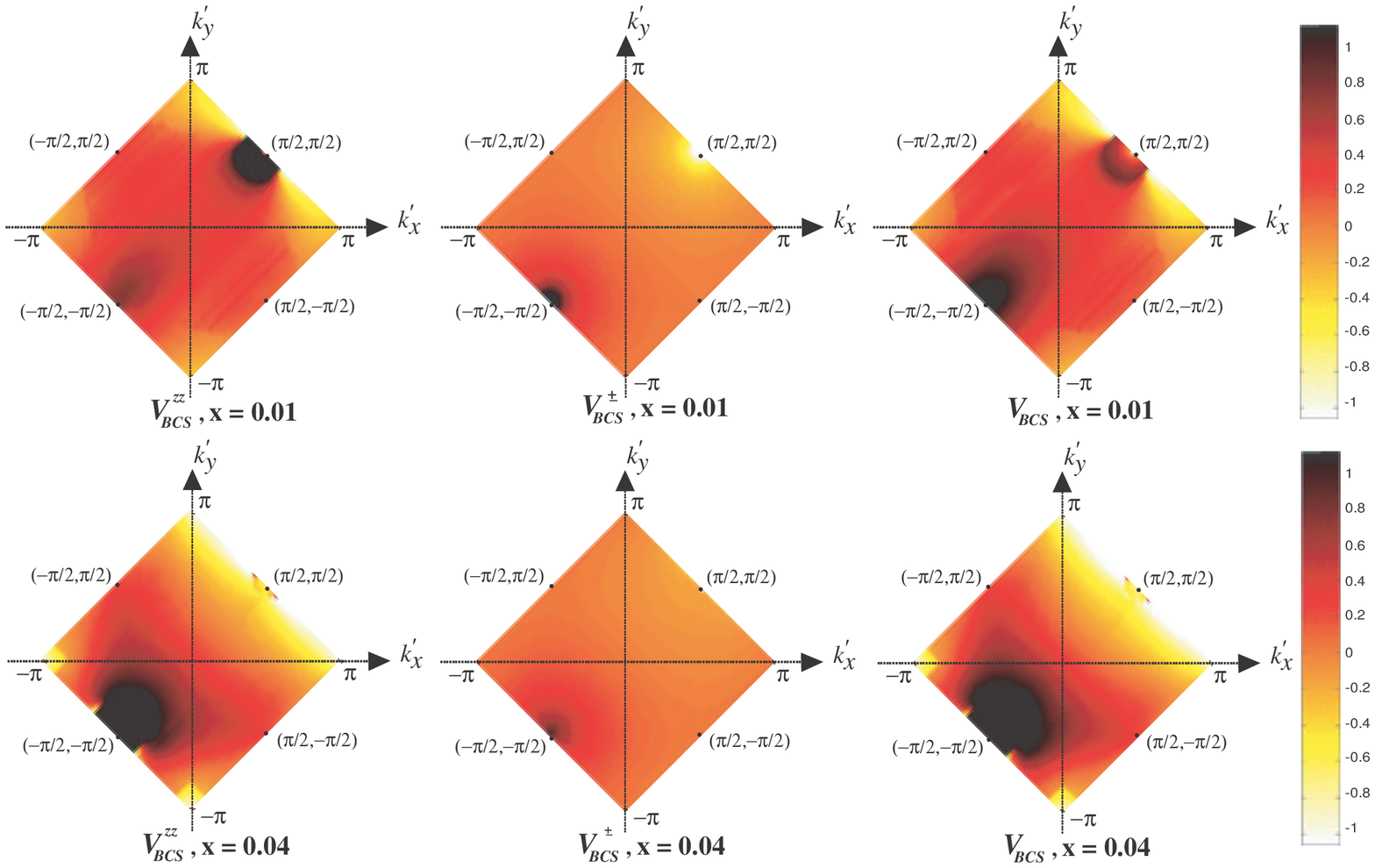}
\caption {(Color online) Pairing potentials ${V}_{BCS}^{zz}(\bm{k},\bm{{k}^{'}})$ (\ref{BCSZZ_PAIRING}), 
${V}_{BCS}^{\pm}(\bm{k},\bm{{k}^{'}})$ (\ref{BCSPM_PAIRING})
and ${V}_{BCS}(\bm{k},\bm{{k}^{'}})$ (\ref{Pairing_tot}) 
as a function of $\bm{{k}^{'}}$, with fixed $\bm{k}=(-\pi/2,-\pi/2)$, for 
two values of the doping: $\mathrm{x}=0.01, 0.04$. 
Numerical datas are plotted in unit of $t$.}
\label{PAIRINGlow-Fig}
\end{figure*}

 \begin{widetext}
 By imposing the configurations (\ref{BCS_criteriaPM1}) and (\ref{BCS_criteriaPM2}) in the effective action,
we get the $\pm$ BCS - type pairing potential
\begin{eqnarray} 
& &\hspace{3mm} {V}_{BCS}^{\pm}(\bm{k},\bm{{k}^{'}}) = 
\sum_{\bm{{q}_{1}},\bm{{q}_{2}}=\bm{k^{'}}+\bm{k},\bm{k^{'}}+\bm{k}+\bm{Q}}  \ \sum_{i,j=1}^{9} 
\left( + \frac{1}{4} \right) \nonumber \\
& &\hspace{5mm} \times \Bigg\{\bigg[ \mathrm{sin}(\theta_{-\bm{k}^{'}})  \ \ \ \mathrm{e}^{\mathrm{i} 
\phi_{-\bm{k}^{'},\downarrow}} \mathrm{cos}(\theta_{-\bm{k}^{'}}) \bigg]
\Big(\sum_{a_{2}^{'}=1,2} 
c_{i,a_{2}^{'}}^{\downarrow,\uparrow}(\bm{k},-\bm{{q}_{2}}) \tilde{s}_{i,a_{2}^{'}}(-\bm{{q}_{2}})\Big)
\left[ \begin{array}{c}  \mathrm{sin}(\theta_{\bm{k}}) \\ 
                             \mathrm{e}^{-\mathrm{i} 
\phi_{\bm{k},\uparrow}} \mathrm{cos}(\theta_{\bm{k}}) \end{array} \right] 
{\mathcal C}_{i,j}(\bm{{k}^{'}}+\bm{k},\bm{{q}_{1}},-\bm{{q}_{2}}) \nonumber \\
% \end{eqnarray} 
% \begin{eqnarray} 
& &\hspace{14mm} \times \bigg[ \mathrm{sin}(\theta_{\bm{{k}^{'}}})  \ \ \ \mathrm{e}^{\mathrm{i} 
\phi_{\bm{{k}^{'}},\uparrow}} \mathrm{cos}(\theta_{\bm{{k}^{'}}}) \bigg]
\Big(\sum_{a_{1}^{'}=1,2} c_{j,a_{1}^{'}}^{\uparrow,\downarrow}(-\bm{k},\bm{{q}_{1}}) 
\tilde{s}_{j,a_{1}^{'}}(\bm{{q}_{1}})\Big)
\left[ \begin{array}{c}  \mathrm{sin}(\theta_{-\bm{k}}) \\ 
                             \mathrm{e}^{-\mathrm{i} \phi_{-\bm{k},\downarrow}} 
\mathrm{cos}(\theta_{-\bm{k}}) \end{array} \right]  \nonumber \\
& &\hspace{7mm} +\bigg[ \mathrm{sin}(\theta_{\bm{{k}^{'}}})  \ \ \ \mathrm{e}^{\mathrm{i} 
\phi_{\bm{{k}^{'}},\uparrow}} \mathrm{cos}(\theta_{\bm{{k}^{'}}}) \bigg]
\Big(\sum_{a_{2}^{'}=1,2} 
c_{i,a_{2}^{'}}^{\uparrow,\downarrow}(-\bm{k},\bm{{q}_{2}}) \tilde{s}_{i,a_{2}^{'}}(\bm{{q}_{2}})\Big)
\left[ \begin{array}{c}  \mathrm{sin}(\theta_{-\bm{k}}) \\ 
                             \mathrm{e}^{-\mathrm{i} 
\phi_{-\bm{k},\downarrow}} \mathrm{cos}(\theta_{-\bm{k}}) \end{array} \right] 
{\mathcal C}_{i,j}(-\bm{k}-\bm{k}^{'},-\bm{{q}_{1}},\bm{{q}_{2}}) \nonumber \\
& &\hspace{14mm} \times \bigg[ \mathrm{sin}(\theta_{-\bm{k}^{'}})  \ \ \ \mathrm{e}^{\mathrm{i} 
\phi_{-\bm{k}^{'},\downarrow}} \mathrm{cos}(\theta_{-\bm{k}^{'}}) \bigg]
\Big(\sum_{a_{1}^{'}=1,2} c_{j,a_{1}^{'}}^{\downarrow,\uparrow}(\bm{k},-\bm{{q}_{1}}) 
\tilde{s}_{j,a_{1}^{'}}(-\bm{{q}_{1}})\Big) 
\left[ \begin{array}{c}  \mathrm{sin}(\theta_{\bm{k}}) \\ 
                             \mathrm{e}^{-\mathrm{i} \phi_{\bm{k},\uparrow}} 
\mathrm{cos}(\theta_{\bm{k}}) \end{array} \right]  \Bigg\}.
\label{BCSPM_PAIRING}
\end{eqnarray}
\end{widetext} 

\begin{figure*}[t]
\includegraphics[width=179mm]{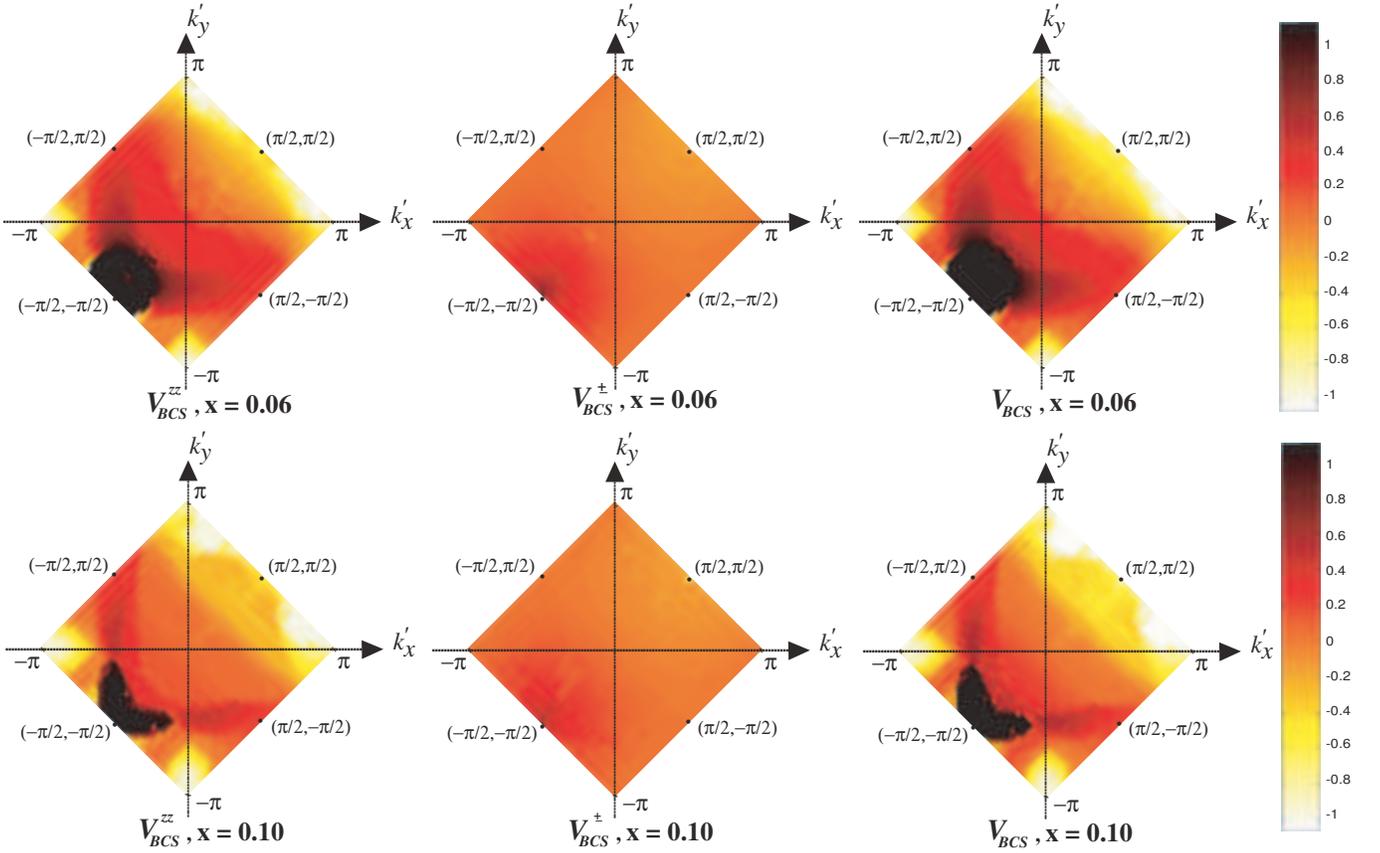}
\caption {(Color online) Pairing potentials ${V}_{BCS}^{zz}(\bm{k},\bm{{k}^{'}})$ 
(\ref{BCSZZ_PAIRING}), 
${V}_{BCS}^{\pm}(\bm{k},\bm{{k}^{'}})$ (\ref{BCSPM_PAIRING})
and ${V}_{BCS}(\bm{k},\bm{{k}^{'}})$ (\ref{Pairing_tot}) 
as a function of $\bm{{k}^{'}}$, with fixed $\bm{k}=(-\pi/2,-\pi/2)$, for 
two values of the doping: $\mathrm{x}=0.06, 0.10$. 
Numerical datas are plotted in unit of $t$.}
\label{PAIRINGhigh-Fig}
\end{figure*}

The correlation functions and pairing interactions have been numerically 
calculated, the results of which are presented in 
Figs.\! \ref{PAIRINGlow-Fig} and \ref{PAIRINGhigh-Fig}.
These figures show the effective interactions bet\-ween two $\gamma_{2}$ fermions, 
one located at the fixed position $\bm{k}=(-\pi/2,-\pi/2)$, while the other one
is at $\bm{{k}^{'}}$ moving inside the ma\-gnetic BZ. 
The intensities of the 
effective interactions are plotted as a function of $\bm{{k}^{'}}$: for
 example the va\-lues given for $\bm{{k}^{'}}=(\pi/2,\pi/2)$ corres\-pond to a 
wave\-vector $\bm{q} _{\bm{d}} =\bm{{k}^{'}}-\bm{k}=(\pi,\pi)$. 
As we assume the static limit 
($\mathrm{i}\omega_{\ell}=0$), \hspace{0.1mm} the  pairing 
potentials 
% \vfill\eject
% \noindent 
take real va\-lues. 
The dark color region means the repulsive interac\-tion
(${V}(\bm{k},\bm{{k}^{'}})>0$), while the light
region displays the attractive interaction
(${V}(\bm{k},\bm{{k}^{'}})<0$).
An interesting feature we can observe in Figs.\! \ref{PAIRINGlow-Fig} and 
\ref{PAIRINGhigh-Fig} is that the total pai\-ring 
potential ${V}_{BCS}$ is almost completely driven by the $zz$ 
(i.e. longitudinal or ''spin-bag'') contribution, without any major 
modication introduced by the 
$\pm$ (i.e. transverse or ''RVB'') correlation. 
As mentioned in Section \ref{Formul-MicroMod}, the results displayed in Figs.\! \ref{PAIRINGlow-Fig} and \ref{PAIRINGhigh-Fig} have been obtained by assuming $t^{'} = -0.3 $, $t^{''} = 0.2 $, and $J = 0.3 $. We have also considered several additional sets of numerical values for the parameters in the range: $-0.35 \leq t^{'} \leq -0.3$, $0.2 \leq t^{''} \leq 0.25$, $0.3 \leq J \leq 0.35$. We do not have observed any significant modification of the results. The robustness of these results with respect to slight variations of the parameter values can be explained as followed. In the present approach, the fermion spectrum contains two Dirac fermions and associated small hole pockets at the nodal points. The appearance of Dirac fermions means that a topologically nontrivial feature is introduced, which does not exist in the  previous spin bag theory. Such a property gives qualitatively specific characteristics to the system which are independent of slight variations of the parameter values, and therefore can explain the robustness of the results.
In the next section, we will discuss the implications of the 
interac\-tion ${V}_{BCS}$ on the possible superconductivity in this model.

\section{Discussion and Conclusions}
\label{Disc-Conc}

In previous sections, we have calculated the 
effective interactions between the two holes
in the background of the AF and RVB orders.
There are four half-pockets of holes in the reduced first BZ as shown in 
Fig.\! 7, or the two hole pockets in the shifted BZ.
These small hole po\-ckets around
$\bm{k} = (\pm \pi/2, \pm \pi/2)$ are different from those of 
the simple SDW
state as assumed in the 
original spin-bag theory. 
\cite{Schrieffer1989,Schrieffer1989long}
Namely the description of the antiferroma\-gnetic state includes the 
singlet formation via the RVB order parameter in addition to the 
N\'eel order, and corres\-pondingly the wavefunctions of the doped holes are 
distinct from the SDW state. Already there opens the gap near 
$\bm{k} = (\pm \pi,0), (0,\pm \pi)$ due to the staggered flux, 
and the superconducting pairings near these points are irrelevant. 
However the pairing force is still dominated by the
$zz$-component of the spin susceptibility, and is
repulsive for the momentum transfer $\bm{q}_{\bm{d}} = \pm (\pi,\pi)$.
When the superconducting gaps at $\bm{k} = (\pm \pi,0), (0,\pm \pi)$ 
are dominant,
this leads to the $d_{x^2-y^2}$ pairing as discussed by 
several authors \cite{moriya} since the sign of the order parameter
is opposite between $\bm{k} = (\pm \pi,0)$ and  $\bm{k} = (0,\pm \pi)$. 
However in the present case, only the states near $\bm{k} = 
(\pm \pi/2, \pm \pi/2)$ could contribute to the pairing.
One of the main predictions of the spin-bag theory 
\cite{Schrieffer1989,Schrieffer1989long} is that the 
superconductivity occurs through the pairing of holes in the small pockets, 
which are delimited by the Fermi surface as we can see in 
Fig.\! \ref{FERMISurf-Fig}. 
The present study shows that the total pairing potential ${V}_{BCS}$ 
is clearly repulsive, or pair breaking, on all the range of doping for a 
wavevector $\bm{q}_{\bm{d}}$ in the $(0,0)$, 
$(\pi,0)$ and $(0,\pi)$ areas of the magnetic BZ.
An attractive behaviour is also found when the momentum transfer is around 
$(\pi,\pi)$. It corresponds to relatively large values of $\bm{q}_{\bm{d}}$.
Namely, the repulsive interaction (dark contribution) near 
$\bm{k} = (-\pi/2, -\pi/2)$ is dominating over the
attractive (light) region with large $\bm{q}_{\bm{d}}$.
Therefore as long as one considers the pairing form which is 
constant over each of the small hole pocket, there occurs no
superconducting instability because the intra-pocket 
pair breaking force is larger than the inter-pocket 
force independently of the relative sign of the 
order parameters between different pockets.

On the other hand, one can consider the pairing with the 
nodes in the small hole pockets as in the case of 
$d_{x^2-y^2}$ pairing realized in real cuprates.
\cite{Tsuei-RMP2000,Tsuei-Nat1997} In this case, however,
most of the interaction cancels within the small pocket 
due to the sign change of the order parameter.
We could not determine the sign of the residual pairing\- 
interac\-tion, but we can safely conclude that
it is very weak even though pair creating.
Therefore we conclude that the doped staggered flux state is stable against 
the superconducting instability when only the magnetic mechanism is considered.
This is in accordance with the recent exact diagonalization studies 
on the $t-J$ model at small doping. \cite{tjhole} Furthermore this suggests 
that the other mechanism of \hspace{0.1mm} superconductivity is active in cuprates at 
% \vfill\eject
% \noindent 
least for small hole doping region. 
Also a related work has been done by Singh and Te\v{s}anovi\'{c}, \cite{SiResa-PRB1990}
where the quantum correction to the spin-bag mean field theory is studied up to one 
loop level and also the doped case is considered. This is along the same line as 
the present study, and they also obtained the repulsive interaction between two 
holes. The new aspect introduced in our study is the RVB correlation represented 
by the flux order parameter. Combining these two works, the pair breaking nature 
of the magnetic interactions seems to be rather robust in the low hole doping limit.

There are two possible routes leading to the superconductivity.
One possibility is that the starting mean field ansatz becomes 
inappropriate for optimal doping case.
According to the SU(2) formulation of the RVB state, the 
staggered flux state is the quantum mechanical mixture of the 
flux state we discussed above and the $d_{x^2-y^2}$ pairing\-
state, and the latter one is more and more weighted as the 
doping proceeds. Therefore our formulation is valid only for 
the small doping region and can not capture the crossover 
to the superconducting state since it is based on the 
perturbative method around the AF + flux state.
On the other hand, starting from the d-wave superconducting state in the SU(2) formalism, the staggered flux fluctuation has been studied.
 \cite{KimLee-AnnPhys1999,HonLee-PRL2003} This is another language describing the 
instability towards the AF ordering, which is represented by the chiral symmetry
breaking in the gauge theory. It is also found the self-energy due to the staggered
flux fluctuation leads to the quasi-particle damping strongly anisotropic
in the momentum space. \cite{HonLee-PRL2003} These works are the approach from the superconducting side, and are complementary to ours.

Taking the view that the magnetic interaction is pair breaking in the low hole doping limit, 
we should look for other forces which are active in the cuprates.
One of the most promising candidates for this 
is the electron-phonon interaction, \cite{NagaosaPhonons,ZeyPhon} 
which already manifests itself in the 
angle-resolved photoemission spectroscopy. \cite{elph}
Howe\-ver, much more work is needed to establish this scenario. 
Especially the interplay between the strong correlation and the electron-phonon 
interaction remains an important issue to be studied.
 
In conclusion, we have studied the extended spin-bag scenario taking into account 
the staggered flux resonating valence bond order in addition to the 
antiferromagnetic order. The pairing potential between the quasi-particles has been 
calculated at the Gaussian level, and it is found that the magnetically mediated 
interaction is pair brea\-king or very weak at the small hole doping limit, suggesting 
the other mechanisms such as electron-phonon interaction are active.

\begin{acknowledgements}

D. B. is thankful to Mireille Lavagna, Andr\'es Jerez and Roland Zeyher for extremely fruitful discussions.

\end{acknowledgements}

\appendix

%%%%%%%%%%%%%%%%%%%%%%%%%%%%%%%%%%%%%%%%%%%%%%%%%%%%%%%%%%%%%%%%%%%%%%%%%%%%%%%%%%%%%%%%%%%%%%%%%%%%%%%%%%%%%%%%%%%%%%%%%%%%%%%%%%%%%%%
%%%%%%%%%%%%%%%%%%%%%%%%%%%%%%%%%%%%%%%%%%%%%%%%%%%%%%%%%%%%%%%%%%%%%%%%%%%%%%%%%%%%%%%%%%%%%%%%%%%%%%%%%%%%%%%%%%%%%%%%%%%%%%%%%%%%%%%

\section{Derivation of the self-consistent equations}
\label{Appen-MFEq}

  We allow the different fields to fluctuate at the first order 
around their saddle point values (\ref{MF_lambda}), (\ref{MF_magne}) and (\ref{MF_chi})

\begin{eqnarray}
& &b_{i} = b_{0}(1+ \delta b_{i}) \ ; \nonumber \\ 
% \nonumber \\ 
& &\ \mathrm{i}\lambda_{i} = 
\lambda_{0}+ \mathrm{i}\delta \lambda_{i} \ ; \nonumber \\ 
% \nonumber \\
% \end{eqnarray} 
% \vfill\eject
% \begin{eqnarray} 
& &m_{i}^{s_{m}} = (m_{i}^{s_{m}})_{0} + \delta  m_{i}^{s_{m}} \ , \ s_{m} = x, y, z \ ; \nonumber \\
% \nonumber \\
% \end{eqnarray} 
% \begin{eqnarray} 
& &\chi_{i,i+s_{\chi}} = \chi_{i}^{s_{\chi}} = (\chi_{i}^{s_{\chi}})_{0} + \delta 
\chi_{i}^{s_{\chi}} \ , \nonumber \\
& &\delta \chi_{i}^{s_{\chi}} = \delta^{'} \chi_{i}^{s_{\chi}} + \mathrm{i} 
\delta^{''} \chi_{i}^{s_{\chi}} \ , \ s_{\chi} = x, y.  \nonumber
\end{eqnarray} 
By expanding the fluctuations at the first order we obtain
\begin{widetext}
\begin{eqnarray}
{\mathcal S}_{1} = \int_{0}^{\beta} d\tau \bigg\{ &-& \sum_{i} 
(b_{0})^{2} (\mathrm{i} \delta \lambda_{i}+ 2 \lambda_{0}\cdot \delta b_{i})
- \alpha J \sum_{\langle i,j \rangle } \big[ (-1)^{i} m \cdot \delta 
m_{j}^{z} + (-1)^{j} m \cdot \delta m_{i}^{z} \big]  
\nonumber \\
&+& \frac{(1-\alpha)}{2} J \sum_{\langle i,j \rangle } \big[ 
(\chi_{ij})_{0} \cdot \delta \chi_{ij}^{*} + (\chi_{ij})_{0}^{*} \cdot \delta \chi_{ij}  
\big]
- \sum_{i,\sigma} \big[ \mathrm{i} \delta \lambda_{i} \cdot
\bar{\Psi}_{i,\sigma} \Psi_{i,\sigma}\big] \nonumber \\
% \end{eqnarray}
% \begin{eqnarray}
&-& \big(\sum_{\langle i,j \rangle} t + \sum_{\langle i,j 
\rangle^{'}}t^{'}  + \sum_{\langle i,j \rangle^{''}} t^{''} \big) 
\sum_{\sigma} (b_{0})^{2}
 ( \delta b_{i}+ \delta b_{j} ) \big[ \bar{\Psi}_{i,\sigma} 
\Psi_{j,\sigma} + \bar{\Psi}_{j,\sigma} \Psi_{i,\sigma} \big] \nonumber \\
&+&\frac{\alpha}{2} J \sum_{\langle i,j \rangle } 
\sum_{\sigma,\sigma^{'}} \sum_{s_{m} = x,y,z} \big[ \delta m_{j}^{s_{m}}(\bar{\Psi}_{i,\sigma} 
\widetilde{\sigma}_{\sigma,\sigma^{'}}^{s_{m}} \Psi_{i,\sigma^{'}}) + \delta 
m_{i}^{s_{m}}(\bar{\Psi}_{j,\sigma} \widetilde{\sigma}_{\sigma,\sigma^{'}}^{s_{m}} 
\Psi_{j,\sigma^{'}}) \big] \nonumber \\ 
&-& \frac{(1-\alpha)}{2} J \sum_{ \langle i,j \rangle } 
\sum_{\sigma}\big[  \delta \chi_{i,j}\cdot \bar{\Psi}_{j,\sigma} \Psi_{i,\sigma} +   \delta 
\chi_{i,j}^{*}\cdot \bar{\Psi}_{i,\sigma} \Psi_{j,\sigma}\big]
\bigg\}.
\label{Def_S1}
\end{eqnarray} 
\end{widetext}
\noindent In the preceding formula, like in those which will follow in the appendices, the 
imaginary time dependence of the Grassmann variables and first order bosonic 
fluctuations follows the Lagrangian (\ref{Lagran_MF}) and is implicit. For clarity in 
the following calculations we decompose ${\mathcal S}_{1}$ into two 
parts
\begin{eqnarray}
{\mathcal S}_{1} = {\mathcal S}_{1}^{bo} + {\mathcal S}_{1}^{fe}.
\label{Aux_S1} \\ \nonumber
\end{eqnarray} 
${\mathcal S}_{1}^{bo}$ contains terms with exclusively auxiliary 
bosonic fields and no Grassmann variables. It corresponds to the three first 
terms of Eq.\! (\ref{Def_S1})
\begin{eqnarray}
\noindent & &{\mathcal S}_{1}^{bo} 
\label{Real_S1_bo}
= \int_{0}^{\beta} d\tau 
\bigg\{
 - \sum_{i} (b_{0})^{2} (\mathrm{i} \delta \lambda_{i}+ 2 
\lambda_{0}\cdot \delta b_{i}) \hspace{18mm} \\
& &\hspace{10mm}- \alpha J \sum_{\langle i,j \rangle } \big[ (-1)^{i} m \cdot \delta 
m_{j}^{z} + (-1)^{j} m \cdot \delta m_{i}^{z} \big] \nonumber \\ 
& &\hspace{10mm}+ \frac{(1-\alpha)}{2} J \sum_{\langle i,j \rangle }
 \big[ (\chi_{ij})_{0} \cdot \delta \chi_{ij}^{*} + (\chi_{ij})_{0}^{*} \cdot
\delta \chi_{ij}  \big] \nonumber 
\bigg\}~. \\ \nonumber
\end{eqnarray} 
${\mathcal S}_{1}^{fe}$ includes all the terms containing fermionic 
varia\-bles $\bar{\Psi}$, $\Psi$. It takes the four last terms of Eq.\! (\ref{Def_S1})
\begin{eqnarray}
\label{Real_S1_fe}
& &{\mathcal S}_{1}^{fe} = \int_{0}^{\beta} d\tau 
\bigg\{
- \sum_{i,\sigma} \big[ \mathrm{i} \delta \lambda_{i} \cdot
\bar{\Psi}_{i,\sigma} \Psi_{i,\sigma}\big]  \hspace{25mm} \\
% \end{eqnarray}
% \vfill\eject
% \begin{eqnarray}
& &\hspace{2mm} - \big(\sum_{\langle i,j \rangle} t + \sum_{\langle i,j 
\rangle^{'}}t^{'}  + \sum_{\langle i,j \rangle^{''}} t^{''} \big) (b_{0})^{2} \nonumber \\ 
& &\hspace{8mm}\times \sum_{\sigma} 
 ( \delta b_{i}+ \delta b_{j} ) \big[ \bar{\Psi}_{i,\sigma} 
\Psi_{j,\sigma} + \bar{\Psi}_{j,\sigma} \Psi_{i,\sigma} \big] \nonumber 
% \\
\end{eqnarray}
\vfill\eject
\begin{eqnarray}
& &\hspace{2mm}+\frac{\alpha}{2} J \sum_{\langle i,j \rangle } 
\sum_{\sigma,\sigma^{'}} \sum_{s_{m} = x,y,z} \big[~ \delta m_{j}^{s_{m}}(\bar{\Psi}_{i,\sigma} 
\widetilde{\sigma}_{\sigma,\sigma^{'}}^{s_{m}} \Psi_{i,\sigma^{'}}) \nonumber \\
& &\hspace{36.5mm}+ \delta 
m_{i}^{s_{m}}(\bar{\Psi}_{j,\sigma} \widetilde{\sigma}_{\sigma,\sigma^{'}}^{s_{m}} 
\Psi_{j,\sigma^{'}}) \big] \nonumber \\
%\end{eqnarray}
%\vfill\eject
%\begin{eqnarray}
& &\hspace{2mm} - \frac{(1-\alpha)}{2} J \sum_{\langle i,j \rangle, \sigma } 
\big[  \delta \chi_{i,j}\bar{\Psi}_{j,\sigma} \Psi_{i,\sigma} 
+ \delta 
\chi_{i,j}^{*}\bar{\Psi}_{i,\sigma} \Psi_{j,\sigma}\big]
\bigg\}~. \nonumber
\end{eqnarray} 

We start by focusing on the bosonic part  ${\mathcal S}_{1}^{bo}$ of 
the action ${\mathcal S}_{1}$ expanded at the first order in 
fluctuations. By passing in the space of moments and Matsubara frequencies we 
obtain
\begin{eqnarray}
\label{Moment_S1_bo}
& &{\mathcal S}_{1}^{bo} = \sqrt{\beta {\mathrm{N_{s}}}} \big[-(b_{0})^{2}  \mathrm{i} \delta 
\lambda_{\bm{q}=0} (\mathrm{i}\omega_{\ell}=0)  \hspace{24mm} \\ 
& &\hspace{21.5mm}- 2 (b_{0})^{2} \lambda_{0} \cdot \delta 
b_{\bm{q}=0} (\mathrm{i}\omega_{\ell}=0) \nonumber \\
& &\hspace{21.5mm}+ 4 \alpha J \cdot m \cdot \delta 
m_{\bm{q}=\bm{Q}}^{z}(\mathrm{i}\omega_{\ell}=0) \big] \nonumber \\
& &\hspace{9mm}+ (1-\alpha) J \sum_{\bm{q},\mathrm{i}\omega_{\ell}}\big[~  
\chi_{-\bm{q}}^{y}(-\mathrm{i}\omega_{\ell}) \cdot \delta \chi_{\bm{q}}^{x}(\mathrm{i}\omega_{\ell}) \nonumber \\
& &\hspace{33.5mm}+\chi_{-\bm{q}}^{x}(-\mathrm{i}\omega_{\ell}) \cdot \delta 
\chi_{\bm{q}}^{y}(\mathrm{i}\omega_{\ell})  \big],  \nonumber
\end{eqnarray} 
% \vfill\eject
\noindent where $\mathrm{i}\omega_{\ell}$ labels the bosonic Matsubara frequencies and 
$\sum_{\bm{q}}$ is expanded over the first BZ.

We study now the fermionic contribution ${\mathcal S}_{1}^{fe}$ and 
decompose it into four contributions
% \vspace{5mm}
% \vfill\eject
\begin{eqnarray}
{\mathcal S}_{1}^{fe} = {\mathcal S}_{1}^{\lambda} + {\mathcal 
S}_{1}^{b} + {\mathcal S}_{1}^{m} + {\mathcal S}_{1}^{\chi}.
\label{S1-contribs} 
\end{eqnarray} 

\noindent By passing in the space of moments and Matsubara frequencies we 
obtain
\begin{widetext}
\begin{eqnarray}
\label{S1-Lamb}
{\mathcal S}_{1}^{\lambda} = \ \frac{1}{\sqrt{\beta {\mathrm{N_{s}}}}} 
\sum_{\bm{k},\bm{q}}~{'} \sum_{\sigma^{'},\sigma} 
\sum_{\mathrm{i}\omega_{m},\mathrm{i}\omega_{n}} 
% &\sigma_{\sigma^{'},\sigma}^{0}& 
\sigma_{\sigma^{'},\sigma}^{0}
 \left[ \bar{\Psi}_{\bm{k}+\bm{q},\sigma^{'}}(\mathrm{i}\omega_{m})  \ 
\bar{\Psi}_{\bm{k}+\bm{q}+\bm{Q},\sigma^{'}}(\mathrm{i}\omega_{m}) \right]
\hspace{50mm}
%\end{eqnarray} 
%\vfill\eject
 \\
%\begin{eqnarray}
\hspace{62mm} \times \left[ \begin{array}{cc} -\mathrm{i} \delta 
\lambda_{\bm{q}}(\mathrm{i}\omega_{m} - \mathrm{i}\omega_{n})  & 
                         -\mathrm{i} \delta 
\lambda_{\bm{q}+\bm{Q}}(\mathrm{i}\omega_{m} - \mathrm{i}\omega_{n}) \\
                         -\mathrm{i} \delta 
\lambda_{\bm{q}+\bm{Q}}(\mathrm{i}\omega_{m} - \mathrm{i}\omega_{n}) & 
                         -\mathrm{i} \delta 
\lambda_{\bm{q}}(\mathrm{i}\omega_{m} - \mathrm{i}\omega_{n})  \end{array} \right] 
 \left[ \begin{array}{c} \Psi_{\bm{k},\sigma}(\mathrm{i}\omega_{n}) \\ 
                             \Psi_{\bm{k}+\bm{Q},\sigma}(\mathrm{i}\omega_{n}) 
\end{array} \right]~,\nonumber
\end{eqnarray} 
% \vfill\eject
\begin{eqnarray}
\label{S1-t}
{\mathcal S}_{1}^{b}&=& \frac{1}{\sqrt{\beta {\mathrm{N_{s}}}}} \sum_{\bm{k},\bm{q}}~{'} 
\sum_{\sigma^{'},\sigma} \sum_{\mathrm{i}\omega_{m},\mathrm{i}\omega_{n}} 
\sigma_{\sigma^{'},\sigma}^{0}
\left[ \bar{\Psi}_{\bm{k}+\bm{q},\sigma^{'}}(\mathrm{i}\omega_{m})  \ 
\bar{\Psi}_{\bm{k}+\bm{q}+\bm{Q},\sigma^{'}}(\mathrm{i}\omega_{m}) \right]  \hspace{68mm} \\
& &\times \left[ \begin{array}{l} 
-2 (b_{0})^{2} 
(\tilde{t}_{\bm{k}}+\tilde{t}_{\bm{k}+\bm{q}}+\tilde{t}_{\bm{k}}^{\hspace{0.05cm}'}+\tilde{t}_{\bm{k}+\bm{q}}^{\hspace{0.05cm}'}+\tilde{t}_{\bm{k}}^{\hspace{0.05cm}''}+\tilde{t}_{\bm{k}+\bm{q}}^{\hspace{0.05cm}''}) \delta 
b_{\bm{q}}(\mathrm{i}\omega_{m} - \mathrm{i}\omega_{n}) \\
\hspace{3.2cm}
-2 (b_{0})^{2} 
(-\tilde{t}_{\bm{k}}+\tilde{t}_{\bm{k}+\bm{q}}+\tilde{t}_{\bm{k}}^{\hspace{0.05cm}'}+\tilde{t}_{\bm{k}+\bm{q}}^{\hspace{0.05cm}'}+\tilde{t}_{\bm{k}}^{\hspace{0.05cm}''}+\tilde{t}_{\bm{k}+\bm{q}}^{\hspace{0.05cm}''}) \delta 
b_{\bm{q}+\bm{Q}}(\mathrm{i}\omega_{m} - \mathrm{i}\omega_{n}) \\
-2 (b_{0})^{2} 
(\tilde{t}_{\bm{k}}-\tilde{t}_{\bm{k}+\bm{q}}+\tilde{t}_{\bm{k}}^{\hspace{0.05cm}'}+\tilde{t}_{\bm{k}+\bm{q}}^{\hspace{0.05cm}'}+\tilde{t}_{\bm{k}}^{\hspace{0.05cm}''}+\tilde{t}_{\bm{k}+\bm{q}}^{\hspace{0.05cm}''}) \delta 
b_{\bm{q}+\bm{Q}}(\mathrm{i}\omega_{m} - \mathrm{i}\omega_{n}) \\
\hspace{3.2cm}
 -2 (b_{0})^{2} 
(-\tilde{t}_{\bm{k}}-\tilde{t}_{\bm{k}+\bm{q}}+\tilde{t}_{\bm{k}}^{\hspace{0.05cm}'}+\tilde{t}_{\bm{k}+\bm{q}}^{\hspace{0.05cm}'}+\tilde{t}_{\bm{k}}^{\hspace{0.05cm}''}+\tilde{t}_{\bm{k}+\bm{q}}^{\hspace{0.05cm}''}) \delta 
b_{\bm{q}}(\mathrm{i}\omega_{m} - \mathrm{i}\omega_{n})   
\end{array} \right] 
\left[ \begin{array}{c} \Psi_{\bm{k},\sigma}(\mathrm{i}\omega_{n}) \\ 
                             \Psi_{\bm{k}+\bm{Q},\sigma}(\mathrm{i}\omega_{n}) 
\end{array}  \right]~,\nonumber
\end{eqnarray}
\begin{eqnarray}
\label{S1-AF}
{\mathcal S}_{1}^{m} &=& \frac{1}{\sqrt{\beta {\mathrm{N_{s}}}}} \sum_{\bm{k},\bm{q}}~{'} 
\sum_{\sigma^{'},\sigma} \sum_{\mathrm{i}\omega_{m},\mathrm{i}\omega_{n}} 
\sum_{s_{m} = x,y,z} \sigma_{\sigma^{'},\sigma}^{s_{m}}
 \left[ \bar{\Psi}_{\bm{k}+\bm{q},\sigma^{'}}(\mathrm{i}\omega_{m})  \ 
\bar{\Psi}_{\bm{k}+\bm{q}+\bm{Q},\sigma^{'}}(\mathrm{i}\omega_{m}) \right] \hspace{52mm} \\
& &\hspace{38mm} \times \left[ \begin{array}{cc}
\alpha \tilde{J}_{\bm{q}} \cdot \delta m_{\bm{q}}^{s_{m}} (\mathrm{i}\omega_{m} - 
\mathrm{i}\omega_{n}) &
- \alpha \tilde{J}_{\bm{q}} \cdot \delta m_{\bm{q}+\bm{Q}}^{s_{m}} (\mathrm{i}\omega_{m} - 
\mathrm{i}\omega_{n}) \\
- \alpha \tilde{J}_{\bm{q}} \cdot \delta m_{\bm{q}+\bm{Q}}^{s_{m}} (\mathrm{i}\omega_{m} - 
\mathrm{i}\omega_{n}) &
\alpha \tilde{J}_{\bm{q}} \cdot \delta m_{\bm{q}}^{s_{m}} (\mathrm{i}\omega_{m} - 
\mathrm{i}\omega_{n})
\end{array} \right]
 \left[ \begin{array}{c} \Psi_{\bm{k},\sigma}(\mathrm{i}\omega_{n}) \\ 
                             \Psi_{\bm{k}+\bm{Q},\sigma}(\mathrm{i}\omega_{n}) 
\end{array}  \right]~,\nonumber
\end{eqnarray}
\begin{eqnarray}
\label{S1-Pi}
 {\mathcal S}_{1}^{\chi} &=& \frac{1}{\sqrt{\beta {\mathrm{N_{s}}}}} \sum_{\bm{k},\bm{q}}~{'} 
\sum_{\sigma^{'},\sigma} 
\sum_{\mathrm{i}\omega_{m},\mathrm{i}\omega_{n}} \sum_{s_{\chi} = x,y} \sigma_{\sigma^{'},\sigma}^{0} 
 \left[ \bar{\Psi}_{\bm{k}+\bm{q},\sigma^{'}}(\mathrm{i}\omega_{m})  \ 
\bar{\Psi}_{\bm{k}+\bm{q}+\bm{Q},\sigma^{'}}(\mathrm{i}\omega_{m}) \right] \hspace{52mm} \\
& &\hspace{6mm} \times \left[ \begin{array}{l} 
(1-\alpha)J \big[ - \delta^{'} \chi_{\bm{q}}^{s_{\chi}}(\mathrm{i}\omega_{m} 
- \mathrm{i}\omega_{n}) \cdot \mathrm{cos}(k_{s_{\chi}}+q_{s_{\chi}}/2) \\
\ \ \ \ \ \ \ \ \ \ \ \ \  
- \delta^{''} \chi_{\bm{q}}^{s_{\chi}}(\mathrm{i}\omega_{m} - 
\mathrm{i}\omega_{n}) \cdot \mathrm{sin}(k_{s_{\chi}}+q_{s_{\chi}}/2) \big] \\
\hspace{4.2cm}
(1-\alpha)J\big[ - \delta^{'} \chi_{\bm{q}+\bm{Q}}^{s_{\chi}}(\mathrm{i}\omega_{m} 
- \mathrm{i}\omega_{n}) \cdot \mathrm{sin}(k_{s_{\chi}}+q_{s_{\chi}}/2) \\
\hspace{4.2cm} \ \ \ \ \ \ \ \ \ \ \ \ \  
                 + \delta^{''} \chi_{\bm{q}+\bm{Q}}^{s_{\chi}}(\mathrm{i}\omega_{m} 
- \mathrm{i}\omega_{n}) \cdot \mathrm{cos}(k_{s_{\chi}}+q_{s_{\chi}}/2)\big] \\
(1-\alpha)J\big[ + \delta^{'} \chi_{\bm{q}+\bm{Q}}^{s_{\chi}}(\mathrm{i}\omega_{m} 
- \mathrm{i}\omega_{n}) \cdot \mathrm{sin}(k_{s_{\chi}}+q_{s_{\chi}}/2) \\
\ \ \ \ \ \ \ \ \ \ \ \ \  
                 - \delta^{''} \chi_{\bm{q}+\bm{Q}}^{s_{\chi}}(\mathrm{i}\omega_{m} 
- \mathrm{i}\omega_{n}) \cdot \mathrm{cos}(k_{s_{\chi}}+q_{s_{\chi}}/2)\big] \\
\hspace{4.2cm}
(1-\alpha)J\big[ + \delta^{'} \chi_{\bm{q}}^{s_{\chi}}(\mathrm{i}\omega_{m} - 
\mathrm{i}\omega_{n}) \cdot \mathrm{cos}(k_{s_{\chi}}+q_{s_{\chi}}/2) \\
\hspace{4.2cm} \ \ \ \ \ \ \ \ \ \ \ \ \  
                 + \delta^{''} \chi_{\bm{q}}^{s_{\chi}}(\mathrm{i}\omega_{m} 
- \mathrm{i}\omega_{n}) \cdot \mathrm{sin}(k_{s_{\chi}}+q_{s_{\chi}}/2)\big] \\
\end{array} \right] 
\left[ \begin{array}{c} \Psi_{\bm{k},\sigma}(\mathrm{i}\omega_{n}) \\ 
                             \Psi_{\bm{k}+\bm{Q},\sigma}(\mathrm{i}\omega_{n}) 
\end{array}  \right]~. \nonumber
\end{eqnarray} 
\end{widetext}
In the previous formulas $\sigma_{\sigma^{'},\sigma}^{0}$ is an element 
of the $(2 \times 2)$ identity matrix, $\mathrm{i} \omega_{m}$ is (like 
$\mathrm{i}\omega_{n}$) a fermionic Matsubara frequency and we have defined 
the following quantities

\begin{eqnarray}
\label{Def-tk}
\tilde{t}_{\bm{k}} &=& t \big[ \mathrm{cos}({k}_{x}) +\mathrm{cos}({k}_{y}) 
\big]~, \\
%\end{eqnarray} 
% \vfill\eject
% \begin{eqnarray}
\label{Def-tkp}
\tilde{t}_{\bm{k}}^{\hspace{0.05cm}'} &=& t^{'} \big[ 2.\mathrm{cos}({k}_{x}).\mathrm{cos}({k}_{y}) \big]~, 
\end{eqnarray} 
\vfill\eject
\begin{eqnarray}
\label{Def-tkpp}
\tilde{t}_{\bm{k}}^{\hspace{0.05cm}''} &=& t^{''} \big[ \mathrm{cos}(2{k}_{x}) + \mathrm{cos}(2{k}_{y}) \big]~, \\
\label{Def-Jq}
\tilde{J}_{\bm{q}} &=& J \big[ \mathrm{cos}(q_{x}) + \mathrm{cos}(q_{y}) 
\big]~. 
\end{eqnarray} 
% \vfill\eject
% \begin{widetext}
By using the Pauli matrices, we can put the different contributions (\ref{S1-Lamb}) - (\ref{S1-Pi}) together, 
and rewrite ${\mathcal S}_{1}^{fe}$ (\ref{S1-contribs}) as
\begin{widetext}
\begin{eqnarray}
{\mathcal S}_{1}^{fe}
=  \sum_{\bm{k},\bm{q}}~{'} \sum_{\sigma^{'},\sigma} 
\sum_{\mathrm{i}\omega_{m},\mathrm{i}\omega_{n}}
\left[ \bar{\Psi}_{\bm{k}+\bm{q},\sigma^{'}}(\mathrm{i}\omega_{m})  \ 
\bar{\Psi}_{\bm{k}+\bm{q}+\bm{Q},\sigma^{'}}(\mathrm{i}\omega_{m}) \right]
\tilde{{\mathcal V}}_{1}(\bm{k}+\bm{q},\sigma^{'},\mathrm{i}\omega_{m};\bm{k},\sigma,\mathrm{i}\omega_{n}) 
\left[ \begin{array}{c} \Psi_{\bm{k},\sigma}(\mathrm{i}\omega_{n}) \\ 
                             \Psi_{\bm{k}+\bm{Q},\sigma}(\mathrm{i}\omega_{n}) 
\end{array}  \right]~, \hspace{10mm} 
% \\
\label{Moment_S1_fe}
% \nonumber \\
% \noindent \mbox{ with $\tilde{{\mathcal V}}_{1}$ the interaction matrix \hspace{125mm}} \nonumber
\end{eqnarray} 
\vfill\eject
with $\tilde{{\mathcal V}}_{1}$ the interaction matrix
\begin{eqnarray}
\label{Def_V1}
& &   \sqrt{\beta {\mathrm{N_{s}}}} \  \tilde{{\mathcal V}}_{1} 
(\bm{k}+\bm{q},\sigma^{'},\mathrm{i}\omega_{m};\bm{k},\sigma,\mathrm{i}\omega_{n})  \\
&=& \sum_{s_{\chi} = x,y}  \bigg\{ \
 \big[ -(1-\alpha)  J \cdot \mathrm{cos}(k_{s_{\chi}}+q_{s_{\chi}}/2) \cdot 
\sigma_{\sigma^{'},\sigma}^{0} \cdot \tilde{\sigma}^{z} \big]
 \delta^{'} \chi_{\bm{q}}^{s_{\chi}}(\mathrm{i}\omega_{m} - 
\mathrm{i}\omega_{n}) \nonumber \\
& &\hspace{12mm} +\big[ -(1-\alpha)  J \cdot \mathrm{sin}(k_{s_{\chi}}+q_{s_{\chi}}/2) \cdot
\sigma_{\sigma^{'},\sigma}^{0} \cdot \mathrm{i} \tilde{\sigma}^{y} \big]
 \delta^{'} \chi_{\bm{q}+\bm{Q}}^{s_{\chi}}(\mathrm{i}\omega_{m} - 
\mathrm{i}\omega_{n}) \nonumber \\
& &\hspace{12mm} +\big[ -(1-\alpha)  J \cdot \mathrm{sin}(k_{s_{\chi}}+q_{s_{\chi}}/2) \cdot 
\sigma_{\sigma^{'},\sigma}^{0} \cdot \tilde{\sigma}^{z} \big]
 \delta^{''} \chi_{\bm{q}}^{s_{\chi}}(\mathrm{i}\omega_{m} - 
\mathrm{i}\omega_{n}) \nonumber \\
& &\hspace{12mm} +\big[ +(1-\alpha)  J \cdot \mathrm{cos}(k_{s_{\chi}}+q_{s_{\chi}}/2) \cdot 
\sigma_{\sigma^{'},\sigma}^{0} \cdot \mathrm{i}\tilde{\sigma}^{y} \big]
 \delta^{''} \chi_{\bm{q}+\bm{Q}}^{s_{\chi}}(\mathrm{i}\omega_{m} - 
\mathrm{i}\omega_{n}) 
 \bigg\} \nonumber \\
& &+ \sum_{s_{m} = x,y,z} \bigg\{
 \big[ \alpha \tilde{J}_{\bm{q}} \cdot \sigma_{\sigma^{'},\sigma}^{s_{m}} 
\cdot \tilde{\sigma}^{0}\big]
 \delta m_{\bm{q}}^{s_{m}} (\mathrm{i}\omega_{m} - \mathrm{i}\omega_{n}) 
+ \big[- \alpha \tilde{J}_{\bm{q}} \cdot  \sigma_{\sigma^{'},\sigma}^{s_{m}} 
\cdot \tilde{\sigma}^{x}\big]
 \delta m_{\bm{q}+\bm{Q}}^{s_{m}} (\mathrm{i}\omega_{m} - \mathrm{i}\omega_{n}) 
\bigg\} \nonumber \\
& &+ \bigg\{
\big[ -\sigma_{\sigma^{'},\sigma}^{0} \cdot \tilde{\sigma}^{0}\big]\mathrm{i} 
\delta \lambda_{\bm{q}}(\mathrm{i}\omega_{m} - \mathrm{i}\omega_{n})
+\big[ -\sigma_{\sigma^{'},\sigma}^{0} \cdot 
\tilde{\sigma}^{x}\big]\mathrm{i} \delta \lambda_{\bm{q}+\bm{Q}}(\mathrm{i}\omega_{m} - \mathrm{i}\omega_{n})
\bigg\} \nonumber \\
& &+ \bigg\{ \
\big[-2  (b_{0})^{2} 
(\tilde{t}_{\bm{k}}^{\hspace{0.05cm}'}+\tilde{t}_{\bm{k}+\bm{q}}^{\hspace{0.05cm}'}+\tilde{t}_{\bm{k}}^{\hspace{0.05cm}''}+\tilde{t}_{\bm{k}+\bm{q}}^{\hspace{0.05cm}''})\sigma_{\sigma^{'},\sigma}^{0} \cdot 
\tilde{\sigma}^{0}- 2 (b_{0})^{2} 
(\tilde{t}_{\bm{k}}+\tilde{t}_{\bm{k}+\bm{q}})\sigma_{\sigma^{'},\sigma}^{0} \cdot \tilde{\sigma}^{z} 
\big] \delta b_{\bm{q}}(\mathrm{i}\omega_{m} - \mathrm{i}\omega_{n}) 
\nonumber \\
& &\hspace{5.5mm} +\big[-2  (b_{0})^{2} 
(\tilde{t}_{\bm{k}}^{\hspace{0.05cm}'}+\tilde{t}_{\bm{k}+\bm{q}}^{\hspace{0.05cm}'}+\tilde{t}_{\bm{k}}^{\hspace{0.05cm}''}+\tilde{t}_{\bm{k}+\bm{q}}^{\hspace{0.05cm}''})\sigma_{\sigma^{'},\sigma}^{0} \cdot 
\tilde{\sigma}^{x}+ 2 (b_{0})^{2} 
(\tilde{t}_{\bm{k}}-\tilde{t}_{\bm{k}+\bm{q}})\sigma_{\sigma^{'},\sigma}^{0} \cdot 
\mathrm{i}\tilde{\sigma}^{y} \big] \delta b_{\bm{q}+\bm{Q}}(\mathrm{i}\omega_{m} - 
\mathrm{i}\omega_{n}) \bigg\}~. \nonumber
\end{eqnarray} 
%\end{widetext}
\indent Finally the mean field equations are obtained after having integrated 
out the fermionic fields by checking the saddle point property
\cite{Nagaosa_books}
\begin{eqnarray}
 \mathrm{Tr}\big[ \tilde{\mathcal G}_{0} \tilde{\mathcal V}_{1} \big] = 0,
\label{MF_Eq}
\end{eqnarray}
which gives by using the previously obtained expressions 
(\ref{Green_0}), (\ref{Aux_S1}), (\ref{Moment_S1_bo}), (\ref{Moment_S1_fe}) the set 
of coupled mean field equations (\ref{MagneAF-MFEq}) \ - \ (\ref{Lambda-MFEq}).

%%%%%%%%%%%%%%%%%%%%%%%%%%%%%%%%%%%%%%%%%%%%%%%%%%%%%%%%%%%%%%%%%%%%%%%%%%%%%%%%%%%%%%%%%%%%%%%%%%%%%%%%%%%%%%%%%%%%%%%%%%%%%%%%%%%%%%
%%%%%%%%%%%%%%%%%%%%%%%%%%%%%%%%%%%%%%%%%%%%%%%%%%%%%%%%%%%%%%%%%%%%%%%%%%%%%%%%%%%%%%%%%%%%%%%%%%%%%%%%%%%%%%%%%%%%%%%%%%%%%%%%%%%%%%

\section{Derivation of the Gaussian Action}
\label{Appen-GaussAct}
We decompose the second order effective action ${\mathcal S}_{2}$ (\ref{Def_S2})
into two parts
\begin{eqnarray}
{\mathcal S}_{2} = {\mathcal S}_{2}^{bo} + {\mathcal S}_{2}^{bub}.
\label{Aux_S2}
\end{eqnarray} 
${\mathcal S}_{2}^{bo}$ contains only contributions due to auxiliary 
bosonic fields. It includes all the terms of Eq.\! (\ref{Def_S2}) with the 
exception of the trace

%\begin{widetext}
\begin{eqnarray}
{\mathcal S}_{2}^{bo} &=& \int_{0}^{\beta} d\tau \bigg\{ - \sum_{i} 
(b_{0})^{2} \big[2 \cdot \mathrm{i} \delta \lambda_{i} \cdot \delta b_{i}+ \lambda_{0} \cdot (\delta 
b_{i})^{2}\big]  
%\\ 
%& &\hspace{13.0mm}
-  \alpha J \sum_{\langle i,j \rangle } \sum_{s_{m} = x,y,z} 
\big[\delta m_{i}^{s_{m}} \cdot \delta m_{j}^{s_{m}}\big] 
%\nonumber  \\ 
%& &\hspace{13.0mm}
+ \frac{(1-\alpha)}{2} J \sum_{\langle i,j \rangle }
\big[  \delta \chi_{ij} \cdot  \delta \chi_{ij}^{*} \big] \nonumber  \\ 
& &\hspace{13.0mm}- \big(\sum_{\langle i,j \rangle} t + \sum_{\langle i,j 
\rangle^{'}}t^{'}  + \sum_{\langle i,j \rangle^{''}} t^{''} \big)\sum_{\sigma} 
(b_{0})^{2} 
%\nonumber  \\ 
%& &\hspace{13.0mm}~~~\times 
\delta b_{i} \cdot \delta b_{j} \big[ \langle\bar{\Psi}_{i,\sigma} 
\Psi_{j,\sigma}\rangle 
+\langle \bar{\Psi}_{j,\sigma} \Psi_{i,\sigma}\rangle \big] \bigg\}.
\label{Real_S2_bo}
\end{eqnarray} 
${\mathcal S}_{2}^{bub}$ takes into account the ``bubble'' 
contributions
\begin{eqnarray}
& &\hspace{55mm}{\mathcal S}_{2}^{bub} = \frac{1}{2} \mathrm{Tr}\big[ \tilde{\mathcal G}_{0} 
\tilde{\mathcal V}_{1} \tilde{\mathcal G}_{0} \tilde{\mathcal V}_{1} \big]. 
\label{Real_S2_bub} \\
\nonumber \\
& &\mbox{We firstly evaluate ${\mathcal S}_{2}^{bo}$ (\ref{Real_S2_bo}) by passing in the space of 
moments and Matsubara frequencies} \hspace{33.5mm} \nonumber
\end{eqnarray} 
\begin{eqnarray}
& &{\mathcal S}_{2}^{bo} = - (b_{0})^{2} \sum_{\bm{q}}~{'} 
\sum_{\mathrm{i}\omega_{\ell}}
\bigg\{ \ 2 \mathrm{i}\big[  \delta \lambda_{\bm{q}}(\mathrm{i}\omega_{\ell}) \cdot
\delta b_{-\bm{q}}(-\mathrm{i}\omega_{\ell}) +
\delta \lambda_{\bm{q}+\bm{Q}}(\mathrm{i}\omega_{\ell}) \cdot \delta 
b_{-\bm{q}-\bm{Q}}(-\mathrm{i}\omega_{\ell}) \big] \hspace{50mm} \nonumber \\
& &\hspace{37mm}+  \lambda_{0} \big[ \ \delta b_{\bm{q}}(\mathrm{i}\omega_{\ell}) \cdot \delta 
b_{-\bm{q}}(-\mathrm{i}\omega_{\ell})
+ \delta b_{\bm{q}+\bm{Q}}(\mathrm{i}\omega_{\ell}) \cdot \delta 
b_{-\bm{q}-\bm{Q}}(-\mathrm{i}\omega_{\ell}) \big]\bigg\} \nonumber \\
& &\hspace{10mm}-  \alpha  \sum_{\bm{q}}~{'} \sum_{\mathrm{i}\omega_{\ell}} \tilde{J}_{\bm{q}} 
\sum_{s_{m} = x,y,z}  \big[ \
 \delta m_{\bm{q}}^{s_{m}} (\mathrm{i}\omega_{\ell}) \cdot \delta m_{-\bm{q}}^{s_{m}} 
(-\mathrm{i}\omega_{\ell}) 
-  \delta m_{\bm{q}+\bm{Q}}^{s_{m}} (\mathrm{i}\omega_{\ell}) \cdot \delta m_{-\bm{q}-\bm{Q}}^{s_{m}} 
(-\mathrm{i}\omega_{\ell}) \big] \nonumber \\
& &\hspace{10mm}+ \frac{(1-\alpha)}{2} J \sum_{\bm{q}}~{'} \sum_{\mathrm{i}\omega_{\ell}} 
\sum_{s_{\chi} = x,y}
\big[ \ \delta^{'} \chi_{\bm{q}}^{s_{\chi}}(\mathrm{i}\omega_{\ell}) \cdot \delta^{'} 
\chi_{-\bm{q}}^{s_{\chi}}(-\mathrm{i}\omega_{\ell})   
+  \delta^{''} \chi_{\bm{q}}^{s_{\chi}}(\mathrm{i}\omega_{\ell}) \cdot \delta^{''} 
\chi_{-\bm{q}}^{s_{\chi}}(-\mathrm{i}\omega_{\ell}) \nonumber \\
& &\hspace{52mm}+  \delta^{'} \chi_{\bm{q}+\bm{Q}}^{s_{\chi}}(\mathrm{i}\omega_{\ell}) \cdot \delta^{'} 
\chi_{-\bm{q}-\bm{Q}}^{s_{\chi}}(-\mathrm{i}\omega_{\ell})
+ \delta^{''} \chi_{\bm{q}+\bm{Q}}^{s_{\chi}}(\mathrm{i}\omega_{\ell}) \cdot \delta^{''} 
\chi_{-\bm{q}-\bm{Q}}^{s_{\chi}}(-\mathrm{i}\omega_{\ell}) \big] \nonumber \\
& &\hspace{10mm}- 4 (b_{0})^{2}\sum_{\bm{q}}~{'} \sum_{\mathrm{i}\omega_{\ell}} \Bigg[ \
\bigg\{ t \cdot l_{1}^{x}\big[ \mathrm{cos}(q_{x})+ 
\mathrm{cos}(q_{y}) \big]
+ 2 t^{'} \cdot l_{2}^{x,y} \mathrm{cos}(q_{x}) \cdot \mathrm{cos}(q_{y}) 
%\nonumber \\
%& &
+ t^{''} \cdot l_{2}^{x,x}\big[ \mathrm{cos}(2q_{x})+ 
\mathrm{cos}(2q_{y}) \big] \bigg\}\nonumber \\
& &\hspace{41.5mm} \times \delta b_{\bm{q}}(\mathrm{i}\omega_{\ell}) \cdot \delta 
b_{-\bm{q}}(-\mathrm{i}\omega_{\ell}) \nonumber \\
& &\hspace{38mm}+\bigg\{- t \cdot l_{1}^{x}\big[ \mathrm{cos}(q_{x})+ 
\mathrm{cos}(q_{y}) \big]
+ 2 t^{'} \cdot l_{2}^{x,y} \mathrm{cos}(q_{x}) \cdot \mathrm{cos}(q_{y}) 
%\nonumber \\
%& &
+ t^{''} \cdot l_{2}^{x,x}\big[ \mathrm{cos}(2q_{x})+ 
\mathrm{cos}(2q_{y}) \big] \bigg\}\nonumber \\
& &\hspace{41.5mm}\times  \delta b_{\bm{q}+\bm{Q}}(\mathrm{i}\omega_{\ell}) \cdot \delta 
b_{-\bm{q}-\bm{Q}}(-\mathrm{i}\omega_{\ell}) 
\Bigg],
\label{Moment_S2_bo}
\end{eqnarray} 
%\end{widetext}
where $\tilde{J}_{\bm{q}}$ is given by Eq.\! (\ref{Def-Jq}) and we have defined for $s_{t}, s_{t^{'}} = x, y$
\begin{eqnarray}
  l_{1}^{s_{t}} 
&=& 
\langle\bar{\Psi}_{i,\sigma} 
\Psi_{i+s_{t},\sigma}\rangle 
= \frac{1}{{\mathrm{N_{s}}}} \sum_{\bm{k}}~{'} 
\frac{(\xi_{\bm{k}+\bm{Q}}-\xi_{\bm{k}})}{E_{\bm{k}}^{up}-E_{\bm{k}}^{low}} \mathrm{cos} (k_{s_{t}}), \nonumber \\
  l_{2}^{s_{t},s_{t^{'}}} 
&=& 
\langle\bar{\Psi}_{i,\sigma} 
\Psi_{i+s_{t}+s_{t^{'}},\sigma}\rangle 
%\nonumber \\
%&=& 
= \frac{1}{{\mathrm{N_{s}}}} \sum_{\bm{k}}~{'} 
\frac{(\xi_{\bm{k}+\bm{Q}}-\xi_{\bm{k}})}{E_{\bm{k}}^{up}-E_{\bm{k}}^{low}} \mathrm{cos} (k_{s_{t}}+ k_{s_{t^{'}}}). \nonumber 
\end{eqnarray} 
In order to write ${\mathcal S}_{2}^{bub}$ (\ref{Real_S2_bub}) and
${\mathcal S}_{2}^{bo}$ (\ref{Moment_S2_bo})
in a compact way which will be useful in the following calculations we 
define the vector of nine components $\delta \vec{X}$ containing the 
first order fluctuations of the bosonic fields 
\begin{eqnarray} 
\delta \vec{X}(\bm{q},\mathrm{i}\omega_{\ell}) = \big[ \delta 
X_{i}(\bm{q},\mathrm{i}\omega_{\ell}) \big]_{1\leq i \leq 9 } =
\left[ 
\begin{array}{c}
\delta^{'} \chi_{\bm{q}}^{x}(\mathrm{i}\omega_{\ell}) \\
\delta^{'} \chi_{\bm{q}}^{y}(\mathrm{i}\omega_{\ell}) \\
\delta^{''} \chi_{\bm{q}}^{x}(\mathrm{i}\omega_{\ell}) \\ 
\delta^{''} \chi_{\bm{q}}^{y}(\mathrm{i}\omega_{\ell}) \\
\delta m_{\bm{q}}^{x}(\mathrm{i}\omega_{\ell}) \\
\delta m_{\bm{q}}^{y}(\mathrm{i}\omega_{\ell}) \\
\delta m_{\bm{q}}^{z}(\mathrm{i}\omega_{\ell}) \\
\delta \lambda_{\bm{q}}(\mathrm{i}\omega_{\ell}) \\
\delta b_{\bm{q}}(\mathrm{i}\omega_{\ell}) 
\end{array} 
\right],~
\delta \vec{X}(\bm{q}+\bm{Q},\mathrm{i}\omega_{\ell}) = \big[ \delta 
X_{i}(\bm{q}+\bm{Q},\mathrm{i}\omega_{\ell}) \big]_{1\leq i \leq 9 } 
=
\left[ 
\begin{array}{c}
\delta^{'} \chi_{\bm{q}+\bm{Q}}^{x}(\mathrm{i}\omega_{\ell}) \\
\delta^{'} \chi_{\bm{q}+\bm{Q}}^{y}(\mathrm{i}\omega_{\ell}) \\
\delta^{''} \chi_{\bm{q}+\bm{Q}}^{x}(\mathrm{i}\omega_{\ell}) \\ 
\delta^{''} \chi_{\bm{q}+\bm{Q}}^{y}(\mathrm{i}\omega_{\ell}) \\
\delta m_{\bm{q}+\bm{Q}}^{x}(\mathrm{i}\omega_{\ell}) \\
\delta m_{\bm{q}+\bm{Q}}^{y}(\mathrm{i}\omega_{\ell}) \\
\delta m_{\bm{q}+\bm{Q}}^{z}(\mathrm{i}\omega_{\ell}) \\
\delta \lambda_{\bm{q}+\bm{Q}}(\mathrm{i}\omega_{\ell}) \\
\delta b_{\bm{q}+\bm{Q}}(\mathrm{i}\omega_{\ell}) 
\end{array} 
\right]~.
\label{Def_VecXq}
\end{eqnarray}
\end{widetext}

Using Eq.\! (\ref{Moment_S2_bo}) we define the matrix $\tilde{\mathcal M}$ so 
that
\begin{eqnarray}
\tilde{\mathcal M}(\bm{q},\bm{{q}_{1}},\bm{{q}_{2}},\mathrm{i}\omega_{\ell}) =
\bigg[
{\mathcal M}_{i,j}(\bm{q},\bm{{q}_{1}},\bm{{q}_{2}},\mathrm{i}\omega_{\ell})
\bigg]_{1\leq i,j \leq 9},  \nonumber 
\end{eqnarray}
\begin{eqnarray}
& &{\mathcal S}_{2}^{bo} =  
\sum_{\bm{q}}~{'} \sum_{\bm{{q}_{1}},\bm{{q}_{2}}=\bm{q},\bm{q}+\bm{Q}} \sum_{\mathrm{i}\omega_{\ell}} 
\sum_{i,j=1}^{9} 
\delta X_{i}(\bm{{q}_{1}},\mathrm{i}\omega_{\ell}) \nonumber \\
& &\hspace{12mm} \times {\mathcal M}_{i,j}(\bm{q},\bm{{q}_{1}},\bm{{q}_{2}},\mathrm{i}\omega_{\ell})
\cdot \delta X_{j}(-\bm{{q}_{2}},-\mathrm{i}\omega_{\ell})~.~~~~~
\label{DefMatrixM}
\end{eqnarray}
We now tackle the calculation of the ``bubble'' part ${\mathcal 
S}_{2}^{bub}$ (\ref{Real_S2_bub}). We have \cite{Nagaosa_books}
%\vfill\eject
\begin{eqnarray}
\label{Def_trace}
& &\mathrm{Tr}\big[ \tilde{\mathcal G}_{0} \tilde{\mathcal V}_{1} \tilde{\mathcal 
G}_{0} \tilde{\mathcal V}_{1} \big]  \\
&=& \sum_{\bm{k},\bm{q}}~{'} \sum_{\sigma,\sigma^{'}} 
\sum_{\mathrm{i}\omega_{n},\mathrm{i}\omega_{\ell}} 
 \mathrm{tr}\big[\ \tilde{\mathcal 
G}_{0}(\bm{k}+\bm{q},\sigma^{'},\mathrm{i}\omega_{n}+\mathrm{i}\omega_{\ell})  \nonumber \\
& &\hspace{20mm} \times \tilde{\mathcal V}_{1} 
(\bm{k}+\bm{q},\sigma^{'},\mathrm{i}\omega_{n}+\mathrm{i}\omega_{\ell};\bm{k},\sigma,\mathrm{i}\omega_{n}) \nonumber \\
& &\hspace{20mm} \times \ \tilde{\mathcal G}_{0} 
(\bm{k},\sigma,\mathrm{i}\omega_{n}) \nonumber \\
& &\hspace{20mm} \times \tilde{\mathcal  V}_{1}(\bm{k},\sigma,\mathrm{i}\omega_{n};\bm{k}+\bm{q},\sigma^{'},\mathrm{i}\omega_{n}+\mathrm{i}\omega_{\ell}) \big]~, \nonumber
\end{eqnarray} 
where $\mathrm{tr}$ is related to the spinor basis (\ref{Def_spinor}). We define the vector 
$\vec{c}^{\hspace{0.07cm}\sigma^{'},\sigma}$ containing the 
coefficients associated with the first order fluctuations in the 
$\tilde{{\mathcal V}}_{1}$ matrix. As we can see in Eq.\! (\ref{Def_V1}) $\delta b_{\bm{q}}$  
(and $\delta b_{\bm{q}+\bm{Q}}$) gives two separated contributions and not 
one like all the other auxiliary bosonic fields whose first order 
fluctuations are contained
%\vfill\eject
\noindent in the vector $\delta \vec{X}$ (\ref{Def_VecXq}). 
Therefore $\vec{c}^{\hspace{0.05cm}\sigma^{'},\sigma}$ has ten 
components and not nine, which is indicated by the index $a^{'}$ being 
equal to one for all the fields except $\delta b_{\bm{q}}$  (and $\delta 
b_{\bm{q}+\bm{Q}}$) for which it takes the values one and two
\begin{widetext}
\begin{eqnarray}
\vec{c}^{\hspace{0.08cm}\sigma^{'},\sigma}(\bm{k},\bm{q}) = \big[ 
c_{i,a^{'}}^{\sigma^{'},\sigma}(\bm{k},\bm{q}) \big]_{1\leq i \leq 9} =
\left[ 
\begin{array}{c}
-(1-\alpha)  J \mathrm{cos}({k}_{x}+q_{x}/2) 
\sigma_{\sigma^{'},\sigma}^{0} \\
-(1-\alpha)  J \mathrm{cos}({k}_{y}+q_{y}/2) 
\sigma_{\sigma^{'},\sigma}^{0} \\
-(1-\alpha)  J \mathrm{sin}({k}_{x}+q_{x}/2) 
\sigma_{\sigma^{'},\sigma}^{0} \\
-(1-\alpha)  J \mathrm{sin}({k}_{y}+q_{y}/2) 
\sigma_{\sigma^{'},\sigma}^{0} \\
\alpha \tilde{J}_{\bm{q}} \sigma_{\sigma^{'},\sigma}^{x} \\
\alpha \tilde{J}_{\bm{q}} \sigma_{\sigma^{'},\sigma}^{y} \\
\alpha \tilde{J}_{\bm{q}} \sigma_{\sigma^{'},\sigma}^{z} \\
-\mathrm{i} \sigma_{\sigma^{'},\sigma}^{0} \\
-2  (b_{0})^{2} 
(\tilde{t}_{\bm{k}}^{\hspace{0.05cm}'}+\tilde{t}_{\bm{k}+\bm{q}}^{\hspace{0.05cm}'}+\tilde{t}_{\bm{k}}^{\hspace{0.05cm}''}+\tilde{t}_{\bm{k}+\bm{q}}^{\hspace{0.05cm}''})\sigma_{\sigma^{'},\sigma}^{0} \\
- 2 (b_{0})^{2} 
(\tilde{t}_{\bm{k}}+\tilde{t}_{\bm{k}+\bm{q}})\sigma_{\sigma^{'},\sigma}^{0} 
\end{array} 
\right],
\label{Def_VecCoeffq}
\end{eqnarray}
\\
\begin{eqnarray}
\vec{c}^{\hspace{0.08cm}\sigma^{'},\sigma}(\bm{k},\bm{q}+\bm{Q}) = \big[ 
c_{i,a^{'}}^{\sigma^{'},\sigma}(\bm{k},\bm{q}+\bm{Q}) \big]_{1\leq i \leq 9} =
\left[ 
\begin{array}{c}
-\mathrm{i}(1-\alpha) J \mathrm{sin}({k}_{x}+q_{x}/2) 
\sigma_{\sigma^{'},\sigma}^{0} \\
-\mathrm{i}(1-\alpha) J \mathrm{sin}({k}_{y}+q_{y}/2) 
\sigma_{\sigma^{'},\sigma}^{0} \\
\mathrm{i}(1-\alpha) J \mathrm{cos}({k}_{x}+q_{x}/2) 
\sigma_{\sigma^{'},\sigma}^{0} \\
\mathrm{i}(1-\alpha) J \mathrm{cos}({k}_{y}+q_{y}/2) 
\sigma_{\sigma^{'},\sigma}^{0} \\
-\alpha \tilde{J}_{\bm{q}} \sigma_{\sigma^{'},\sigma}^{x} \\
-\alpha \tilde{J}_{\bm{q}} \sigma_{\sigma^{'},\sigma}^{y} \\
-\alpha \tilde{J}_{\bm{q}} \sigma_{\sigma^{'},\sigma}^{z} \\
-\mathrm{i} \sigma_{\sigma^{'},\sigma}^{0} \\
-2  (b_{0})^{2} 
(\tilde{t}_{\bm{k}}^{\hspace{0.05cm}'}+\tilde{t}_{\bm{k}+\bm{q}}^{\hspace{0.05cm}'}+\tilde{t}_{\bm{k}}^{\hspace{0.05cm}''}+\tilde{t}_{\bm{k}+\bm{q}}^{\hspace{0.05cm}''})\sigma_{\sigma^{'},\sigma}^{0} \\
2 \mathrm{i} (b_{0})^{2} 
(\tilde{t}_{\bm{k}}-\tilde{t}_{\bm{k}+\bm{q}})\sigma_{\sigma^{'},\sigma}^{0} 
\end{array} 
\right]~,
\label{Def_VecCoeffqQ}
\end{eqnarray}
%\end{widetext}
where the terms $\tilde{t}_{\bm{k}}$, $\tilde{t}_{\bm{k}}^{\hspace{0.05cm}'}$, $\tilde{t}_{\bm{k}}^{\hspace{0.05cm}''}$ and $\tilde{J}_{\bm{q}}$
 are defined in Eqs.\! (\ref{Def-tk}), (\ref{Def-tkp}), (\ref{Def-tkpp}) and (\ref{Def-Jq}), respectively. In the same manner
 we define the vector $\vec{s}$ to associate each 
fluctuation element $\delta X_{i}$ to its corresponding Pauli matrix 
appearing in the expression of the $\tilde{{\mathcal V}}_{1}$ matrix (\ref{Def_V1})
\begin{eqnarray}
& & \hspace{30mm} \vec{s}(\bm{q}) = \big[ \tilde{s}_{i,a^{'}}(\bm{q}) \big]_{1\leq i \leq 9} =
\left[ 
\begin{array}{c}
\tilde{\sigma}^{z} \\
\tilde{\sigma}^{z} \\
\tilde{\sigma}^{z} \\
\tilde{\sigma}^{z} \\
\tilde{\sigma}^{0} \\
\tilde{\sigma}^{0} \\
\tilde{\sigma}^{0} \\
\tilde{\sigma}^{0} \\
\tilde{\sigma}^{0} \\
\tilde{\sigma}^{z} 
\end{array} 
\right]~,~
\vec{s}(\bm{q}+\bm{Q}) = \big[ \tilde{s}_{i,a^{'}}(\bm{q}+\bm{Q}) \big]_{1\leq i \leq 9} =
\left[ 
\begin{array}{c}
\tilde{\sigma}^{y} \\
\tilde{\sigma}^{y} \\
\tilde{\sigma}^{y} \\
\tilde{\sigma}^{y} \\
\tilde{\sigma}^{x} \\
\tilde{\sigma}^{x} \\
\tilde{\sigma}^{x} \\
\tilde{\sigma}^{x} \\
\tilde{\sigma}^{x} \\
\tilde{\sigma}^{y} 
\end{array} 
\right]~. 
\label{Def_VecPauliq} 
\end{eqnarray}
The summation over the fermionic Matsubara frequencies $\mathrm{i}\omega_{n}$ is incorporated in \\
\begin{eqnarray}
& & \hspace{5mm} F_{\tilde{s}_{i,a^{'}}(\bm{{q}_{1}}), 
\tilde{s}_{j,a^{''}}(\bm{{q}_{2}})}^{\sigma^{'},\sigma}(\bm{k},\bm{q},\mathrm{i}\omega_{\ell}) 
% \nonumber \\
= \frac{1}{\beta}\sum_{\mathrm{i}\omega_{n}} 
\mathrm{tr} \big[\ \tilde{\mathcal 
G}_{0}(\bm{k}+\bm{q},\sigma^{'},\mathrm{i}\omega_{n}+\mathrm{i}\omega_{\ell})
\tilde{s}_{i,a^{'}}(\bm{{q}_{1}}) 
% \nonumber \\ & &\hspace{13.5mm} \times \ 
\tilde{\mathcal G}_{0} 
(\bm{k},\sigma,\mathrm{i}\omega_{n}) 
\tilde{s}_{j,a^{''}}(\bm{{q}_{2}}) \
 \big]~, 
\label{Def_Finteg}
\end{eqnarray}
with $\bm{{q}_{1}},\bm{{q}_{2}} = \bm{q}, \bm{q}+\bm{Q}$. Considering the identity matrix and the 
three $SU(2)$ Pauli matrices we get from Eq.\! (\ref{Def_Finteg}) sixteen 
different thermal integrals which are de\-tailed in Appendix \ref{Appen-ThermalInt}.
\vfill\eject
\end{widetext}

%\vfill\eject
  Using the preceding expressions (\ref{Def_VecXq}), 
(\ref{Def_VecCoeffq}), (\ref{Def_VecCoeffqQ}), (\ref{Def_VecPauliq}), 
and the formulas associated with the 
$F$ - integrals, we define the matrix $\tilde{\Pi}$ which explicitly 
incorporates at a RPA (``bubble'') level the contributions of the 
different fields fluctuations to the correlation functions
\begin{eqnarray}
\tilde{\Pi}(\bm{q},\bm{{q}_{1}},\bm{{q}_{2}},\mathrm{i}\omega_{\ell}) = 
\bigg[
{\Pi}_{i,j}(\bm{q},\bm{{q}_{1}},\bm{{q}_{2}},\mathrm{i}\omega_{\ell})
\bigg]_{1\leq i,j \leq 9}~,~~
\label{Def_PiMatrix}
\end{eqnarray}
with the matricial elements
\begin{eqnarray}
& &{\Pi}_{i,j}(\bm{q},\bm{{q}_{1}},\bm{{q}_{2}},\mathrm{i}\omega_{\ell}) \hspace{50mm} \nonumber \\
&=&
\frac{1}{{\mathrm{N_{s}}}} \sum_{\bm{k}}~{'} \sum_{\sigma^{'},\sigma} \ 
\sum_{a^{'},a^{''}=1}^{2}
c_{i,a^{'}}^{\sigma^{'},\sigma}(\bm{k},\bm{{q}_{1}}) \nonumber \\
& &\hspace{15mm} \times F_{\tilde{s}_{i,a^{'}}(\bm{{q}_{1}}), 
\tilde{s}_{j,a^{''}}(\bm{{q}_{2}})}^{\sigma^{'},\sigma}(\bm{k},\bm{q},\mathrm{i}\omega_{\ell})
c_{j,a^{''}}^{\sigma,\sigma^{'}}(\bm{k},\bm{{q}_{2}})~. \nonumber
\end{eqnarray}

From Eqs.\! (\ref{Def_trace}), (\ref{Def_Finteg}) and (\ref{Def_PiMatrix}) we get
\begin{eqnarray}
\label{Result_trace}
& &\mathrm{Tr}\big[ \tilde{\mathcal G}_{0} \tilde{\mathcal V}_{1} \tilde{\mathcal 
G}_{0} \tilde{\mathcal V}_{1} \big] \hspace{55mm} \\
&=&
\sum_{\bm{q}}~{'} \sum_{\bm{{q}_{1}},\bm{{q}_{2}}=\bm{q},\bm{q}+\bm{Q}} \sum_{\mathrm{i}\omega_{\ell}} 
\sum_{i,j=1}^{9}
\delta X_{i}(\bm{{q}_{1}},\mathrm{i}\omega_{\ell}) \nonumber \\
& &\hspace{19mm} \times {\Pi}_{i,j}(\bm{q},\bm{{q}_{1}},\bm{{q}_{2}},\mathrm{i}\omega_{\ell}) 
 \delta X_{j}(-\bm{{q}_{2}},-\mathrm{i}\omega_{\ell})~. \nonumber
\end{eqnarray}
With Eqs.\! (\ref{Aux_S2}), (\ref{DefMatrixM}), (\ref{Real_S2_bub}) and 
(\ref{Result_trace}) we finally obtain $\mathcal{S}_{2}$ given by Eq.\! (\ref{Result_S2}).

%%%%%%%%%%%%%%%%%%%%%%%%%%%%%%%%%%%%%%%%%%%%%%%%%%%%%%%%%%%%%%%%%%%%%%%%%%%%%%%%%%%%%%%%%%%%%%%%%%%%%%%%%%%%%%%%%%%%%%%%%%%%%%%%%%%%%%
%%%%%%%%%%%%%%%%%%%%%%%%%%%%%%%%%%%%%%%%%%%%%%%%%%%%%%%%%%%%%%%%%%%%%%%%%%%%%%%%%%%%%%%%%%%%%%%%%%%%%%%%%%%%%%%%%%%%%%%%%%%%%%%%%%%%%%

\section{Thermal integrals expressions}
\label{Appen-ThermalInt}

We give here the analytic expressions of the sixteen different $F$ 
integrals (\ref{Def_Finteg}) useful to calculate at a RPA level the 
two (bosonic) fields correlation functions (\ref{RPA_Corre_f}). We define 

\begin{eqnarray*}
& &E_{u,l}^{4}=
\big[E_{\bm{k}+\bm{q}}^{up}-(\mathrm{i}\omega_{n}+\mathrm{i}\omega_{\ell})\big]
\big[ (\mathrm{i}\omega_{n}+\mathrm{i}\omega_{\ell})- E_{\bm{k}+\bm{q}}^{low}\big] \\
& &\hspace{10mm} \times
\big[E_{\bm{k}}^{up}-\mathrm{i}\omega_{n}\big]
\big[\mathrm{i}\omega_{n}-E_{\bm{k}}^{low}  \big]~, \\
\\
& &g_{0,0} = 2(\mathrm{i}\omega_{n})^{2}-\mathrm{i}\omega_{\ell}(\xi_{\bm{k}}+\xi_{\bm{k}+\bm{Q}}) \\
& &\hspace{4mm}+\mathrm{i}\omega_{n} \big[2 \mathrm{i}\omega_{\ell} - 
(\xi_{\bm{k}}+\xi_{\bm{k}+\bm{Q}}+\xi_{\bm{k}+\bm{q}}+\xi_{\bm{k}+\bm{q}+\bm{Q}})  \big]~, \\
\\
& &g_{0,x} =
-2\mathrm{i}\omega_{n} 
(\Delta_{\bm{k},\sigma}^{m}+\Delta_{\bm{k}+\bm{q},\sigma^{'}}^{m})-2\mathrm{i}\omega_{\ell}\Delta_{\bm{k},\sigma}^{m} \nonumber \\
& &\hspace{4mm}+\Delta_{\bm{k},\sigma}^{m}(\xi_{\bm{k}+\bm{q}}+\xi_{\bm{k}+\bm{q}+\bm{Q}})
+\Delta_{\bm{k}+\bm{q},\sigma^{'}}^{m}(\xi_{\bm{k}}+\xi_{\bm{k}+\bm{Q}})~, \\
\\
& &g_{0,y} =
-2\mathrm{i}\omega_{n} 
(\mathrm{i}\Delta_{\bm{k},\sigma}^{\chi}+\mathrm{i}\Delta_{\bm{k}+\bm{q},\sigma^{'}}^{\chi})
-2\mathrm{i}\omega_{\ell}\mathrm{i}\Delta_{\bm{k},\sigma}^{\chi} \nonumber \\
& &\hspace{4mm}+\mathrm{i} 
\Delta_{\bm{k},\sigma}^{\chi}(\xi_{\bm{k}+\bm{q}}+\xi_{\bm{k}+\bm{q}+\bm{Q}})
+\mathrm{i}\Delta_{\bm{k}+\bm{q},\sigma^{'}}^{\chi}(\xi_{\bm{k}}+\xi_{\bm{k}+\bm{Q}})~, \\
\\
& &g_{0,z} =
\mathrm{i}\omega_{n}(\xi_{\bm{k}}-\xi_{\bm{k}+\bm{Q}}+\xi_{\bm{k}+\bm{q}}-\xi_{\bm{k}+\bm{q}+\bm{Q}}) \\
& &\hspace{4mm}+ \mathrm{i}\omega_{\ell}(\xi_{\bm{k}}-\xi_{\bm{k}+\bm{Q}})
- (\xi_{\bm{k}}\xi_{\bm{k}+\bm{q}}-\xi_{\bm{k}+\bm{Q}}\xi_{\bm{k}+\bm{q}+\bm{Q}})~,
\end{eqnarray*}
\vfill\eject
\begin{eqnarray*}
& &g_{x,y} =
\mathrm{i}\omega_{n}(-\mathrm{i}\xi_{\bm{k}}+\mathrm{i}\xi_{\bm{k}+\bm{Q}}
+\mathrm{i}\xi_{\bm{k}+\bm{q}}-\mathrm{i}\xi_{\bm{k}+\bm{q}+\bm{Q}}) \nonumber \\
& &\hspace{4mm}-\mathrm{i}\omega_{\ell}(\mathrm{i}\xi_{\bm{k}}-\mathrm{i}\xi_{\bm{k}+\bm{Q}})
%\nonumber \\
%& &\hspace{6mm}
+ (\mathrm{i}\xi_{\bm{k}}\xi_{\bm{k}+\bm{q}+\bm{Q}}-\mathrm{i}\xi_{\bm{k}+\bm{Q}}\xi_{\bm{k}+\bm{q}})
~, \\
\\
& &g_{x,z} =
2\mathrm{i}\omega_{n} 
(\Delta_{\bm{k},\sigma}^{\chi}-\Delta_{\bm{k}+\bm{q},\sigma^{'}}^{\chi})+2\mathrm{i}\omega_{\ell}\Delta_{\bm{k},\sigma}^{\chi} \nonumber \\
& &\hspace{4mm}- \Delta_{\bm{k},\sigma}^{\chi}(\xi_{\bm{k}+\bm{q}}+\xi_{\bm{k}+\bm{q}+\bm{Q}})
+\Delta_{\bm{k}+\bm{q},\sigma^{'}}^{\chi}(\xi_{\bm{k}}+\xi_{\bm{k}+\bm{Q}})~, \\
\\
& &g_{y,z} =
2\mathrm{i}\omega_{n} 
(\mathrm{i}\Delta_{\bm{k},\sigma}^{m}-\mathrm{i}\Delta_{\bm{k}+\bm{q},\sigma^{'}}^{m})
+2\mathrm{i}\omega_{\ell}\mathrm{i}\Delta_{\bm{k},\sigma}^{m} \nonumber \\
& &\hspace{4mm}- \mathrm{i}\Delta_{\bm{k},\sigma}^{m}(\xi_{\bm{k}+\bm{q}}+\xi_{\bm{k}+\bm{q}+\bm{Q}})
+\mathrm{i}\Delta_{\bm{k}+\bm{q},\sigma^{'}}^{m}(\xi_{\bm{k}}+\xi_{\bm{k}+\bm{Q}})~,
\end{eqnarray*}
and we obtain
\begin{eqnarray*}
& &F_{\tilde{\sigma}^{0},\tilde{\sigma}^{0}}^{{\sigma^{'},\sigma}}(\bm{k},\bm{q},\mathrm{i}\omega_{\ell}) 
%\nonumber \\
=
\frac{1}{\beta}\sum_{\mathrm{i}\omega_{n}}
\frac{1}{E_{u,l}^{4}} 
%\nonumber \\
%& &\hspace{2mm}\times 
\bigg\{
 g_{0,0} \hspace{25mm} \nonumber \\
& &\hspace{19mm}+ (\xi_{\bm{k}}\xi_{\bm{k}+\bm{q}}+\xi_{\bm{k}+\bm{Q}}\xi_{\bm{k}+\bm{q}+\bm{Q}}) \nonumber \\
& &\hspace{19mm}+ 2( \Delta_{\bm{k},\sigma}^{m}\Delta_{\bm{k}+\bm{q},\sigma^{'}}^{m}- 
\Delta_{\bm{k},\sigma}^{\chi}\Delta_{\bm{k}+\bm{q},\sigma^{'}}^{\chi})\bigg\}~, \nonumber \\
\\
& &F_{\tilde{\sigma}^{x},\tilde{\sigma}^{x}}^{{\sigma^{'},\sigma}}(\bm{k},\bm{q},\mathrm{i}\omega_{\ell}) 
%\nonumber \\
=
\frac{1}{\beta}\sum_{\mathrm{i}\omega_{n}}
\frac{1}{E_{u,l}^{4}} 
%\nonumber \\
%& &\hspace{2mm} \times 
\bigg\{
g_{0,0} \hspace{25mm} \nonumber \\
& &\hspace{19mm}+ (\xi_{\bm{k}}\xi_{\bm{k}+\bm{q}+\bm{Q}}+\xi_{\bm{k}+\bm{Q}}\xi_{\bm{k}+\bm{q}}) \nonumber \\
& &\hspace{19mm}+ 2( \Delta_{\bm{k},\sigma}^{m}\Delta_{\bm{k}+\bm{q},\sigma^{'}}^{m} +
\Delta_{\bm{k},\sigma}^{\chi}\Delta_{\bm{k}+\bm{q},\sigma^{'}}^{\chi})\bigg\}~, \nonumber \\
\\
& &F_{\tilde{\sigma}^{y},\tilde{\sigma}^{y}}^{{\sigma^{'},\sigma}}(\bm{k},\bm{q},\mathrm{i}\omega_{\ell}) 
%\nonumber \\
=
\frac{1}{\beta}\sum_{\mathrm{i}\omega_{n}}
\frac{1}{E_{u,l}^{4}} 
%\nonumber \\
%& &\hspace{2mm} \times 
\bigg\{
g_{0,0} \hspace{25mm} \nonumber \\
& &\hspace{19mm}+ (\xi_{\bm{k}}\xi_{\bm{k}+\bm{q}+\bm{Q}}+\xi_{\bm{k}+\bm{Q}}\xi_{\bm{k}+\bm{q}}) \nonumber \\
& &\hspace{19mm}- 2( \Delta_{\bm{k},\sigma}^{m}\Delta_{\bm{k}+\bm{q},\sigma^{'}}^{m} + 
\Delta_{\bm{k},\sigma}^{\chi}\Delta_{\bm{k}+\bm{q},\sigma^{'}}^{\chi})\bigg\}~, \nonumber \\
\\
& &F_{\tilde{\sigma}^{z},\tilde{\sigma}^{z}}^{{\sigma^{'},\sigma}}(\bm{k},\bm{q},\mathrm{i}\omega_{\ell}) 
%\nonumber \\
=
\frac{1}{\beta}\sum_{\mathrm{i}\omega_{n}}
\frac{1}{E_{u,l}^{4}} 
%\nonumber \\
%& &\hspace{2mm} \times 
\bigg\{
g_{0,0} \hspace{25mm} \nonumber \\
& &\hspace{19mm}+ (\xi_{\bm{k}}\xi_{\bm{k}+\bm{q}}+\xi_{\bm{k}+\bm{Q}}\xi_{\bm{k}+\bm{q}+\bm{Q}}) \nonumber \\
& &\hspace{19mm}- 2( \Delta_{\bm{k},\sigma}^{m}\Delta_{\bm{k}+\bm{q},\sigma^{'}}^{m}-
\Delta_{\bm{k},\sigma}^{\chi}\Delta_{\bm{k}+\bm{q},\sigma^{'}}^{\chi})
\bigg\}, \nonumber \\
\\
%%%%%%%%%%%%%%%%%%%%%%%%%%%%%%%%%%%%%%%%%%%%%%%%%%%%%%%%%%%%%%%%%%%%%%%%%%%%%%%%%%%%%%%%%%%%%%%%%%%%%%%%%%%%%%%%%
& &F_{\tilde{\sigma}^{0},\tilde{\sigma}^{x}}^{{\sigma^{'},\sigma}}(\bm{k},\bm{q},\mathrm{i}\omega_{\ell}) 
%\nonumber \\
=
\frac{1}{\beta}\sum_{\mathrm{i}\omega_{n}}
\frac{1}{E_{u,l}^{4}} 
%\nonumber \\
%& &\hspace{2mm} \times 
\bigg\{g_{0,x} \nonumber \\
& &\hspace{6mm}-\Delta_{\bm{k},\sigma}^{\chi}(\xi_{\bm{k}+\bm{q}}-\xi_{\bm{k}+\bm{q}+\bm{Q}}) 
+\Delta_{\bm{k}+\bm{q},\sigma^{'}}^{\chi}(\xi_{\bm{k}}-\xi_{\bm{k}+\bm{Q}})
\bigg\}, \nonumber \\
\\
& &F_{\tilde{\sigma}^{x},\tilde{\sigma}^{0}}^{{\sigma^{'},\sigma}}(\bm{k},\bm{q},\mathrm{i}\omega_{\ell}) 
%\nonumber \\
=
\frac{1}{\beta}\sum_{\mathrm{i}\omega_{n}}
\frac{1}{E_{u,l}^{4}} 
%\nonumber \\
%& &\hspace{2mm} \times 
\bigg\{g_{0,x} \nonumber \\
& &\hspace{6mm}+\Delta_{\bm{k},\sigma}^{\chi}(\xi_{\bm{k}+\bm{q}}-\xi_{\bm{k}+\bm{q}+\bm{Q}}) 
-\Delta_{\bm{k}+\bm{q},\sigma^{'}}^{\chi}(\xi_{\bm{k}}-\xi_{\bm{k}+\bm{Q}})
\bigg\}, \nonumber \\
\\
%%%%%%%%%%%%%%%%%%%%%%%%%%%%%%%%%%%%%%%%%%%%%%%%%%%%%%%%%%%%%%%%%%%%%%%%%%%%%%%%%%%%%%%%%%%%%%%%%%%%%%%%%%%%%%%%%
& &F_{\tilde{\sigma}^{0},\tilde{\sigma}^{y}}^{{\sigma^{'},\sigma}}(\bm{k},\bm{q},\mathrm{i}\omega_{\ell}) 
%\nonumber \\
=
\frac{1}{\beta}\sum_{\mathrm{i}\omega_{n}}
\frac{1}{E_{u,l}^{4}} 
%\nonumber \\
%& &\hspace{2mm} \times 
\bigg\{g_{0,y} \nonumber \\
& &\hspace{2mm}-\mathrm{i}\Delta_{\bm{k},\sigma}^{m}(\xi_{\bm{k}+\bm{q}}-\xi_{\bm{k}+\bm{q}+\bm{Q}}) 
+\mathrm{i}\Delta_{\bm{k}+\bm{q},\sigma^{'}}^{m}(\xi_{\bm{k}}-\xi_{\bm{k}+\bm{Q}})
\bigg\}~, \nonumber
\end{eqnarray*}
\vfill\eject
\begin{eqnarray*}
& &F_{\tilde{\sigma}^{y},\tilde{\sigma}^{0}}^{{\sigma^{'},\sigma}}(\bm{k},\bm{q},\mathrm{i}\omega_{\ell})
%\nonumber \\
=
\frac{1}{\beta}\sum_{\mathrm{i}\omega_{n}}
\frac{1}{E_{u,l}^{4}} 
%\nonumber \\
%& &\hspace{2mm} \times 
\bigg\{g_{0,y} \nonumber \\
& &\hspace{2mm}+\mathrm{i}\Delta_{\bm{k},\sigma}^{m}(\xi_{\bm{k}+\bm{q}}-\xi_{\bm{k}+\bm{q}+\bm{Q}})  
-\mathrm{i}\Delta_{\bm{k}+\bm{q},\sigma^{'}}^{m}(\xi_{\bm{k}}-\xi_{\bm{k}+\bm{Q}})
\bigg\}~, \nonumber \\
\\
%%%%%%%%%%%%%%%%%%%%%%%%%%%%%%%%%%%%%%%%%%%%%%%%%%%%%%%%%%%%%%%%%%%%%%%%%%%%%%%%%%%%%%%%%%%%%%%%%%%%%%%%%%%%%%%%%%%%
& &F_{\tilde{\sigma}^{0},\tilde{\sigma}^{z}}^{{\sigma^{'},\sigma}}(\bm{k},\bm{q},\mathrm{i}\omega_{\ell}) 
%\nonumber \\
=
\frac{1}{\beta}\sum_{\mathrm{i}\omega_{n}}
\frac{1}{E_{u,l}^{4}} 
%\nonumber \\
%& &\hspace{2mm} \times 
\bigg\{
g_{0,z} \hspace{25mm} \nonumber \\
& &\hspace{19mm}+  2( \Delta_{\bm{k},\sigma}^{m}\Delta_{\bm{k}+\bm{q},\sigma^{'}}^{\chi}- 
\Delta_{\bm{k},\sigma}^{\chi}\Delta_{\bm{k}+\bm{q},\sigma^{'}}^{m})
\bigg\}~, \nonumber \\
\\
& &F_{\tilde{\sigma}^{z},\tilde{\sigma}^{0}}^{{\sigma^{'},\sigma}}(\bm{k},\bm{q},\mathrm{i}\omega_{\ell}) 
%\nonumber \\
=
\frac{1}{\beta}\sum_{\mathrm{i}\omega_{n}}
\frac{1}{E_{u,l}^{4}} 
%\nonumber \\
%& &\hspace{2mm} \times 
\bigg\{
g_{0,z} \hspace{25mm} \nonumber \\
& &\hspace{19mm}- 2( \Delta_{\bm{k},\sigma}^{m}\Delta_{\bm{k}+\bm{q},\sigma^{'}}^{\chi}- 
\Delta_{\bm{k},\sigma}^{\chi}\Delta_{\bm{k}+\bm{q},\sigma^{'}}^{m})
\bigg\}~, \nonumber \\
\\
& &F_{\tilde{\sigma}^{x},\tilde{\sigma}^{y}}^{{\sigma^{'},\sigma}}(\bm{k},\bm{q},\mathrm{i}\omega_{\ell}) 
%\nonumber \\
=
\frac{1}{\beta}\sum_{\mathrm{i}\omega_{n}}
\frac{1}{E_{u,l}^{4}} 
%\nonumber \\
%& &\hspace{2mm} \times 
\bigg\{
g_{x,y} \hspace{25mm} \nonumber \\
& &\hspace{19mm}+2\mathrm{i}(\Delta_{\bm{k},\sigma}^{m}\Delta_{\bm{k}+\bm{q},\sigma^{'}}^{\chi}+\Delta_{\bm{k},\sigma}^{\chi}\Delta_{\bm{k}+\bm{q},\sigma^{'}}^{m})
\bigg\}~, \nonumber \\
\\
& &F_{\tilde{\sigma}^{y},\tilde{\sigma}^{x}}^{{\sigma^{'},\sigma}}(\bm{k},\bm{q},\mathrm{i}\omega_{\ell}) 
%\nonumber \\
=
\frac{1}{\beta}\sum_{\mathrm{i}\omega_{n}}
\frac{1}{E_{u,l}^{4}} 
%\nonumber \\
%& &\hspace{2mm} \times 
\bigg\{
-g_{x,y} \hspace{20mm} \nonumber \\
& &\hspace{16mm}+2\mathrm{i}(\Delta_{\bm{k},\sigma}^{m}\Delta_{\bm{k}+\bm{q},\sigma^{'}}^{\chi}+\Delta_{\bm{k},\sigma}^{\chi}\Delta_{\bm{k}+\bm{q},\sigma^{'}}^{m})
\bigg\}~, \nonumber \\
\\
& &F_{\tilde{\sigma}^{x},\tilde{\sigma}^{z}}^{{\sigma^{'},\sigma}}(\bm{k},\bm{q},\mathrm{i}\omega_{\ell}) 
%\nonumber \\
=
\frac{1}{\beta}\sum_{\mathrm{i}\omega_{n}}
\frac{1}{E_{u,l}^{4}} 
%\nonumber \\
%& &\hspace{2mm} \times 
\bigg\{g_{x,z} \hspace{25mm} \nonumber \\
& &\hspace{5mm} -\Delta_{\bm{k},\sigma}^{m}(\xi_{\bm{k}+\bm{q}}-\xi_{\bm{k}+\bm{q}+\bm{Q}})
- \Delta_{\bm{k}+\bm{q},\sigma^{'}}^{m}(\xi_{\bm{k}}-\xi_{\bm{k}+\bm{Q}}) 
\bigg\}~, \nonumber 
\end{eqnarray*}
\vfill\eject
\begin{eqnarray*}
& &F_{\tilde{\sigma}^{z},\tilde{\sigma}^{x}}^{{\sigma^{'},\sigma}}(\bm{k},\bm{q},\mathrm{i}\omega_{\ell}) 
%\nonumber \\
=
\frac{1}{\beta}\sum_{\mathrm{i}\omega_{n}}
\frac{1}{E_{u,l}^{4}} 
%\nonumber \\
%& &\hspace{2mm} \times 
\bigg\{-g_{x,z} \hspace{20mm} \nonumber \\
& &\hspace{5mm} -\Delta_{\bm{k},\sigma}^{m}(\xi_{\bm{k}+\bm{q}}-\xi_{\bm{k}+\bm{q}+\bm{Q}})
- \Delta_{\bm{k}+\bm{q},\sigma^{'}}^{m}(\xi_{\bm{k}}-\xi_{\bm{k}+\bm{Q}}) 
\bigg\}~, \nonumber \\
\\
& &F_{\tilde{\sigma}^{y},\tilde{\sigma}^{z}}^{{\sigma^{'},\sigma}}(\bm{k},\bm{q},\mathrm{i}\omega_{\ell}) 
%\nonumber \\
=
\frac{1}{\beta}\sum_{\mathrm{i}\omega_{n}}
\frac{1}{E_{u,l}^{4}} 
%\nonumber \\
%& &\hspace{2mm} \times 
\bigg\{
g_{y,z} \nonumber \\
& &\hspace{4mm} -\mathrm{i} 
\Delta_{\bm{k},\sigma}^{\chi}(\xi_{\bm{k}+\bm{q}}-\xi_{\bm{k}+\bm{q}+\bm{Q}})
-\mathrm{i}\Delta_{\bm{k}+\bm{q},\sigma^{'}}^{\chi}(\xi_{\bm{k}}-\xi_{\bm{k}+\bm{Q}})
\bigg\}~, \nonumber \\
\\
& &F_{\tilde{\sigma}^{z},\tilde{\sigma}^{y}}^{{\sigma^{'},\sigma}}(\bm{k},\bm{q},\mathrm{i}\omega_{\ell}) 
%\nonumber \\
=
\frac{1}{\beta}\sum_{\mathrm{i}\omega_{n}}
\frac{1}{E_{u,l}^{4}} 
%\nonumber \\
%& &\hspace{2mm} \times 
\bigg\{
-g_{y,z} \nonumber \\
& &\hspace{4mm} -\mathrm{i} 
\Delta_{\bm{k},\sigma}^{\chi}(\xi_{\bm{k}+\bm{q}}-\xi_{\bm{k}+\bm{q}+\bm{Q}})
-\mathrm{i}\Delta_{\bm{k}+\bm{q},\sigma^{'}}^{\chi}(\xi_{\bm{k}}-\xi_{\bm{k}+\bm{Q}})
\bigg\}~. \nonumber
\end{eqnarray*}

%%%%%%%%%%%%%%%%%%%%%%%%%%%%%%%%%%%%%%%%%%%%%%%%%%%%%%%%%%%%%%%%%%%%%%%%%%%%%%%%%%%%%%%%%%%%%%%%%%%%%%%%%%%%%%%%%%%%%%%%%%%%%%%%%%%%%%
%%%%%%%%%%%%%%%%%%%%%%%%%%%%%%%%%%%%%%%%%%%%%%%%%%%%%%%%%%%%%%%%%%%%%%%%%%%%%%%%%%%%%%%%%%%%%%%%%%%%%%%%%%%%%%%%%%%%%%%%%%%%%%%%%%%%%%

\section{Existence of a Goldstone mode}
\label{Appen-Haldane}

In  this appendix we examine the contribution to the second order action ${\mathcal S}_{2}$ given by $m_{\bm{q}+\bm{Q}}^{x}$ and $m_{\bm{q}+\bm{Q}}^{y}$ in the half filled, uniform and static limit
\begin{eqnarray}
\mathrm{x} = 0, \ \bm{q} = \bm{0}, \ \mathrm{i}\omega_{\ell} = 0.
\label{Golstone-limit}
\end{eqnarray}
We write it as a part of ${\mathcal S}_{2}$ (\ref{Result_S2}) 
\begin{widetext} 
\begin{eqnarray}
& &\hspace{1.5mm} {\mathcal S}_{2}^{m_{x,y}} 
%\nonumber \\
%& &
=\hspace{1.5mm}\Bigg({\mathcal M}_{5,5}(\bm{q}=\bm{0},\bm{Q},\bm{Q},\mathrm{i}\omega_{\ell}=0) 
%\nonumber \\
%\hspace{6.5mm}
+\frac{1}{2} {\Pi}_{5,5}(\bm{q}=\bm{0},\bm{Q},\bm{Q},\mathrm{i}\omega_{\ell}=0)\Bigg)
\big[ \delta m_{\bm{Q}}^{x}(\mathrm{i}\omega_{\ell}=0) \big]^{2} \nonumber \\
& &\hspace{14.0mm}+ \Bigg({\mathcal M}_{6,6}(\bm{q}=\bm{0},\bm{Q},\bm{Q},\mathrm{i}\omega_{\ell}=0) 
%\nonumber \\
%& &\hspace{6.5mm}
+\frac{1}{2} {\Pi}_{6,6}(\bm{q}=\bm{0},\bm{Q},\bm{Q},\mathrm{i}\omega_{\ell}=0)\Bigg)
\big[ \delta m_{\bm{Q}}^{y}(\mathrm{i}\omega_{\ell}=0) \big]^{2} \nonumber \\
% \end{eqnarray}
% \begin{eqnarray}
& &\hspace{12.0mm}=~\alpha \tilde{J}_{\bm{q}=\bm{0}}\big(\delta m_{\bm{Q}}^{x}\big)^{2} + \alpha \tilde{J}_{\bm{q}=\bm{0}}\big(\delta m_{\bm{Q}}^{y}\big)^{2} 
%\nonumber \\
%& &\hspace{16.0mm}
+\frac{1}{2{\mathrm{N_{s}}}} \sum_{\bm{k}}~{'} \sum_{\sigma,\sigma^{'}}
\Big[-\alpha \tilde{J}_{\bm{q}=\bm{0}}  \sigma_{\sigma^{'},\sigma}^{x}\Big]
F_{\tilde{\sigma}^{x},\tilde{\sigma}^{x}}^{{\sigma^{'},\sigma}} 
%\nonumber \\
%& &\hspace{27.5mm} \times
\Big[-\alpha \tilde{J}_{\bm{q}=\bm{0}}  \sigma_{\sigma,\sigma^{'}}^{x}\Big]  \big(\delta m_{\bm{Q}}^{x}\big)^{2} \nonumber \\
& &\hspace{16.0mm}+\frac{1}{2{\mathrm{N_{s}}}} \sum_{\bm{k}}~{'} \sum_{\sigma,\sigma^{'}}
\Big[-\alpha \tilde{J}_{\bm{q}=\bm{0}}  \sigma_{\sigma^{'},\sigma}^{y}\Big] 
F_{\tilde{\sigma}^{x},\tilde{\sigma}^{x}}^{{\sigma^{'},\sigma}} 
%\nonumber \\
%& &\hspace{27.5mm}\times 
\Big[-\alpha \tilde{J}_{\bm{q}=\bm{0}}  \sigma_{\sigma,\sigma^{'}}^{y}\Big]  \big(\delta m_{\bm{Q}}^{y}\big)^{2}.
\end{eqnarray}
\end{widetext}
By using Eqs.\! (\ref{DefMatrixM}) and (\ref{Def_PiMatrix}) we get in the simplified limit (\ref{Golstone-limit})
\begin{eqnarray}
{\mathcal S}_{2}^{m_{x,y}} &=& \left(2 \alpha J - \frac{1}{{\mathrm{N_{s}}}}
\sum_{\bm{k}}~{'}\frac{8\alpha^{2}J^{2}}{E_{\bm{k}}^{up}-E_{\bm{k}}^{low}}\right) 
\nonumber \\
& &\hspace{0.5mm}\times
\Big[\big(\delta m_{\bm{Q}}^{x}\big)^{2}+\big(\delta m_{\bm{Q}}^{y}\big)^{2} \Big].
\label{Goldstone-S2AF}
\end{eqnarray}
The mean field equation (\ref{MagneAF-MFEq}) related to the magnetization gives
\begin{eqnarray}
\frac{1}{{\mathrm{N_{s}}}}\sum_{\bm{k}}~{'}\frac{8\alpha^{2}J^{2}}{E_{\bm{k}}^{up}-E_{\bm{k}}^{low}}=2 \alpha J,
\hspace{19mm} & & \nonumber \\
\mbox{then with Eq.\! (\ref{Goldstone-S2AF}) we have finally: ${\mathcal S}_{2}^{m_{x,y}} = 0$.} 
\hspace{9mm} & & \nonumber
\end{eqnarray}

\vfill\eject
The behaviour of the second order action concerning the $x$ and $y$ directions of the magnetization implies that the transverse spin-spin correlation function $\chi^{\pm}$ contains a gapless pole. It is a consequence of the Goldstone theorem, \cite{Goldstone-Th} which has to be applied in the present case because of the rotational symmetry breaking in spin space due to the imposed AF order. It is in agreement with the effective field theory of quantum antiferromagnets, which was built by Haldane for one-dimensional chains \cite{Haldane1983} and extended later to square lattice systems. \cite{AF2D} The existence of this Goldstone mode has already been observed in the original spin-bag approach. \cite{Schrieffer1989,Schrieffer1989long} 

%%%%%%%%%%%%%%%%%%%%%%%%%%%%%%%%%%%%%%%%%%%%%%%%%%%%%%%%%%%%%%%%%%%%%%%%%%%%%%%%%%%%%%%%%%%%%%%%%%%%%%%%%%%%%%%%%%%%%%%%%%%%%%%%%%%%%%
%%%%%%%%%%%%%%%%%%%%%%%%%%%%%%%%%%%%%%%%%%%%%%%%%%%%%%%%%%%%%%%%%%%%%%%%%%%%%%%%%%%%%%%%%%%%%%%%%%%%%%%%%%%%%%%%%%%%%%%%%%%%%%%%%%%%%%

%%%%%%%%%%%%%%%%%%%%%%%%%%%%%%%%%%%%%%%%%%%%%%%%%%%%%%%%%%%%%%%%%%%%%%%%%%%%%%%%%%%%%%%%%%%%%%%%%%%%%%%%%%%%%%%%%%%%%%%%%%%%%%%%%%%%%%
%%%%%%%%%%%%%%%%%%%%%%%%%%%%%%%%%%%%%%%%%%%%%%%%%%%%%%%%%%%%%%%%%%%%%%%%%%%%%%%%%%%%%%%%%%%%%%%%%%%%%%%%%%%%%%%%%%%%%%%%%%%%%%%%%%%%%%


\begin{thebibliography}{90}

\bibitem{Bednorz1986}
  J.G. Bednorz, and K.A. Muller, Z. Phys. B \textbf{64}, 189 
(1986). 

\bibitem{RMP} 
For a review see P.A. Lee, N. Nagaosa, and X.-G.Wen, 
Rev. Mod. Phys. \textbf{78}, 17 (2006).


\bibitem{Schrieffer1989}
J.R. Schrieffer, X.-G. Wen, and S.-C. Zhang, 
Phys. Rev. Lett. \textbf{60}, 944 (1988).

\bibitem{Schrieffer1989long}
J.R. Schrieffer, X.-G. Wen, and S.-C. Zhang, 
Phys. Rev. B \textbf{39}, 11663 (1989).

\bibitem{Anderson1987}
  P.W. Anderson, Science \textbf{235}, 1196 (1987).

\bibitem{Hsu1990}
 T.C. Hsu, Phys. Rev. B \textbf{41}, 11379 (1990).

\bibitem{tjhole}
T.K. Lee, C.-M. Ho, and N. Nagaosa, Phys. Rev. Lett. \textbf{90}, 067001 (2003).

\bibitem{ogatahimeda}
A. Himeda and M. Ogata, Phys. Rev. B \textbf{60}, R9935 (1999). 

\bibitem{tjexact}
P.W. Leung, and Y.F. Cheng, Phys. Rev. B \textbf{69}, 180403 (2004).


\bibitem{Shen-PRB2003}
F. Ronning, T. Sasagawa, Y. Kohsaka, K.M. Shen, A. Damascelli, 
C. Kim, T. Yoshida, N.P. Armitage, D.H. Lu, D.L. Feng, L.L. Miller, H. Takagi, and Z.-X. Shen,
Phys. Rev. B \textbf{67}, 165101 (2003).
 

\bibitem{Shen-JPSJ2003}
Y. Kohsaka, T. Sasagawa, F. Ronning, T. Yoshida, C. Kim, T. Hanaguri, 
M. Azuma, M. Takano, Z.-X. Shen, and H. Takagi,
 J. Phys. Soc. Jpn. \textbf{72}, 1018 (2003).
 

\bibitem{note}The superconductivity occurs through 
the pairing of holes in the small pockets. 

\bibitem{Rice1988}
  F.C. Zhang, and T.M. Rice, Phys. Rev. B \textbf{37}, 3759 
(1988).

\bibitem{Fukuyama1993}
T. Tanamoto, H. Kohno, and H. Fukuyama,
J. Phys. Soc. Jpn. \textbf{62}, 717 (1993).

\bibitem{ARPES-1998}
C. Kim, P.J. White, Z.-X. Shen, T. Tohyama, Y. Shibata, 
S. Maekawa, B.O. Wells, Y.J. Kim, R.J. Birgeneau, and M.A. Kastner,
Phys. Rev. Lett. \textbf{80}, 4245 (1998).

\bibitem{Barnes1976}
  S.E. Barnes, J. Phys. F \textbf{6}, 1375 (1976); J. 
Phys. F \textbf{7}, 2637 (1977).

\bibitem{Coleman1984}
  P. Coleman, Phys. Rev. B \textbf{29}, 3035 (1984).

\bibitem{AndersontSL}
  Z. Zou, and P.W. Anderson, Phys. Rev. B \textbf{37}, 627 (1988).

\bibitem{Fukuyama1988}
  Y. Suzumura, Y. Hasegawa, and H. Fukuyama, J. Phys. Soc. Jpn.
\textbf{57}, 2768 (1988).

\bibitem{Lavagna1994}
  G. Stemmann, \textit{Ph.D Thesis}, Universit\'e Joseph Fourier, Grenoble, 
1994 (unpublished) ; G. Stemmann,  C. P\'epin, and M. Lavagna, Phys. Rev. B
\textbf{50}, 4075 (1994).

\bibitem{KivRokSet}
  S.A. Kivelson, D.S. Rokhsar, and J.A. Sethna, 
Phys. Rev. B \textbf{35}, 8865 (1987).

\bibitem{AndersonPRL1990}
  P.W. Anderson, Phys. Rev. Lett.
\textbf{64}, 1839 (1990).

\bibitem{Affleck1989}
I. Affleck, and J.B. Marston, Phys. Rev. B \textbf{37}, 3774 
(1988).

 \bibitem{Marston1989}
J.B. Marston, and I. Affleck, Phys. Rev. B \textbf{39}, 11538 
(1989).

\bibitem{Strato}
  R.L. Stratonovitch, Sov. Phys. Dokl. \textbf{2}, 416 
(1958). 

\bibitem{Hub}
  J. Hubbard, Phys. Rev. Lett. \textbf{3}, 77  (1959).

\bibitem{RiWieg1989}
  Y. Hasegawa, P. Lederer, T.M. Rice, and P.B. Wiegmann, Phys. 
Rev. Lett. \textbf{63}, 907 (1989).

\bibitem{Bogo-trans}
N.N. Bogoliubov, Nuovo Cimento \textbf{7}, 794 (1958).

\vfill\eject

\bibitem{Val-trans}
J. Valatin, Nuovo Cimento \textbf{7}, 843 (1958).

\bibitem{QFTPAnomaly}
  A.N. Redlich, Phys. Rev. Lett. \textbf{52}, 18 (1984); Phys. Rev. D \textbf{29}, 2366 (1984).

\bibitem{Fukuyama1992}
 H. Matsukawa, and H. Fukuyama, J. Phys. Soc. Jpn. \textbf{61}, 1882 (1992).

\bibitem{HsuAffl-1991}
 T.C. Hsu, J.B. Marston, and I. Affleck, Phys. Rev. B \textbf{43}, 2866 (1991).

\bibitem{BCS}
   L.N. Cooper, Phys. Rev. \textbf{104}, 1189 (1956);
   J. Bardeen, L.N. Cooper, and J.R. Schrieffer, Phys. Rev. 
\textbf{108}, 1175 (1957).

\bibitem{Schrief-book}
J.R. Schrieffer, \textit{Theory of Superconductivity, revised prin\-ting}
(Advanced Book Classics, Perseus Books, Reading, Massachusetts, 1999).

\bibitem{Nagaosa_books}
  N. Nagaosa, \textit{Quantum Field Theory in Condensed Matter 
Physics}, chap. 5 (Springer-Verlag, Berlin, 1999); \textit{Quantum Field Theory 
in Strongly Correlated Electronic Systems}, chap. 3 (Springer-Verlag, Berlin, 
1999).

\bibitem{Kubo_book}
  R. Kubo, M. Toda, N. Hashitsume, \textit{Statistical Physics II - 
Nonequilibrium Statistical Mechanics} (Springer-Verlag, Berlin, 1998).

\bibitem{Eliashberg}
  G.M. Eliashberg, J. Exptl. Theoret. Phys. (USSR)
\textbf{38}, 966 (1960), translated in Soviet Phys. JETP \textbf{11}, 
696 (1960).

\bibitem{Nernst}
Y. Wang, Z.A. Xu, T. Kakeshita, S. Uchida,
S. Ono, Y. Ando, and N.P. Ong, 
Phys. Rev. B \textbf{64}, 224519 (2001), and references therein.

\bibitem{Tsuei-RMP2000}
C.C. Tsuei, and J.R. Kirtley, Rev. Mod. Phys. \textbf{72}, 969 (2000).

\bibitem{Tsuei-Nat1997}
C.C. Tsuei, J.R. Kirtley, Z.F. Ren, J.H. Wang, H. Raffy, and Z.Z. Li, 
Nature \textbf{387}, 481 (1997).

\bibitem{moriya}
For a review see for example T. Moriya and K. Ueda, Rep. Prog. Phys. \textbf{66}, 1299 (2003).

\bibitem{SiResa-PRB1990}
A. Singh, and Z. Te\v{s}anovi\'{c}, Phys. Rev. B \textbf{41}, 614 (1990);
 \textit{ibid.} \textbf{41}, 11604 (1990).

\bibitem{KimLee-AnnPhys1999}
D.H. Kim, and P.A. Lee, 
Ann. Phys. (N.Y.) \textbf{272}, 130 (1999).

\bibitem{HonLee-PRL2003}
C. Honerkamp, and P.A. Lee, 
Phys. Rev. Lett. \textbf{90}, 246402 (2003).

\bibitem{NagaosaPhonons}
S. Ishihara, and N. Nagaosa, Phys. Rev. B \textbf{69}, 144520 (2004).  

\bibitem{ZeyPhon}
R. Zeyher, and A. Greco, Phys. Rev. B \textbf{64}, 140510 (2001).  

\bibitem{elph}
A. Lanzara, , P.V. Bogdanov, X.J. Zhou, S.A. Kellar, D.L. Feng, E.D. Lu, T. Yoshida, H. Eisaki, A. Fujimori, K. Kishio, J.-I. Shimoyama,
 T. Noda, S. Uchida, Z. Hussain, and Z.-X. Shen, Nature (London) \textbf{412}, 510 (2001); 
Z.-X. Shen, A. Lanzara, S. Ishihara, and N. Nagaosa, Phil. Mag. B \textbf{82}, 1349 (2002).


\bibitem{Goldstone-Th}
J. Goldstone, Nuovo Cimento \textbf{19}, 154 (1961).

\bibitem{Haldane1983}
  F.D.M. Haldane, Phys. Lett. \textbf{93A}, 464 (1983); Phys. Rev. Lett. \textbf{50}, 1153 (1983).

\bibitem{AF2D}
  S. Chakravarty, B.I. Halperin, and D.R. Nelson, Phys. Rev. Lett. \textbf{60}, 1057 (1988);
Phys. Rev. B \textbf{39}, 2344 (1989).
  

\end{thebibliography}
\end{document}